\documentclass[11pt, prd, aps, showpacs, nofootinbib,superscriptaddress]{revtex4-1}

\usepackage{amsmath,amssymb}
\usepackage{graphicx}
\usepackage{textcomp}
\usepackage{enumitem}
\usepackage{graphicx}
\usepackage{dsfont}
\usepackage{bm}
\usepackage{xcolor}
\usepackage{subfig}
\usepackage{caption,placeins}
\usepackage{comment}
\usepackage{bbold}
\usepackage{cancel}

\newcommand{\be}{\begin{equation}}
\newcommand{\ee}{\end{equation}}
\newcommand{\beq}{\begin{equation}}
\newcommand{\eeq}{\end{equation}}
\newcommand{\bea}{\begin{eqnarray}}
\newcommand{\eea}{\end{eqnarray}}
\newcommand{\ket}[1]{\ensuremath{| {#1} \rangle }}
\newcommand{\bra}[1]{\ensuremath{\langle {#1} |}}

\newcommand{\beqn}{\begin{eqnarray}}   
\newcommand{\eeqn}{\end{eqnarray}}

\newcommand{\dthree}{d^{\hspace{1pt}3}}
\newcommand{\dfour}{d^{\hspace{1.35pt}4}}

\renewcommand{\vec}[1]{\bm{#1}}

\begin{document}

\title{First lattice calculation of radiative leptonic decay rates of pseudoscalar mesons}
%chrischrisendxxxxxxxxxxxxxxxxxxxxxxxxxxxxxxxxxxxxxxxxxxxxxxxxx
\newcommand{\Romatre}{Dipartimento di Fisica, Universit\`a  Roma Tre and INFN, Sezione di Roma Tre, Via della Vasca Navale 84, I-00146 Rome, Italy}
\newcommand{\RomatreINFN}{Istituto Nazionale di Fisica Nucleare, Sezione di Roma Tre,\\ Via della Vasca Navale 84, I-00146 Rome, Italy}
\newcommand{\Sissa}{SISSA, Via Bonomea 265, I-34136, Trieste,  and INFN Sezione di Roma La Sapienza Piazzale Aldo Moro 5, 00185 Roma, Italy}
\newcommand{\soton}{Department of Physics and Astronomy, University of Southampton,  Southampton SO17 1BJ, UK}
\newcommand{\Romadue}{Dipartimento di Fisica and INFN, Universit\`a di Roma ``Tor Vergata", Via della Ricerca Scientifica 1, I-00133 Roma, Italy}
\newcommand{\LaSapienza}{Physics Department and INFN Sezione di Roma La Sapienza Piazzale Aldo Moro 5, 00185 Roma, Italy}
\newcommand{\Regensburg}{Universit\"at Regensburg, Fakult\"at f\"ur Physik,  Universit\"atsstrasse 31,  93040 Regensburg, Germany}
\newcommand{\cern}{TH Department, CERN, CH-1211, Geneva 23, Switzerland}
\newcommand{\odense}{SDU eScience Center, University of Southern Denmark, Campusvej 55, DK-5230 Odense M, Denmark}
%\center{R.\,Frezzotti$^a$, D. Giusti$^b$}\\
%${}^a\!\!$
%{\em Laboratoire de Physique Th\'eorique, Universit\'e Paris Sud, \\
%           Centre d'Orsay, F-91405 Orsay-Cedex, France} \\
%\vspace{.3cm}
%${}^b\!\!$
%{\em Dipartimento di Fisica, Universit\`a di Roma ``La Sapienza'', } \\ 
% {\em          and INFN, Sezione di Roma, P.le A. Moro 2, I-00185 Rome, Italy} \\
%\vspace{.3cm}

%\begin{document}
%\title{Finite Volume QED Corrections to Hadronic Masses and Amplitudes in Lattice QCD}
\author{A.\,Desiderio}\affiliation{\Romadue} 
\author{R.\,Frezzotti}\affiliation{\Romadue} 
\author{M.\,Garofalo}\affiliation{\Romatre}
\author{D.\,Giusti}\affiliation{\Regensburg}\affiliation{\RomatreINFN}
\author{M.\,Hansen}\affiliation{\odense}
\author{V.\,Lubicz}\affiliation{\Romatre} 
\author{G.\,Martinelli}\affiliation{\LaSapienza}
\author{C.T.\,Sachrajda}\affiliation{\soton}
\author{F.\,Sanfilippo}\affiliation{\RomatreINFN}
\author{S.\,Simula}\affiliation{\RomatreINFN}
\author{N.\,Tantalo}\affiliation{\Romadue}

\begin{abstract}
We present a non-perturbative lattice calculation of the form factors which contribute to the amplitudes for the radiative decays $P\to \ell \bar \nu_\ell   \gamma$, where $P$ is a pseudoscalar meson and $\ell$ is a charged lepton.   Together with the non-perturbative determination of the corrections to the processes $P\to \ell \bar \nu_\ell$ due to the exchange of a virtual photon, this allows accurate predictions  at $O(\alpha_{em})$ to be made for  leptonic decay rates for pseudoscalar mesons ranging from the pion to the $D_s$ meson.  We are able to separate unambiguously and non-pertubatively  the point-like contribution, from the structure-dependent, infrared-safe, terms in the amplitude.   The fully non-perturbative $O(a)$ improved  calculation  of the inclusive leptonic decay rates will lead to the determination of the corresponding Cabibbo-Kobayashi-Maskawa  (CKM) matrix elements also at $O(\alpha_{em})$. Prospects for a precise evaluation of leptonic decay rates with emission of a hard photon are also very interesting, especially for the decays of heavy $D$ and $B$ mesons for which currently only model-dependent predictions are available to compare with existing experimental data.
\end{abstract}

\maketitle

%%%
%%%
\section{introduction}
%%%
%%%
The unitarity  of the CKM matrix is one of the most precise tests of the Standard Model.   Indeed, CKM unitarity may  rule out many theoretically well motivated models for new physics and put severe constraints on the energy scale where new phenomena might occur,  well beyond the range accessible to direct experimental searches. In this respect,   leptonic decay rates of light and heavy pseudoscalar mesons are essential ingredients for the extraction of the CKM matrix elements. A first-principles calculation of these quantities requires non-perturbative accuracy and hence numerical lattice simulations. Moreover, in order to fully exploit the presently  available experimental information and to perform the next generation of flavour-physics tests, $O(\alpha_{em})$ electromagnetic corrections must be included. In this endeavour, the radiative leptonic decays $P\to \ell \bar \nu_\ell (\gamma)$ (where $P$ is a negatively charged pseudoscalar meson,  $\ell$ a lepton,  $\bar \nu_\ell$ the corresponding anti-neutrino and  $\gamma$ a photon) are particularly important, see\,\cite{Tanabashi:2018oca}. 

Knowledge of the radiative leptonic decay rate in the region of small (soft) photon energies is required in order to properly define the infrared-safe measurable decay rate for the process $P\to \ell \bar \nu_\ell (\gamma)$. Indeed, according to the well-known Bloch-Nordsieck mechanism\,\cite{Bloch:1937pw}, the integral of the radiative decay rate in the phase space region corresponding to soft photons must be added to the decay rate with no real photons in the final states (the so-called virtual rate) in order to cancel infrared divergent contributions appearing in unphysical quantities at intermediate stages of the calculations. 

On the one hand, in the limit of ultra-soft photon energy the radiative decay rate can be reliably calculated in an effective theory in which the meson is treated as a point-like particle. 
This is a manifestation of the well-known mechanism known as the ``universality of infrared divergences'' (see for example Ref.\,\cite{Weinberg:1995mt,Low:1954kd}) that finds its physical explanation in the fact that ultra-soft photons cannot resolve the internal structure of the meson. 
On the other hand, the ultra-soft limit is an idealisation and experimental measurements, particularly in the case of heavy mesons, are inclusive up to photon energies that may be too large to safely neglect the Structure-Dependent (SD) corrections to the point-like approximation.    

In the region of hard (experimentally detectable) photon energies, radiative leptonic decays represent important probes of the internal structure of the mesons. Moreover, radiative decays can provide independent determinations of CKM matrix elements with respect to the purely leptonic channels. 
A non-perturbative calculation of the radiative decay rates can be particularly important for heavy mesons  since, 
unlike the case of pions and kaons where such decays have been studied using Chiral Perturbation Theory (ChPT)\,\cite{Bijnens:1996wm,Geng:2003mt,Mateu:2007tr,Unterdorfer:2008zz,Cirigliano:2011ny}, no model-independent calculations have ever been performed. Even in the case of light mesons, although the quoted ChPT calculations represent a first-principles approach to the problem, the low-energy constants entering in the final results at $O(p^6)$ have been estimated in phenomenological analyses relying in part on model-dependent assumptions.          

In Ref.\,\cite{Carrasco:2015xwa}  a strategy to compute QED radiative corrections to the $P\to \ell \bar \nu_\ell (\gamma)$ decay rates at $O(\alpha_{em})$ by starting from first-principles lattice calculations was proposed. The strategy has subsequently been applied in Refs.\,\cite{Lubicz:2016xro,Lubicz:2016mpj,Tantalo:2016vxk,Giusti:2017dwk,DiCarlo:2019thl}, within the RM123 approach\,\cite{deDivitiis:2011eh,deDivitiis:2013xla},  to provide the first non-perturbative model-independent calculation of the decay rates $\pi^-\to \mu^- \bar \nu_\mu (\gamma)$ and  $K^-\to \mu^- \bar \nu_\mu (\gamma)$. In these calculations the real soft-photon contributions have been evaluated in the point-like effective theory and, using the ChPT results quoted above, the SD corrections have been estimated to be negligible for these processes (see\,\cite{Carrasco:2015xwa}). In the same phenomenological analysis it has been shown that the SD corrections might instead be relevant for the decays of pions and kaons into electrons. Moreover, by using the same single-pole dominance approximation as originally used in Ref.\,\cite{Becirevic:2009aq}, SD contributions have been estimated to be phenomenologically important for decays of heavy-flavour mesons. 

In this paper we present the first non-perturbative lattice calculation of the rates for the radiative decays $P\to \ell \bar \nu \gamma$, where $P$ is a pion, kaon, $D$ or $D_s$ meson. We use the $N_f=2+1+1$  gauge ensembles generated by the European Twisted Mass Collaboration (ETMC) and analysed for mesonic observables 
in Ref.\,\cite{Carrasco:2014cwa}. Preliminary results from this study were presented in Ref.\,\cite{deDivitiis:2019uzm}; 
the decays of bottom mesons will be studied in future papers. Note also that Kane et al. have presented preliminary results 
for the decays $D_s^+\to\ell^+\nu\gamma$ and $K^-\to\ell^-\bar\nu\gamma$, where $\ell^\pm$ represents the charged leptons and $\gamma$ is a hard photon with energy in the range of about 0.5-1\,GeV in Ref.\,\cite{Kane:2019jtj}. 

The plan of the remainder of this paper is as follows. In Section \ref{sec:general} we introduce the basic quantities  which enter in the amplitude for  the leptonic decay  of a pseudoscalar meson with the emission of a real photon; in particular we define the axial and vector form factors $F_A$ and $F_V$. We express the decay rates in terms of these quantities in 
Appendix\,\ref{sec:decayrateformulae}.  In Section\,\ref{sec:euclid} we describe the general strategy  that we followed  to extract the amplitudes from suitable Euclidean correlation functions and  discuss finite-time effects. 
The presence of discretisation effects which diverge at small photon momenta is demonstrated 
in Section\,\ref{sec:fsf} and Appendix\,\ref{sec:acorr}, together with a strategy for subtracting them non-perturbatively.   In Section\,\ref{sec:numerical} we present the numerical results for pions, kaons, $D$ and $D_s$ mesons.  Many formulae which are used in the paper are discussed  and derived in Appendices\,\ref{sec:decayrateformulae}-\ref{sec:acorr}. 
Finally, in Appendix\,\ref{app:numres} we present some of our numerical results, including the correlation matrices, in a way which we hope may be useful
to readers who wish to use them in phenomenological applications.

%%%
%%%
\section{Definition of the form factors}
\label{sec:general}
%%%
%%%
%
\begin{figure}[!t]
\begin{center}
\includegraphics[width=0.4\textwidth]{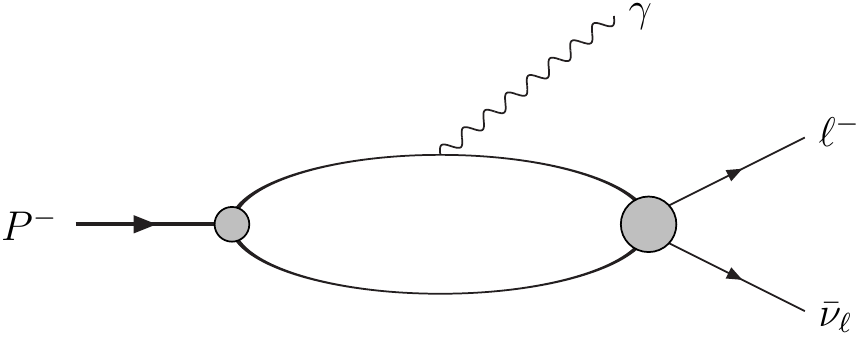}\hspace{0.5in}
\includegraphics[width=0.4\textwidth]{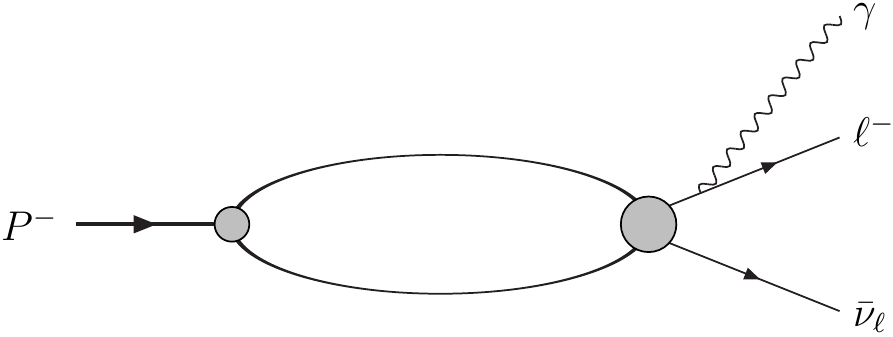}
\end{center}
\caption{
%\footnotesize 
{\it Feynman diagrams representing the amplitudes with the emission of a real photon from the $P^-$ meson (left panel) or from the final-state charged lepton $\ell^-$ (right panel). \hspace*{\fill} }
%}
%The  diagram corresponding to disconnected emissions from sea quarks is not shown.  
\label{fig:emissions}
}
\end{figure}
The non-perturbative contribution to the radiative leptonic decay rate for the
processes $P\to \ell \bar \nu_\ell \gamma$ is encoded in the
following hadronic matrix-element, see left panel of~Fig.\,\ref{fig:emissions},  
\begin{flalign}
H^{\alpha r}_W(k,\vec p)=\epsilon_\mu^r(k)\, H^{\alpha \mu}_W(k, \vec p)
=
\epsilon_\mu^r(k)\, \int \dfour y\, e^{ik\cdot y}\,  \mathtt{T}\bra{0} j_W^\alpha(0) j^\mu_{em}(y)\ket{P(\vec p)}\;,
\label{eq:starting}
\end{flalign}
where $\epsilon_\mu^r(k)$ is the polarisation vector of the outgoing  photon with four-momentum $k$, $\vec p$ is the momentum  of the ingoing pseudoscalar meson of mass $m_P$ ($p\equiv (E,\vec p)$, $E=\sqrt{m_P^2+\vec p^2}$ and $p^2=m_P^2$). The operators 
\begin{flalign}
j^\mu_{em}(x)=\sum_f q_f \bar \psi_f(x) \gamma^\mu \psi_f(x)\;,
\qquad
j_W^\alpha(x)=j_V^\alpha(x)-j_A^\alpha(x) = \bar \psi_U(x) \, (\gamma^\alpha - \gamma^\alpha \gamma_5) \, \psi_D(x)\;,
\end{flalign}
are respectively the electromagnetic hadronic current and the hadronic weak current expressed in terms of the quark fields $\psi_f$ having electric charge $q_f$ in units of the charge of the positron;  $\psi_{U}$ and $\psi_{D}$ indicate the fields of an up-type or a  down-type quark and for the mesons considered in this study $U$ can be either an up or a charm quark and $D$ a down or a strange quark. In order to calculate the full amplitude one has also to consider the contribution in which the photon is emitted from the final-state charged lepton, see the right panel of~Fig.\,\ref{fig:emissions}.  This latter contribution however, can be computed  in perturbation theory using the meson's decay constant $f_P$. Both contributions are included in the formulae for the decay rate given in appendix\,\ref{sec:decayrateformulae}.

The decomposition of $H^{\alpha r}_W(k,\vec p)$ in terms of scalar form factors has been discussed in Ref.\,\cite{Carrasco:2015xwa} (see also~\cite{Bijnens:1992en}). Here we adopt the same basis used in that paper to write
\begin{flalign}
H^{\alpha r}_W(k,\vec p) &=\epsilon_\mu^r(k)\Bigg\{
H_1\,\left[k^2 g^{\mu\alpha}-k^\mu k^\alpha\right]
+
H_2\, \left[(p\cdot k-k^2)k^\mu-k^2(p-k)^\mu\right](p-k)^\alpha
\nonumber \\
\nonumber \\
&
-i\frac{F_V}{m_P}\varepsilon^{\mu\alpha\gamma\beta}k_\gamma p_\beta
+\frac{F_A}{m_P}\left[(p\cdot k-k^2)g^{\mu\alpha}-(p-k)^\mu k^\alpha\right]
\nonumber \\
\nonumber \\
&
+
f_P\left[g^{\mu\alpha}+\frac{(2p-k)^\mu(p-k)^\alpha}{2p\cdot k-k^2}\right]
\Bigg\}\;.
\label{eq:ffdef}
\end{flalign}
The term in the last line of Eq.\,(\ref{eq:ffdef}), which we write as $H^{\alpha \mu}_{pt}(k,\vec p)$,
is the point-like infrared-divergent contribution.  The other terms correspond to the so called SD contribution, $H^{\alpha \mu}_{SD}(k,\vec p)$. $H^{\alpha \mu}_{pt}(k,\vec p)$ saturates the Ward Identity  (WI)  satisfied by $H^{\alpha \mu}_W(k,\vec p)$
\begin{flalign} 
k_\mu\, H^{\alpha \mu}_W(k,\vec p)= k_\mu\,H^{\alpha \mu}_{pt}(k,\vec p)= i\bra{0}j_W^\alpha(0)\ket{P(\vec p)} = f_P\, p^\alpha\;, \qquad k_\mu\,H^{\alpha \mu}_{SD}(k,\vec p) =0\, ,
\label{eq:contWI}
\end{flalign} 
as explained in detail in Appendix\,\ref{sec:acorr}.
The four form factors $H_{1,2}$ and $F_{V,A}$ are scalar functions of Lorentz invariants, $m^2_P$,  $p\cdot k$ and $k^2$.   Eq.\,(\ref{eq:ffdef}) is valid for generic (off-shell) values of the photon momentum and for generic choices of the polarisation vectors. The knowledge of  the four form factors in the case of off-shell photons ($k^2\neq 0$) gives access to the study of decays in which the pseudoscalar meson decays into four leptons. These processes are very interesting in the search of physics beyond the Standard Model and will be the subject of a future work.  In this paper we concentrate on the case in which the photon is on-shell. 

By setting $k^2 =0$,  at fixed meson mass,  the form factors are functions of  $p\cdot k$ only.  Moreover, by choosing a \emph{physical} basis for the polarisation vectors so that
%, i.e. such that  (see Eqs.\,(\ref{eq:pol1}) and Eqs.\,(\ref{eq:pol2}))
%
\begin{flalign}
\epsilon_r(\vec k)\cdot k=0\;,
\end{flalign}
one has
\begin{flalign}
H^{\alpha r}_W(k,\vec p) =
\epsilon^r_\mu(\vec k)\Bigg\{
-i\frac{F_V}{m_P}\varepsilon^{\mu\alpha\gamma\beta}k_\gamma p_\beta
+
\left[\frac{F_A}{m_P}+\frac{f_P}{p\cdot k}\right]
\left(p\cdot k\, g^{\mu\alpha}-p^\mu k^\alpha\right)
+
\frac{f_P}{p\cdot k}\, p^\mu p^\alpha
\Bigg\}\;.
\end{flalign}
Once the decay constant $f_P$ and the two SD axial and vector form factors $F_A$ and $F_V$ are known, the radiative decay rate can be calculated by using the formulae given in appendix\,\ref{sec:decayrateformulae}.  These formulae  are  expressed in terms of the convenient dimensionless variable
\begin{flalign} 
x_\gamma= \frac{2p\cdot k}{m_P^2}\;
\qquad {\rm with} \qquad 
0\le x_\gamma \le 1-\frac{m_\ell^2}{m_P^2}\;,\label{eq:xgamma}
\end{flalign} 
where $m_\ell$ is the mass of the outgoing lepton in the $P\to \ell  \bar \nu_\ell \gamma$ decay.

Our definition of the form factor $F_A$ differs from the definition, $F_A^{\rm B}$,   of refs.\,\cite{Beneke:2018wjp,Kane:2019jtj}
\beq F_A^{\rm B} = F_A + \frac{m_P\, f_P}{p\cdot k} \, . \eeq
We note that $F_A^{\rm B}$   includes the point-like infrared divergent  contribution  which totally dominates at low values of $x_\gamma$ thus obscuring the interesting   structure-dependent contribution. For this reason we strongly advocate the use of our definition\,\cite{Carrasco:2015xwa}. Moreover the sign of $F_V$ used in this paper is opposite to the one used in Ref.\,\cite{Kane:2019jtj}.
%%%
%%%
\section{Form  factors from Euclidean correlation functions }
\label{sec:euclid}
%%%
%%%
In order to relate the hadronic matrix element to Euclidean correlation functions, the primary quantities computed in lattice calculations, it is useful to express the $H^{\alpha r}_W(k, \vec p)$, defined in Eq.\,(\ref{eq:starting}) in Minkowski space in terms of the contributions coming from the different time orderings. To this end, we define
\begin{flalign}
H^{\alpha r}_W(k,\vec p)=H^{\alpha r}_{W,1}(k, \vec p)+H^{\alpha r}_{W,2}(k,\vec p)\;,
\qquad
j^r(\vec k)=   \epsilon^r_\mu(\vec k)\, \int \dthree y\, e^{-i\vec k\cdot \vec y}\, j^\mu_{em}(0,\vec y)\;,
\end{flalign}
and perform the $t_y$ integral,
\bea
H^{\alpha r}_{W,1}(k,\vec p)
&=&
\int_{-\infty}^0 dt_y\, e^{iE_\gamma t_y}\, \bra{0} j_W^\alpha(0) e^{i(\hat H-E-i\varepsilon)t_y} j^r(\vec k)\ket{P(\vec p)}
\nonumber \\ &=&-i\bra{0} j_W^\alpha(0) \frac{1}{\hat H+E_\gamma-E-i\varepsilon}j^r(\vec k)\ket{P(\vec p)}\;,
\nonumber \\
H^{\alpha r}_{W,2}(k,\vec p)
&=&
\int_0^{\infty} dt_y\, e^{iE_\gamma t_y}\, \bra{0} j^r(\vec k) e^{-i(\hat H-i\varepsilon)t_y} j_W^\alpha(0) \ket{P(\vec p)}
\nonumber \\ &=&-i\bra{0} j^r(\vec k) \frac{1}{\hat H-E_\gamma-i\varepsilon}j_W^\alpha(0)\ket{P(\vec p)}\;,
\label{eq:mink}
\eea
where $\hat H$ is the QCD  Hamiltonian operator, $E=\sqrt{m_P^2+\vec{p}^2}$ is the energy of the decaying meson $P$ and  $E_\gamma=\vert \vec k\vert$ is  the energy of  the outgoing real photon.

The important observation that makes the lattice calculation possible by using standard effective-mass/residue techniques is that the integrals over $t_y$ appearing in the definition of $H_W^{\alpha r}(k,\vec p)$ can be Wick rotated to the Euclidean space  without encountering any obstruction. Such obstructions arise whenever  there are states propagating between the operators in the $\mathtt{T}$-products   that have energies smaller than the energy of the external states\,\cite{Maiani:1990ca}. This doesn't happen in our case. For this reason $H^{\alpha r}_{W,1,2}(k,\vec p)$ can be rewritten in terms of  Euclidean integrals, 
\begin{flalign}
H^{\alpha r}_{W,1}(k,\vec p)
&=
-i\int_{-\infty}^0 dt_y\, \bra{0} j_W^\alpha(0) e^{(\hat H+E_\gamma-E)t_y} j^r(\vec k)\ket{P(\vec p)}
\nonumber \\
\nonumber \\
H^{\alpha r}_{W,2}(k,\vec p)
&=
-i\int_0^{\infty} dt_y\,  \bra{0} j^r(\vec k) e^{-(\hat H-E_\gamma)t_y} j_W^\alpha(0) \ket{P(\vec p)}\;,
\label{eq:HW12}
\end{flalign}
both of which are convergent for physical (non-vanishing) photon energies. In Eqs.\,(\ref{eq:HW12}) and below $t_y$ is a Euclidean time variable. Indeed, the hadronic state of lowest energy that can propagate between the two currents is the pseudoscalar meson with spatial momentum $\vec p-\vec k$ (it appears in the time-ordering $H^{\alpha r}_{W,1}$) and we have
\begin{flalign}
\sqrt{m_P^2+(\vec p-\vec k)^2}+E_\gamma > \sqrt{m_P^2+\vec p^2}\;,
\qquad
\vert \vec k\vert \neq0\;.
\end{flalign}
As a consequence, $H_W^{\alpha r}(k,\vec p)$ can be rewritten as
\begin{flalign}
H^{\alpha r}_W(k,\vec p)=-i\int \dfour y\, e^{E_\gamma t_y-i\vec k\cdot \vec y}\, \epsilon_\mu^r(\vec k) \mathtt{T}\bra{0} j_W^\alpha(0) j^\mu_{em}(y)\ket{P(\vec p)}\;.
\label{eq:euctimecorr}
\end{flalign}
From this observation it follows that the hadronic matrix-element can be extracted from the Euclidean correlation functions 
\begin{flalign}
C^{\alpha r}_W(t;\vec k,\vec p)= -i\, 
\epsilon^r_\mu(\vec k)\, \int \dfour y\, \dthree \vec{x}\, e^{t_y E_\gamma-i\vec k \cdot \vec y+i\vec p\cdot \vec x}\, 
\mathtt{T} \bra{0}j_W^\alpha(t) j^\mu_{em}(y) P(0,\vec x) \ket{0}\;,
\label{eq:correlatorinf}
\end{flalign}
where $P=i\bar{\psi}_D\gamma_5\psi_U$ is a Hermitian pseudoscalar interpolating operator having the flavour quantum numbers of the incoming meson. In Eq.\,(\ref{eq:correlatorinf}),   using the translational invariance   of the correlation function,  we have moved the origin in time to the pseudoscalar source, $P(0, \vec x)$, and placed the weak current at $t$. 

In the large-$t$ limit one has
\begin{flalign}
R^{\alpha r}_W(t;\vec k,\vec p) 
&= \frac{2E}{e^{-t(E-E_\gamma)}\, \bra{P(\vec p)}P(0)\ket{0}}\, C^{\alpha r}_W(t;\vec k,\vec p)
=
H^{\alpha r}_W(k,\vec p) + \cdots
\label{eq:Rinf}
\end{flalign}
where the ellipsis represents the sub-leading exponentials. 

\begin{figure}[!t]
\begin{center}
\includegraphics[width=0.45\textwidth]{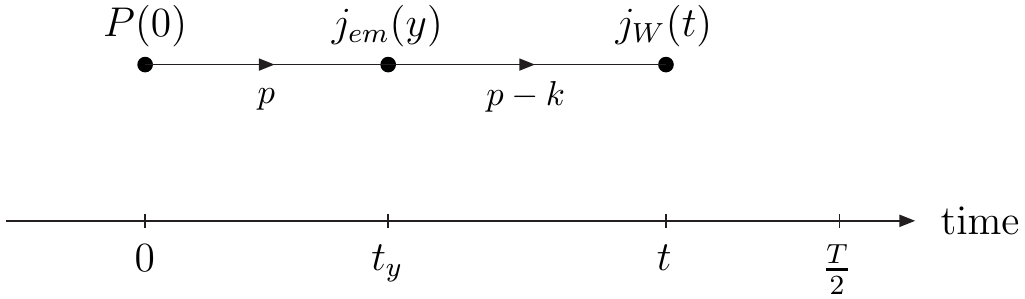}
\hspace{0.6in}
\includegraphics[width=0.45\textwidth]{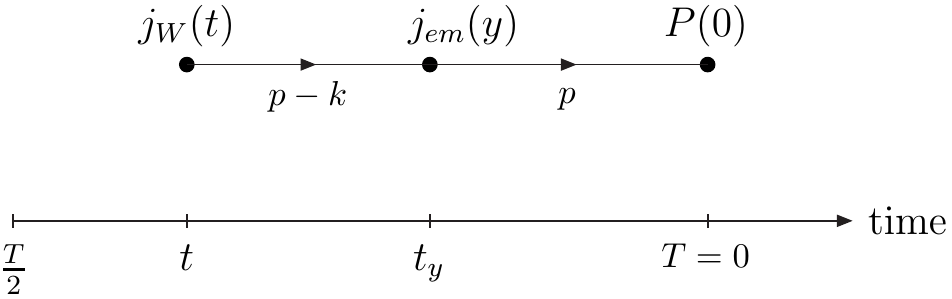}
\end{center}
%\footnotesize{
\caption{\it Schematic diagrams representing the correlation function $\mathcal{C}^{\alpha r}_W(t,T/2;\vec k,\vec p)$ used to extract the form factors, see Appendix\,\ref{sec:eucorr}.  The interpolating operator for the meson $P$ and the weak current $j_W$  are placed at fixed times 0 and $t$, and the electromagnetic current $j_{em}$ is inserted at $t_y$ which is integrated over $0 \le t_y\le T$, where $T$ is the temporal extent of the lattice. The left and right panels correspond to the leading contributions to the correlation functions for $t_y<T/2$ and $t_y>T/2$ respectively, with mesons propagating with momenta $p$ or $p-k$.  \hspace*{\fill}}
\label{fig:correlator}
%}
\end{figure}
The expressions for the correlation function $C^{\alpha r}_W(t;\vec k,\vec p)$ in Eq.\,(\ref{eq:correlatorinf}) and for the ratio $R^{\alpha r}_W(t;\vec k,\vec p)$ in Eq.\,(\ref{eq:Rinf}) refer to the ideal case of a lattice with infinite time-extent.  The extraction of the matrix elements from correlation functions computed on a finite lattice in our numerical simulations is discussed in Appendix\,\ref{sec:eucorr}.  Although some of the details of the appendix refer to our specific lattice procedures (the choice of lattice Fermions, renormalisation of the operators, etc.) the strategy itself is general and can be directly translated to other lattice  discretisations of QCD and  of QED.   Here in the main text,  we use Fig.\,\ref{fig:correlator}
 to illustrate the strategy used in our numerical simulations, performed with (anti-) periodic boundary conditions in time for the (fermionic) bosonic fields, to extract the form factors. 
 The two panels in Fig.\,\ref{fig:correlator} represent the \emph{forward} ($0\ll t\ll T/2$) and \emph{backward} ($T/2\ll t\ll T$) halves of the lattice. In both cases, the $t_y$ integral is dominated by the region in which $t_y$ is close to $t$, allowing for the propagation of the lightest state over the longest time interval.

\begin{figure}[!t]
\begin{center}
\includegraphics[width=0.35\textwidth]{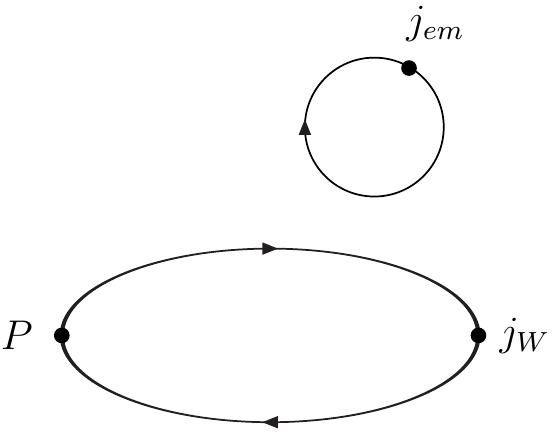}\hspace{0.8in}
\includegraphics[width=0.35\textwidth]{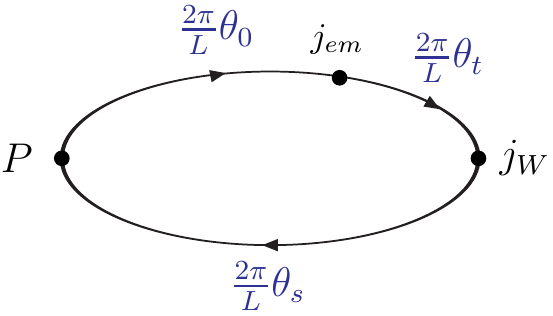}
\end{center}
%\footnotesize{
\caption{\it The diagram on the left represents the contributions to the correlation functions arising from the emission of the photon by the sea quarks. In our numerical simulations we work in the electroquenched approximation and neglect such diagrams. The diagram on the right explains our choice of the spatial boundary conditions, which allow us to set arbitrary values for the meson and photon spatial momenta.
The spatial momenta of the valence quarks, modulo $2\pi/L$, in terms of the twisting angles are as indicated. Each diagram implicitly includes all orders in QCD.
\hspace*{\fill}
%!TEX encoding = UTF-8 Unicode}
\label{fig:disctheta}}
\end{figure}

In Fig.\,\ref{fig:disctheta} we show two more diagrams to illustrate two important points concerning our numerical calculation of the correlation functions  and of the form factors. The diagram in the left panel shows a {\it quark-disconnected} contribution to the correlation function originating from the possibility that the external real photon is emitted from sea quarks. In this work we have been using the so-called electroquenched approximation in which the sea-quarks are electrically neutral. In practice this means that we have \emph{neglected} the contributions represented by the diagram in the left panel of Figure\,\ref{fig:disctheta}. 

The quark-connected diagram in the right panel of Figure\,\ref{fig:disctheta} is shown in order to explain the strategy we have used to set the values of the spatial momenta. We exploited the fact that, by working within the electroquenched approximation, i.e.\! in the absence of the contributions illustrated in the left panel of the figure, it is possible to choose arbitrary values of the spatial momenta by using different spatial boundary conditions for the quark fields\,\cite{deDivitiis:2004kq}. More precisely, we set the boundary conditions for the ``spectator'' quark such that $\psi(x+\vec{n} L)=\exp(2\pi i \vec{n} \cdot \vec \theta_s/L) \psi(x)$, where $L$ is the spatial extent of the lattice in each spatial direction. We treat the two propagators that are connected to the electromagnetic current as the results of the Wick contractions of two different fields having the same mass and electric charge but satisfying different boundary conditions\,\cite{Boyle:2007wg}. This is possible at the price of accepting tiny violations of unitarity that are exponentially suppressed with the volume. By setting the boundary conditions as illustrated in the figure we have thus been able to choose arbitrary (non-quantised) values for the meson and photon spatial momenta
\begin{flalign}
\vec p = \frac{2\pi}{L}\left( \vec \theta_0-\vec \theta_s\right)\;,
\qquad
\vec k = \frac{2\pi}{L}\left( \vec \theta_0-\vec \theta_t\right)\;,
\label{eq:momenta}
\end{flalign}
by tuning the real three-vectors $\vec \theta_{0,t,s}$. We find that the most precise results are obtained with small values of $|\vec{p}|$ and in particular with $\vec p=\vec 0$.  

The numerical results presented in the following sections have been obtained by setting the non-zero components of the spatial momenta along the third-direction, i.e.
\begin{flalign}\label{eq:momenta}
\vec p=(0,0,\vert \vec p\vert)\;,\qquad
\vec k=(0,0,E_\gamma)\;.
\end{flalign}
With this particular choice of the kinematical configuration, a convenient basis for the polarisation vectors of the photon (see Appendix\,\ref{sec:eucorr} for more details) is the one in which the \emph{two} physical polarisation vectors are given by
\begin{flalign}\label{eq:epsilons}
&
\epsilon_\mu^1=\left(0, -\frac{1}{\sqrt{2}},
-\frac{1}{\sqrt{2}}, 0\right)\;,
\qquad
\epsilon_\mu^2=\left(0, \frac{1}{\sqrt{2}},
-\frac{1}{\sqrt{2}}, 0\right)\;,
\end{flalign}
while the unphysical polarisation vectors vanish identically, $\epsilon_\mu^0=\epsilon_\mu^3=0$. Notice that in this basis we have
\begin{flalign}
\epsilon^r\cdot p =\epsilon^r\cdot k=0\;, 
\end{flalign}
and, consequently,
\begin{flalign}
%H^{ir}_A(k,p)=
%\epsilon_i^r\, p\cdot k\, \left[\frac{F_A}{m_P}+\frac{f_P}{p\cdot k}\right]\;,
%\qquad
H^{jr}_A(k,\vec p)=
\frac{\epsilon_j^r\, m_P}{2}\, x_\gamma\left[F_A+\frac{2f_P}{m_P x_\gamma}\right]\;,
\qquad
H^{jr}_V(k,\vec p)=\frac{i \left(E_\gamma\, \vec \epsilon^r \wedge \vec p-E\, \vec \epsilon^r \wedge \vec k\right)^j}{m_P}\, 
F_V\;.\label{eq:Hir}
\end{flalign}
Using these formulae, we have built the following numerical estimators
\bea   R_A(t) & =    &
\frac{1}{2 m_P}\sum_{r=1,2}\sum_{j=1,2} \frac{R^{j r}_A(t,T/2;\vec k,\vec p)}{\epsilon_j^r}\quad  
\to \,   x_\gamma F_A(x_\gamma)+\frac{2f_P}{m_P}  \,   ,
\label{eq:estimatorsAA} \eea
\bea R_V(t) &=&
\frac{m_P}{4}\sum_{r=1,2}\sum_{j=1,2} \frac{R^{jr}_V(t,T/2;\vec k,\vec p)}{
i \left(E_\gamma\, \vec \epsilon^r \wedge \vec p-E\, \vec \epsilon^r \wedge \vec k\right)^j}\quad  \to \, F_V(x_\gamma)\, ,  
\label{eq:estimators}
\eea
for the form factors, which we determine by fitting to the plateaux in the region $0\ll t \ll T/2$. The discussion here and below corresponds explicitly to the forward half of the lattice ($0\ll t \ll T/2$). We combine the results with those from the backward half ($T/2\ll t\ll T$) by exploiting time-reversal symmetry as explained in Appendix\,\ref{sec:eucorr}\,.

The ratios $R^{j r}_W(t,T/2;\vec k,\vec p)$ appearing in Eqs.\,(\ref{eq:estimatorsAA}) and (\ref{eq:estimators}), which we evaluate separately for the axial ($W=A$) and vector ($W=V$) components of the weak current, are the finite-$T$ generalisations (see Eq.\,(\ref{eq:plateaux})) of the ratios $R^{\alpha r}_W(t;\vec k,\vec p)$ defined above in Eq.\,(\ref{eq:Rinf}). The values of the meson energies and of the matrix elements $\bra{P}P\ket{0}$ needed to build these estimators have been obtained from standard effective-mass/residue analyses of pseudoscalar-pseudoscalar two-point functions. We have also computed the pseudoscalar-axial two-point functions from which we have extracted the decay constants $f_P$ on our data sets in order to be able to separate the SD axial form factor $F_A$ from the point-like contribution $2 f_P/(m_P x_\gamma)$.

\section{Non-perturbative subtraction of infrared divergent discretisation effects}
\label{sec:fsf}

In this section we want to  stress a very important issue associated with infrared divergent cutoff effects which can jeopardise the extraction of $F_A$ at small values of $x_\gamma$. We  also  introduce a strategy to overcome this problem. 
\begin{figure}[!t]
\begin{center}
\includegraphics[width=0.5\textwidth]{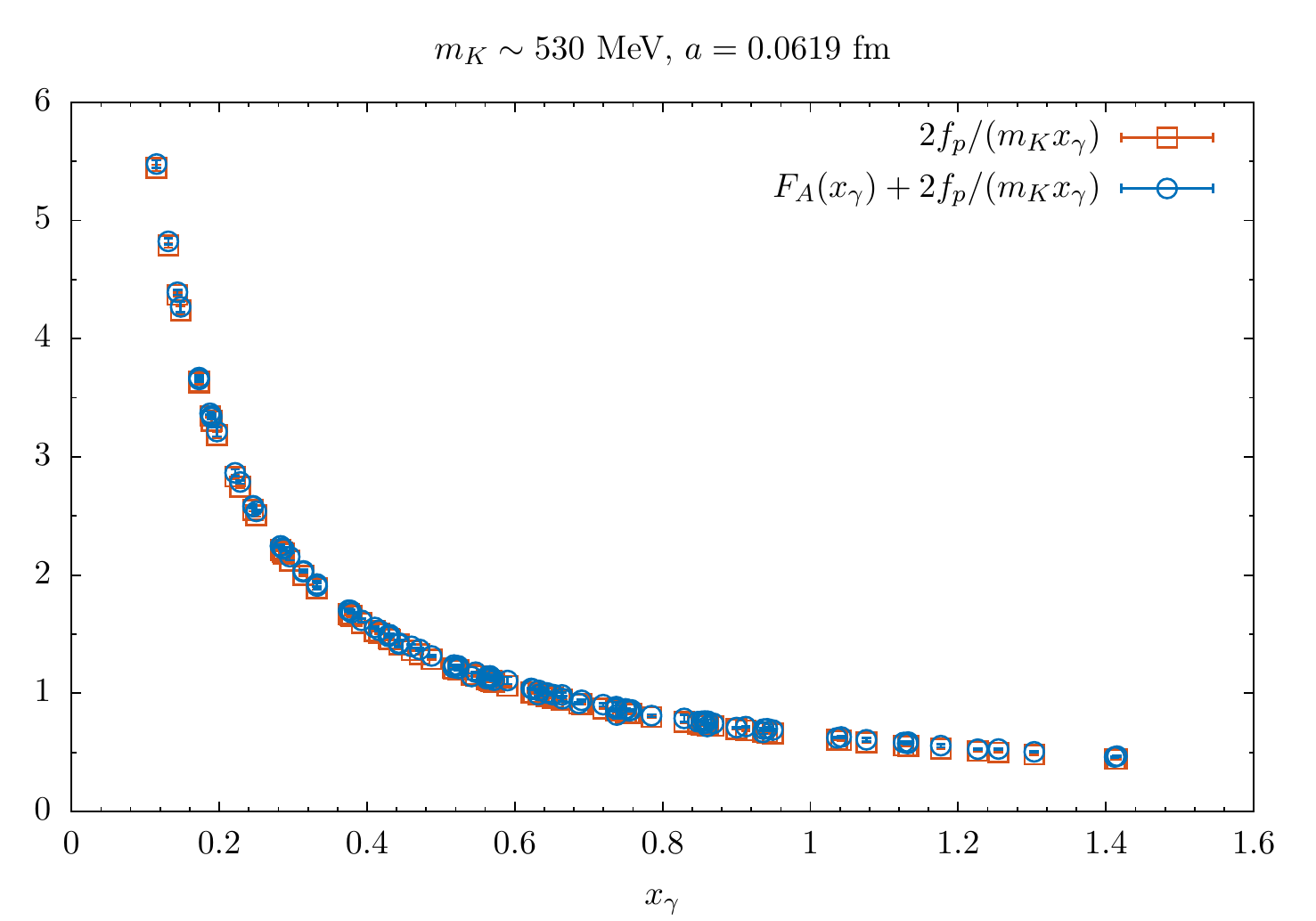}\hfill
\includegraphics[width=0.5\textwidth]{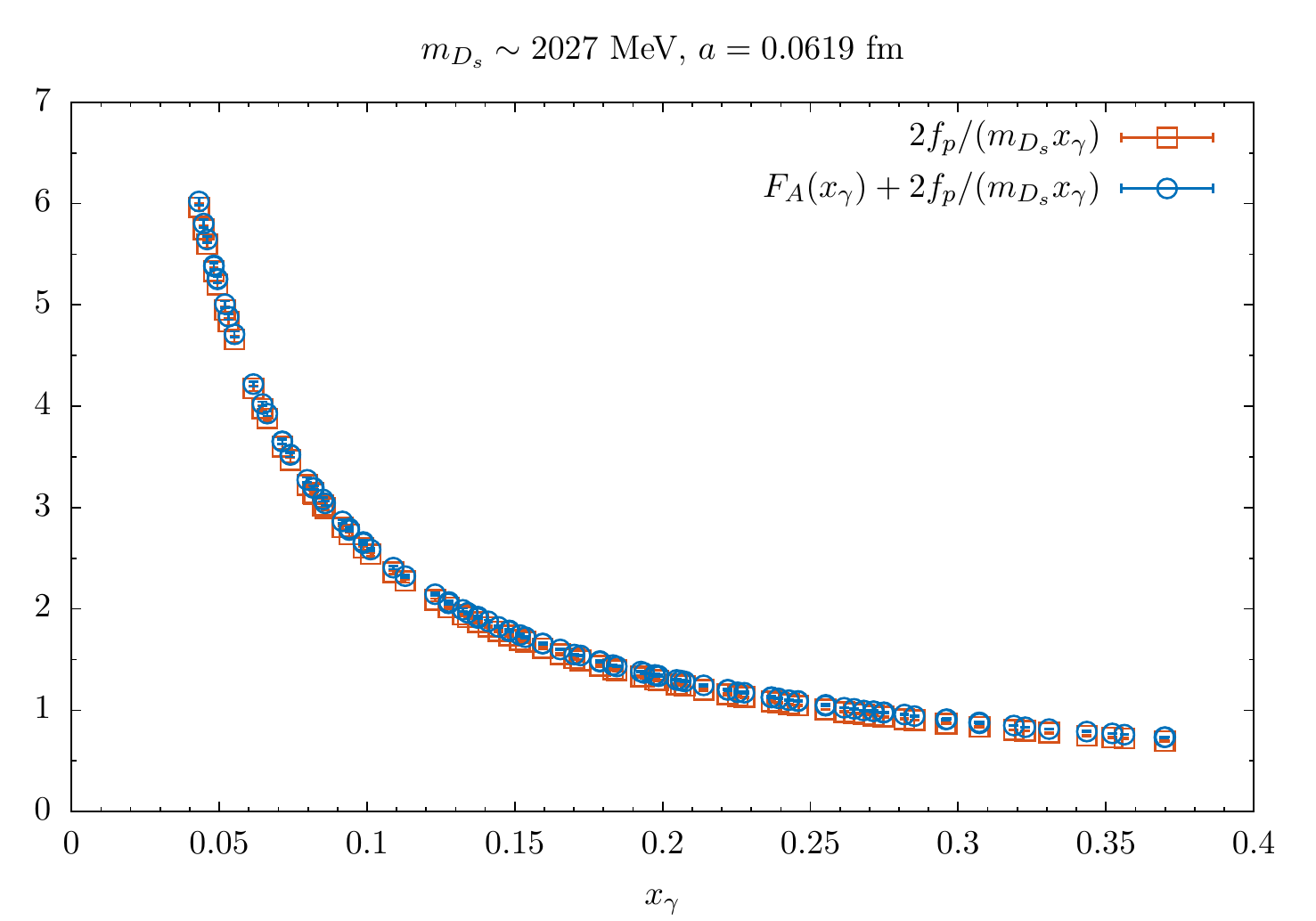}
\end{center}
\caption{
%\footnotesize {
\it  The blue circles represent $F_A(x_\gamma)+2f_P/(m_P x_\gamma)$, extracted directly from $R_A(t)$,  as a function of $x_\gamma$ for the $K$ meson (left) and for the $D_s$ meson (right).  
%In the  figure the form factor $F_A(x_\gamma)=F_A^{NP\, sub}$ has been extracted from $\bar R_A(x_\gamma)$,  Eq.\,(\ref{eq:fsf}),  and the meson decay constant and the mass are computed  in the standard way from the two-point functions.  
The red  squares  represent the point-like  contribution given by  $2f_P/(m_P x_\gamma)$.  The data are taken from the ensemble    D15.48  of Ref.\,\cite{DiCarlo:2019thl}. 
\hspace*{\fill}}
%}
\label{fig:AAa}
\end{figure}

In Fig.\,\ref{fig:AAa}  we plot $F_A(x_\gamma)+2f_P/(m_P x_\gamma)$, the sum of the point-like and SD axial form factors which is extracted directly from the correlation functions using $R_A(t)$ (see Eq.\,(\ref{eq:estimatorsAA})), as a function of $x_\gamma$  for the $K$ (left panel) and  the $D_s$ (right panel) mesons. 
%In this figure  the structure dependent form factor $F_A(x_\gamma)$  is $F^{NPsub}_A(x_\gamma)$ which has been  extracted from $\bar R_A(x_\gamma)$ in  Eq.\,(\ref{eq:fsf}).    
The point-like contribution, $2f_P/(m_P x_\gamma)$, dominates the axial  form factor in the full physical range of photon energies and is overwhelming at small $x_\gamma$.  Using the decay constant and mass, computed  in the standard way from the two-point functions, we can in principle  subtract the point-like term and extract $F_A(x_\gamma)$.   However, this turns out to be very difficult because of the possible presence of discretisation effects  which cannot be excluded by the WI of the lattice action. Moreover, these lattice artefacts diverge as $x_\gamma \to 0$.  We now propose a non-perturbative method to eliminate this problem.  
%value of  $F_A(x_\gamma)$  in the  figure, $FA^{NPsub}(x_\gamma)$,  has been obtained using the non-perturbative subtraction of the $O(a^2)/x_\gamma$ artefacts   defined in Eq.\,(\ref{eq:estimators3}), see also Figure\,\ref{fig:FA_as}, for the $K$ meson and for the $D_s$ meson.  This will be discussed in the following.  

\begin{figure}[!t]
\begin{center}
\includegraphics[width=0.70\textwidth]{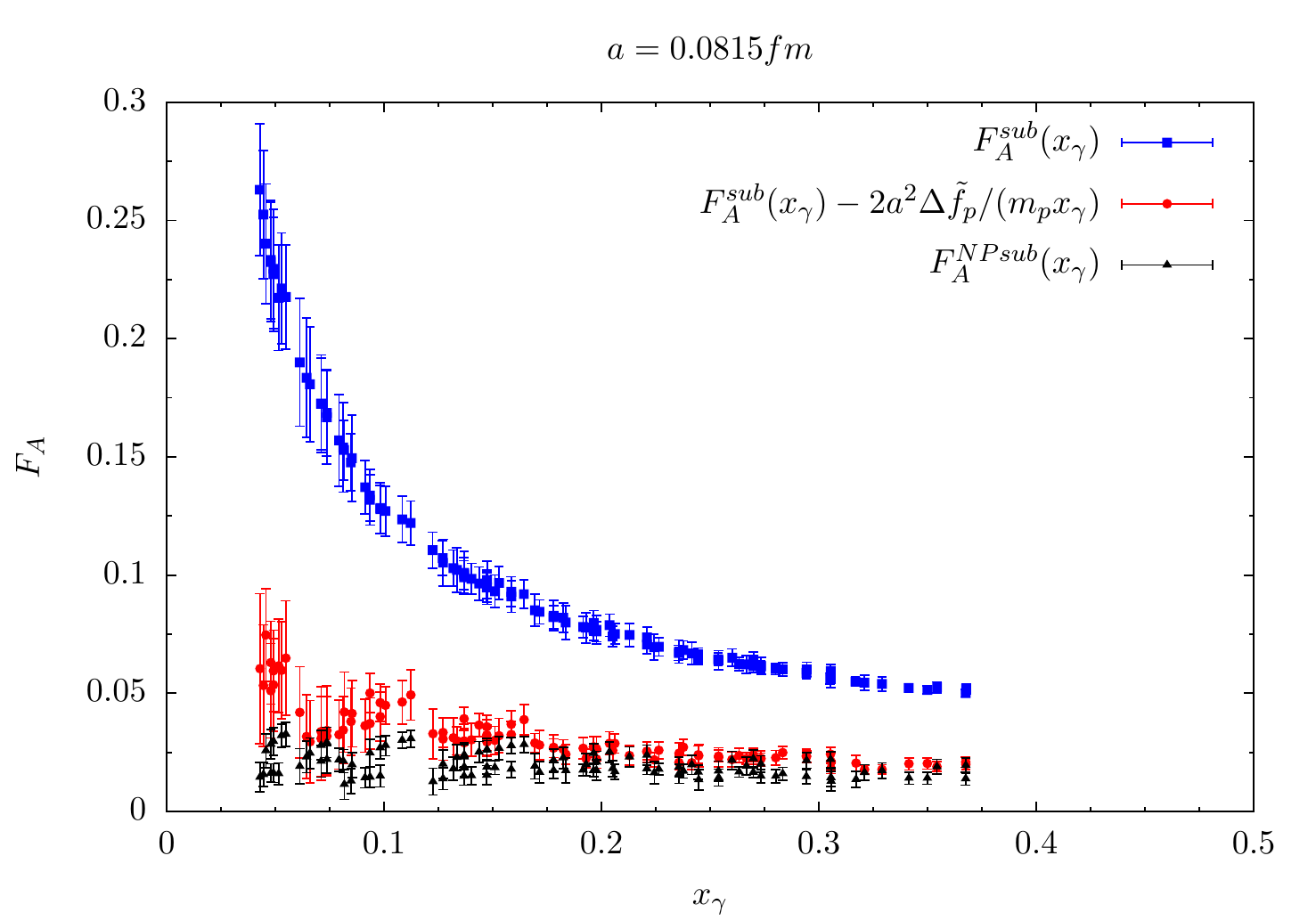}
\end{center}
%\hfill
%\includegraphics[width=0.45\textwidth]{pics/cutafterT2}
\caption{
%\footnotesize{
\it  Study of $F_A$ for the $D_s$ meson. The upper (blue) points show $F_A^{sub}(x_\gamma)$ obtained from Eq.\,(\ref{eq:estimators3}). The divergence at small $x_\gamma$ is reduced by
fitting and subtracting the $O(a^2/x_\gamma)$ artefacts, at the price of increased uncertainties at small $x_\gamma$; these are the intermediate (red) points.  The most accurate results are given by $F_A^{NP\, sub}$, obtained by the non-perturbative subtraction of these artefacts as in Eq.\,(\ref{eq:fsf}) and are shown by the lower (black) points. The data are obtained using the ensemble B55.32 of Ref.\,\cite{DiCarlo:2019thl}. 
\hspace*{\fill}}
%}
%\label{fig:correlator}
%}
%\begin{center}
%\includegraphics[width=0.45\textwidth]{pics/disc}\hfill
%\includegraphics[width=0.45\textwidth]{pics/thetabc}
%\end{center}
%\caption{
%\footnotesize {\it The diagram on the left represents the contributions to the correlators and, consequently, to the form-factors associated with the possibility that the photon is emitted by sea-quarks. In our numerical simulations we have been working within the so-called electroquenched approximation in which the sea-quarks are electrically neutral. In practice this means that in our numerical results we have \emph{neglected} the quark-disconnected contributions represented in the the left panel. The diagram on the right explains our choice of the spatial boundary conditions. By treating the two propagators attached to the electromagnetic current (blue and red lines) as two differente flavours, having the same mass and electric charge but different boundary conditions, we managed to choose arbitrary values for the meson and photon spatial momenta.}
\label{fig:FA_as}
\end{figure}

At finite lattice spacing the axial form factor is constrained, as in the continuum (see Eq.\,(\ref{eq:contWI})), by an exact lattice WI 
\begin{equation} \frac{2\sin(k_\mu a /2)}{a}\,  H^{\alpha \mu}_{L}(k,\vec p) = - \, \langle 0\vert j_A^\alpha(0)  \vert P(\vec p) \rangle=-f_P^L p_L^\alpha  \, , \label{eq:llWI} \end{equation}
that  is true  at all orders in the lattice spacing $a$ (see Appendix\,\ref{sec:acorr}).  The label $L$ here, and in the remainder of this section, stands for ``Lattice" as the discussion concerns the Ward Identity in a discrete space-time. It should not be confused with the spatial extent of the Lattice. 
 This however does not exclude the presence of cutoff effects in Eq.\,(\ref{eq:estimatorsAA}). These  are terms of $O(a^2)$\,\footnote{We assume here that we are using a lattice discretisation in which the leading artefacts are $O(a^2)$. For Wilson Fermions in which they are $O(a)$, the discussion has to be modified accordingly.} and, in particular, include contributions of $O(a^2/ x_\gamma)$
%.  At finite lattice spacing Eq.\,(\ref{eq:estimatorsAA}) reads  
%
\begin{flalign}  \frac{R_A(t)}{x_\gamma}\to  \frac{1}{4\, x_\gamma}\sum_{r=1,2}\sum_{j=1,2}
 \frac{ 2  \, H^{jr}_A(k,\vec p)} {\epsilon_j^r\, m_P}=
 \left[F_A(x_\gamma) + a^2 \Delta F_A(x_\gamma) \right]+\frac{2}{m_P \, x_\gamma}\left(f_P+a^2 \Delta f_P \right) + \cdots\,  ,
\label{eq:estimators2b}
\end{flalign}
where the ellipsis represents higher orders in $a^2$, while the quantities $\Delta F_A$ and $\Delta f_P$ depend upon the parameters of the theory regularised on the lattice, on the  light and heavy quark  masses  and upon $\Lambda_{QCD}$. Discretisation effects in the pseudoscalar masses are also absorbed into $\Delta F_A$ and $\Delta f_P$. The crucial point to notice is  that  the lattice decay constant of the WI in Eq.\,(\ref{eq:llWI}) $f_P^L \neq f_P+a^2 \Delta f_P $. This implies  the presence of the extra term of $O(a^2 /x_\gamma)$ which  appears, in spite of the naive expectations based on  the exact lattice WI.  Thus the coefficient of the last term in Eq.\,(\ref{eq:estimators2b}) is not in general given by 
$2f^L_P/(m_P\, x_\gamma)$,  
where $f^L_P$ is the quantity extracted from the axial-pseudoscalar lattice correlation functions  at finite lattice spacing.  
More precisely, once the matrix element $\langle 0|j_A^\alpha(0)|P(p)\rangle$
is parametrised as in Eq.\,(\ref{eq:llWI}), the definition of $f_P^L$ at fixed cut-off depends upon the choice of the index $\alpha$ and of the lattice momentum $p_L^\alpha$ and, for this reason, is not unique. Therefore, given a generic definition of $f_P^L$, one cannot expect a complete cancellation of the infrared divergent term
on the right-hand side of Eq.\,(\ref{eq:estimators2b}), because a residual lattice artefact will survive
%It is  therefore not possible to cancel completely the infrared divergent  term   contributing to the right hand side of  Eq.\,(\ref{eq:estimators2b})  by using the  decay constant $f^L_P$ obtained from two-point functions because a residual lattice artefact will survive
%
\begin{flalign}
F_A^{sub}(x_\gamma)=  \frac{1}{4\, x_\gamma}\sum_{r=1,2}\sum_{j=1,2}
 \frac{ 2  \, H^{jr}_A(k,\vec p)} {\epsilon_j^r\, m_P}-\frac{2 f^L_P}{x_\gamma\, m_P}= 
F_A(x_\gamma) + a^2 \Delta F_A(x_\gamma) +\frac{2  a^2 \Delta \tilde  f_P}{ x_\gamma \, m_P}\:,
\label{eq:estimators3}
\end{flalign}
generating an effective,  unphysical infrared divergent contribution to $F_A^{sub}(x_\gamma)$ at finite cutoff ($a^2 \Delta \tilde  f_P=f_P-f^L_P+a^2 \Delta f_P$). This phenomenon is illustrated for the $D_s$ meson in Fig.\,\ref{fig:FA_as}  where $F_A^{sub}$ is plotted as function of $x_\gamma$. Since the subtraction of the potentially divergent term is incomplete we observe a fast rise of the effective $F_A^{sub}(x_\gamma)$ at small values of $x_\gamma$. For this reason, even if one has data at different values of the lattice spacing, it is particularly difficult to extract the continuum form factor $F_A(x_\gamma)$ from $F_A^{sub}(x_\gamma)$, especially at small $x_\gamma$ and for heavy mesons. This is illustrated by the intermediate (red) points in Fig.\,\ref{fig:FA_as} which were obtained by fitting and subtracting the $O(a^2/x_\gamma)$ artefacts. The divergence at small $x_\gamma$ is reduced but the relative statistical uncertainties are increased.

We now present an alternative strategy that avoids this problem. In Appendix\,\ref{sec:acorr} we show that the correlation function $C^{\alpha r}_A(t;\vec k,\vec p)$ has a smooth behaviour as a function of $\vec k$ and that from $C^{\alpha r}_A(t;\vec 0,\vec p)$ it is possible to extract directly $H^{ir}_A(0,\vec p)=\epsilon^{r}_i\,f_P$ (see Eq.\,(\ref{eq:Hir})). We can then construct the quantity
\begin{equation}
\bar R_A(t) =
e^{-tE_\gamma}\,  \frac{\sum_{r=1,2}\sum_{j=1,2}\frac{C^{j r}_A(t,T/2;\vec k,\vec p)}{\epsilon_j^r}}{\sum_{r=1,2}\sum_{j=1,2}\frac{C^{j r}_A(t,T/2;\vec 0,\vec p)}{\epsilon_j^r}}-1  
%= \frac{H^{ir}_A(k,\vec p)}{H^{ir}_A(0,\vec p)}- 1 + \dots \,  ,
\label{eq:estimators3b}
\end{equation}
that, by construction, vanishes identically at $x_\gamma=0$. 
Up to statistical uncertainties, each term in the sums in the numerator and denominator of Eq.\,(\ref{eq:estimators3b}) is independent of the indices $j,r$. For the study of the constraints imposed by the electromagnetic Ward identity, it is helpful to view the right-hand side as 
$H^{jr}_A(k,p)/H^{jr}_A(0,p)-1$ (which is also independent of $j,r$).
From the improved estimator $\bar R_A(t)$ we can extract the structure dependent form factor $F_A$ using
\beq   \frac{2f_P}{  m_P x_\gamma }\bar R_A(t)\to F_A^{NPsub}(x_\gamma)
%=\frac{2f_P}{  m_P x_\gamma }\left( \frac{H^{ir}_A(k,\vec p)}{H^{ir}_A(0,\vec p)}-1\right)
= F_A(x_\gamma) + O(a^2)\,  ,   \label{eq:fsf}
\eeq
a quantity that we also show in Fig.\,\ref{fig:FA_as} and that, in contrast to $F_A^{sub}$, does not show any divergent behaviour at small $x_\gamma$.   The reduction of   the uncertainty  on $F_A(x_\gamma)$ using  $\bar R_A(t)$, with respect to a fit  to the right-hand side of Eq.\,(\ref{eq:estimators3}), as shown in Fig.\,\ref{fig:FA_as},  is impressive, particularly at small $x_\gamma$ and also for heavy mesons where there are discretisation effects of $O(a^2m_{D_{(s)}}^2)$.  In the following we will only present results obtained with this method.

The knowledge of $C^{j r}_A(t,T/2;\vec 0,\vec p)$ allows us also to define an alternative estimator for the form factor $F_V(x_\gamma)$, namely
%!TEX encoding = UTF-8 Unicode{\footnotesize
 \bea
\bar R_V(t)
=
f_P m_P \,  \frac{
\left( \sum_{r=1,2}\sum_{j=1,2}\frac{ C^{jr}_V(t,T/2;\vec k,\vec p)-C^{jr}_V(t,T/2;\vec 0,\vec p)}{i \left(E_\gamma\, \vec \epsilon^r \wedge \vec p-E\, \vec \epsilon^r \wedge \vec k\right)^j}\, e^{-tE_\gamma}\right)}{
\left( \sum_{r=1,2}\sum_{j=1,2} \, \frac{C^{j r}_A(t,T/2;\vec 0,\vec p) }{\epsilon_j^r}\, \right)}
%^{-1}
  \to F_V(x_\gamma)\, ,
\label{eq:estimators4}
\eea
%}
that we find has reduced statistical errors compared to $R_V(t)$. Note that because of parity symmetry the correlation function $C_V^{jr}(t, T/2;\vec 0,\vec p)=0$, but this is only approximately true when it is estimated using a finite statistical sample. We find that taking the difference $C^{jr}_V(t,T/2;\vec k,\vec p)-C^{jr}_V(t,T/2;\vec 0,\vec p)$ in the numerator of Eq.\,(\ref{eq:estimators4}) results in a significant reduction of the statistical uncertainty for physical values of $x_\gamma$ (see Eq.\,(\ref{eq:xgamma})).
%The knowledge of $C^{j r}_A(t;\vec 0,\vec p)$ allows also to define an alternative estimator for the form factor $F_V(x_\gamma)$, namely
%
%\begin{flalign}
%\bar R_V(t)
%=
%\frac{f_P m_P}{4}
%\frac{ \sum_{r=1,2}\sum_{j=1,2} C^{jr}_V(t;\vec k,\vec p)-C^{jr}_V(t;\vec 0,\vec p)\epsilon_j^r\, e^{-tE_\gamma}}{ \sum_{r=1,2}\sum_{j=1,2} \, C^{j r}_A(t;\vec 0,\vec p)\, i \left(E_\gamma\, \vec \epsilon^r \wedge \vec p-E\, \vec \epsilon^r \wedge \vec k\right)^j}
%\quad  \to \quad F_V(x_\gamma)\;,
%\label{eq:estimators4}
%\end{flalign}
%

\section{Numerical results}
\label{sec:numerical}
%{\bf figura 8 e seguenti dire qualcosa sui punti che corrispondonio a masse diverse} 
The results presented in this paper were obtained using the ETMC gauge ensembles with $N_f= 2+1+1$ dynamical quarks at three different values of the lattice spacing,  $a =0.0885(36),0.0815(30)$ and $0.0619(18)$\,fm,  with meson masses in the range $220$-$2110$\,MeV.  Details about these  ensembles are given in table II of Ref.\,\cite{DiCarlo:2019thl},  see also Table\,\ref{tab:masses&simudetails} in Appendix\,\ref{app:numres}.  In total we have included 125  different combinations of momenta obtained by assigning to each of the $\theta_{i=0,t,s}$ five different values; making  the same assignments for all  choices of the quark masses. 
In the figures below we illustrate the quality and features of our results by showing examples of plots for light and heavy mesons. The plots used for illustration correspond to  unphysical values of the  $\overline{\rm MS}$ renormalised light-quark mass, $m_{ud}(2$\,GeV$)=11.7$\,MeV.  The corresponding meson  masses  are $m_{D_s}=2027\, (3)$\,MeV, $m_D=1929\, (6)$\,MeV,   $m_K=530\, (2)$\,MeV and $m_\pi= 228\, (2)$\,MeV.   Similar plots can be shown for other values of the simulation parameters.    

The scale setting is taken from Ref.\,\cite{Carrasco:2014cwa}, where the continuum value of $r_0$\,\cite{Sommer:1993ce} was obtained imposing
$m_\pi^{\rm exp}  = m_{\pi^0} = 134.98$\,MeV  and  $f_\pi^{\rm exp} = 130.41(20)$\,MeV. The values of the strange and charm quark masses, obtained by extrapolating the kaon and $D$ meson masses  to the continuum and at the physical point in  the light quark masses,  are $m_s(2\,{\rm GeV}) = 99.6\,(4.3)$\,MeV and   $m_c(2\,{\rm GeV}) = 1.176\, (39)$\,GeV.
In the following  for the renormalised quark mass we shall use $m =\mu/Z_P$, where  $\mu$ is the twisted mass of the given quark and   $Z_P$ is the renormalisation constant  of the pseudoscalar density in  the $\overline{\mathrm{MS}}$ scheme, at 2\,GeV,  computed with method M2\,\cite{Carrasco:2014cwa}.  The values of $\mu$ used in our simulation can  also be found in Tables\,\ref{tab:masses&simudetails} and \ref{tab:MpiL} in Appendix\,\ref{app:numres} (see also
Table II of Ref.\,\cite{DiCarlo:2019thl}).  Renormalisation of the corresponding axial-vector and vector currents with Twisted Mass Fermions gives $F_A = Z_V F^0_A$ and $F_V =Z_A F^0_V$  where $F^0_A$ and $F^0_V$ are the unrenormalised quantities as explained in Eq.\,(\ref{eq:jWVA}), $Z_A$ has been computed with method M2 and $Z_V$ with the WI\,\cite{Carrasco:2014cwa}.
In Table\,\ref{tab:MpiL} of  Appendix\,\ref{app:numres}  we give further details of our simulation including the values of the angles $\theta_{i=0,s,t}$ used to fix the hadron and photon momenta, see Eq.\,(\ref{eq:momenta}).

\begin{figure}[!t]
\begin{center}
\includegraphics[width=0.5\textwidth]{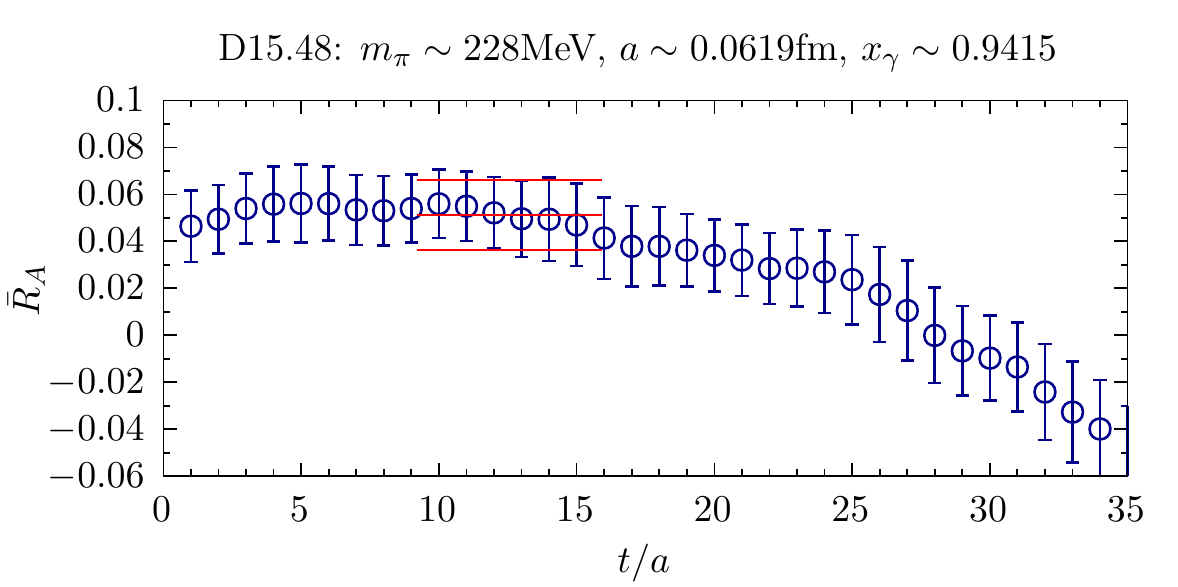}\hfill
\includegraphics[width=0.5\textwidth]{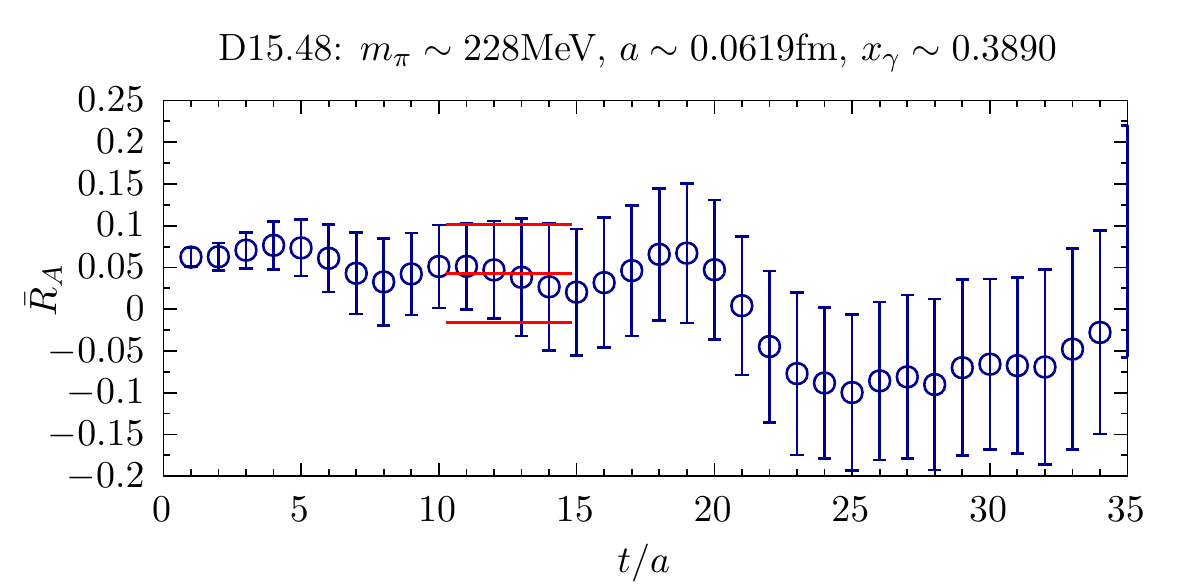}\hfill
\includegraphics[width=0.5\textwidth]{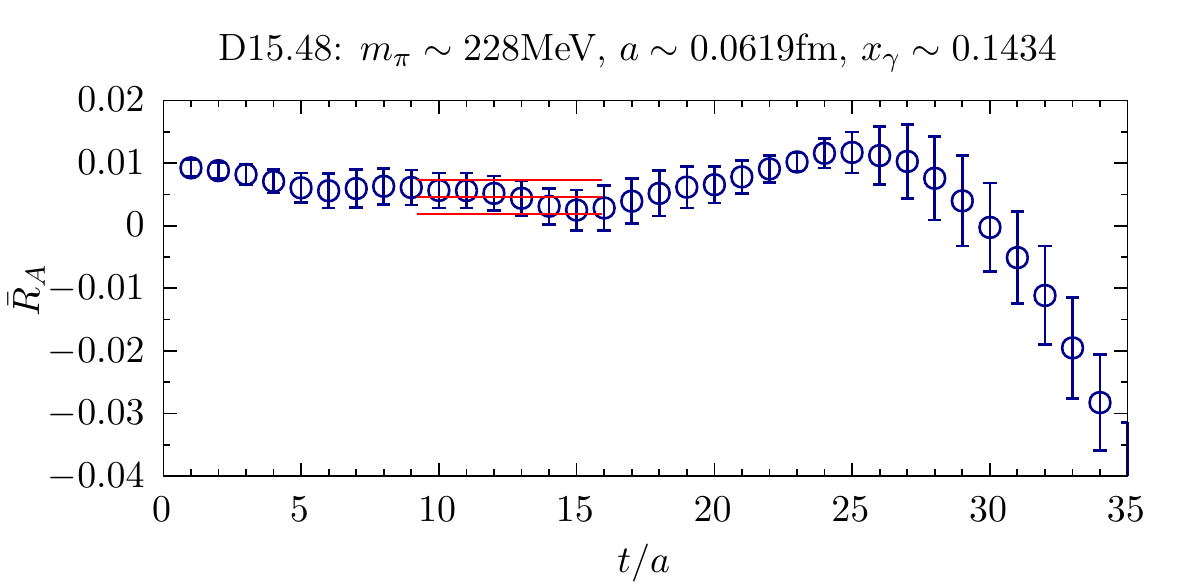}\hfill
\includegraphics[width=0.5\textwidth]{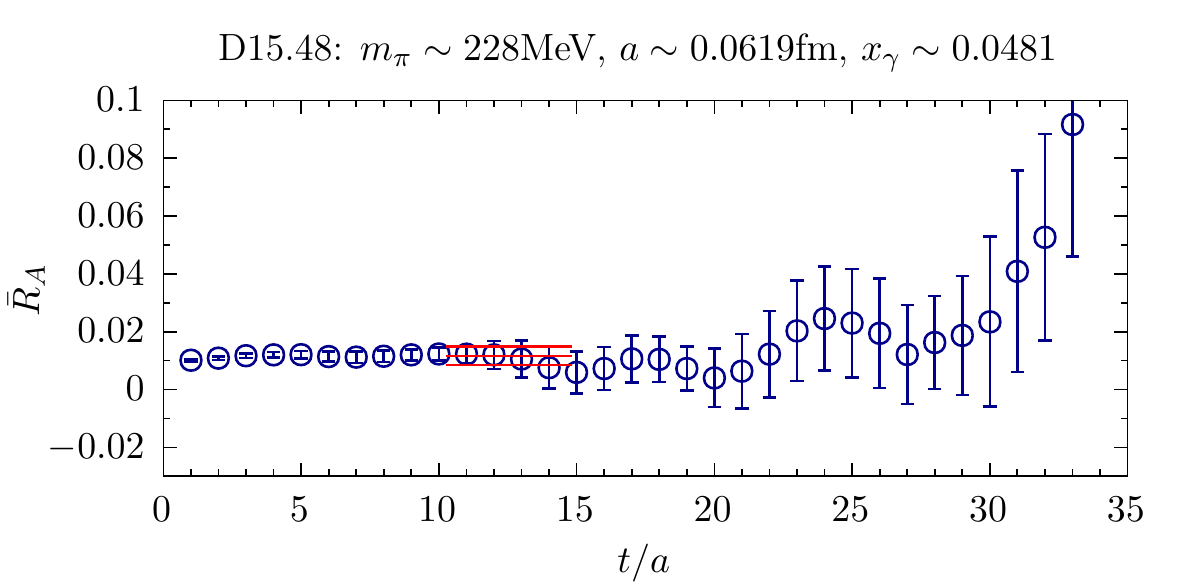}\hfill
\end{center}
\caption{
%\footnotesize
{\it Examples of fits to plateaux for the ratio $\bar R_A(t)$ for the kaon (left) and $D$ meson (right)  at  larger (upper panels) or smaller (lower panels) values of $x_\gamma$. The values obtained from the fits, together with their uncertainties, are indicated by the horizontal (red) bands.\hspace*{\fill}}
}
\label{fig:AA}
\end{figure}
\begin{figure}[!t]
\begin{center}
\includegraphics[width=0.5\textwidth]{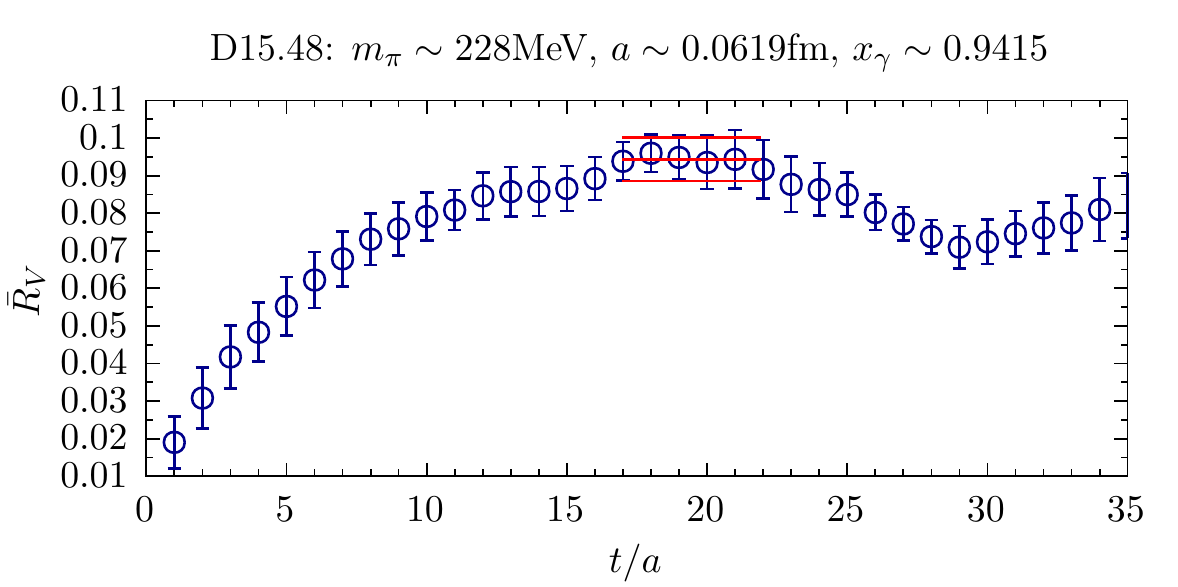}\hfill
\includegraphics[width=0.5\textwidth]{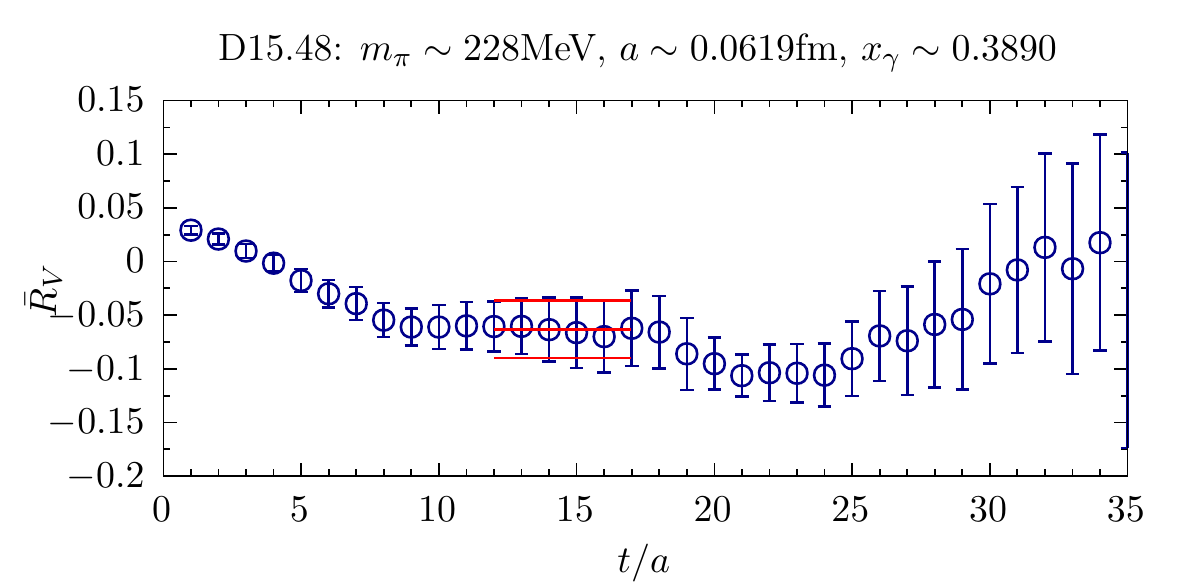}\hfill
\includegraphics[width=0.5\textwidth]{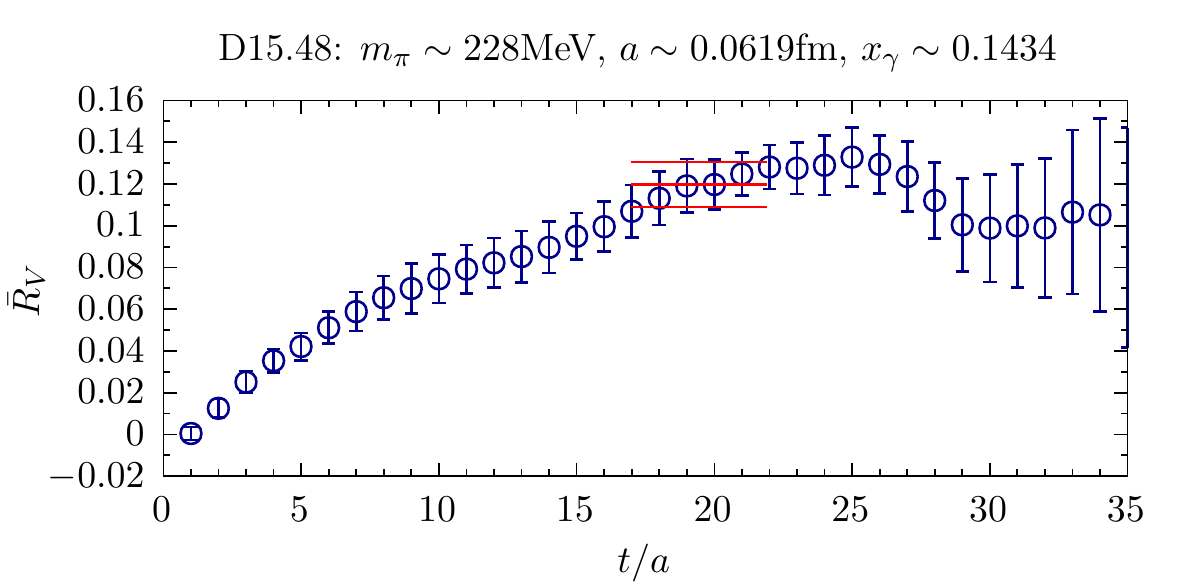}\hfill
\includegraphics[width=0.5\textwidth]{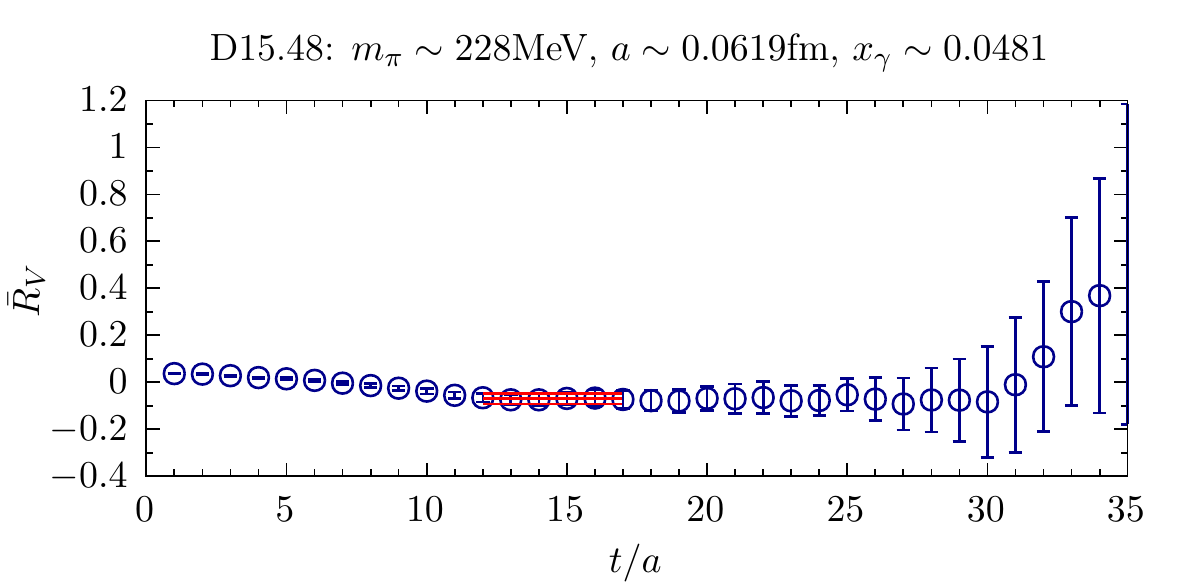}\hfill
\end{center}
\caption{
%\footnotesize{
\it Examples of fits to plateaux for the ratio $\bar R_V(t)$ for the kaon (left) and $D$ meson (right)  at  larger (upper panels) or smaller (lower panels) values of $x_\gamma$. The values obtained from the fits, together with their uncertainties, are indicated by the horizontal (red) bands.\hspace*{\fill}}
%}
\label{fig:VV}
\end{figure}

In Figs.\,\ref{fig:AA}  and\,\ref{fig:VV} we show examples of plateaux for the ratios $\bar R_{A,V}(t)$,  defined in  Eqs.\,(\ref{eq:estimators3b}) and (\ref{eq:estimators4}) respectively,  for $K$ and $D$ mesons. These figures are representative of the signal quality also for other values of masses,  momenta and lattice spacings.  
The values of all  the form factors discussed in the following   have been extracted from the plateaux obtained by  using Eqs.\,(\ref{eq:estimators3b}) and (\ref{eq:estimators4}). The time interval used  for the extraction has been chosen, for each of the data ensembles and for  each of the mesons,   in such a way as to observe a reasonable plateau for all values of the meson/photon momenta. 
In order to extract the  form factors  $F_{A,V}$  at physical values of the quark masses and in the continuum limit we have used a variety of fitting formulae for light and heavy mesons as discussed below.

For pions and kaons, we have covered the full  physical range of  $x_\gamma$,  $0\le x_\gamma\le 1-m_\ell^2/m_{\pi,K}^2$  (indeed we even have data for  unphysical values corresponding to $x_\gamma > 1$).  

For the pion, guided by ChPT,  we fit to the formula
\begin{eqnarray} F_{A,V}(x_\gamma)= \frac{m_\pi}{f_\pi} \left[ \left(c_0+c^\prime_0\, \frac{m^2_\pi}{\left(4\pi f_\pi\right)^2}+   \tilde c_0\, \frac{a^2}{r_0^2}\right)+ \left( c_1^\prime\, \frac{m^2_\pi}{\left(4\pi f_\pi\right)^2}+  \tilde c_1 \, \frac{a^2}{r_0^2} \right) x_\gamma    \right]\, . \label{eq:extrlight}\end{eqnarray} 
This is certainly not the most general formula to include higher orders terms in ChPT, for example it does not contain chiral logarithms,
% or  flavour $SU(3)$   symmetry breaking terms,  
 but it is sufficiently simple and adequate to describe the pion data.  The two coefficients $\tilde c_0$ and $\tilde c_1$ take into account  possible mass-independent discretisation effects.  $c_1^\prime$ is multiplied by $m_\pi^2$ because it arises in higher orders in ChPT. On the other hand the discretisation term  proportional to  $\tilde c_1$ is not multiplied by the mass of the meson because at this order in $a$  there is an explicit violation of chiral invariance in the lattice Fermion Lagrangian.  

When using the simpler expression in  Eq.\,(\ref{eq:extrlight}) we exclude data at  pion masses  $m_\pi \gtrsim 350$\,MeV. Since in our data we have pion masses up to about $500$\,MeV, we have also performed fits in the full range by modifying Eq.\,(\ref{eq:extrlight}) to include higher order terms as follows:
\begin{eqnarray}&& F_{A,V}(x_\gamma)= \frac{m_\pi}{f_\pi} \left[ \left(c_0+c^\prime_0\, \frac{m^2_\pi}{\left(4\pi f_\pi\right)^2}+  \tilde c_0 \, \frac{a^2}{r_0^2}+  \Delta c^\prime_0 \frac{m^4_\pi}{\left(4\pi f_\pi\right)^4}+  \Delta \tilde c_0 \, a^2 m_\pi^2\right)  \right.\nonumber \\&& \hspace{0.2in}\left.+ \left( c_1^\prime\, \frac{m^2_\pi}{\left(4\pi f_\pi\right)^2}+ \tilde c_1 \, \frac{a^2}{r_0^2}+\Delta c_1^\prime \frac{m^4_\pi}{\left(4\pi f_\pi\right)^4} +   \Delta \tilde c_1 \, a^2 m_\pi^2  \right) x_\gamma   \right]\, . \label{eq:extrlightb}\end{eqnarray} 
  The higher-order coefficients  $\Delta c^\prime_0$,  $\Delta \tilde c_0$, $\tilde c_1$, $\Delta c_1^\prime$ and $\Delta \tilde c_1$  have very little effect on the extrapolated results and for this reason they are not well determined.  Indeed they only contribute to a slight increase in the  uncertainty in the value of the pion  form factors at $x_\gamma=0$ and in the slope in $x_\gamma$. 
Similarly, in the different fits that we performed, we also added some of the possible lattice artefacts that break Lorentz  invariance, for example those  proportional to $a^2 \vert \vec k\vert^2$  (in the frame where the meson is at rest),  where $\vec k$  is the momentum of the photon.  We found that  their effect is very small  and this was only taken into account in the evaluation of the final uncertainties.

Since $SU(3)$ breaking effects may be important, and we only have results obtained at two values of the {\it strange} quark mass, for the kaon we first interpolate the form factors to the physical kaon mass and then fit them to the formula  
\begin{eqnarray} F_{A,V}(x_\gamma)= \frac{m_K}{f_K} \left[ \left(c_0+c^\prime_0\, \frac{m^2_\pi}{\left(4\pi f_\pi\right)^2}+   \tilde c_0\, \frac{a^2}{r_0^2}\right)+ \left( c_1+\,c_1^\prime\frac{m^2_\pi}{\left(4\pi f_\pi\right)^2}+  \tilde c_1 \, \frac{a^2}{r_0^2} \right) x_\gamma    \right]\, . \label{eq:extrlightc}\end{eqnarray} 
with pion masses $m_\pi < 350$\,MeV and 
\begin{eqnarray}&& F_{A,V}(x_\gamma)= \frac{m_K}{f_K} \left[ \left(c_0+c^\prime_0\, \frac{m^2_\pi}{\left(4\pi f_\pi\right)^2}+  \tilde c_0 \, \frac{a^2}{r_0^2}+  \Delta c^\prime_0 \frac{m^4_\pi}{\left(4\pi f_\pi\right)^4}+  \Delta \tilde c_0 \, a^2m_\pi^2\right) \right.\nonumber \\&&\hspace{0.2in} \left.+ \left( c_1+c_1^\prime\,\frac{m^2_\pi}{\left(4\pi f_\pi\right)^2}+ \tilde c_1 \, \frac{a^2}{r_0^2}+   \Delta c_1^\prime \frac{m^4_\pi}{\left(4\pi f_\pi\right)^4}+ \Delta \tilde c_1 \, a^2m_\pi^2 \right) x_\gamma   \right]\, , \label{eq:extrlightd}\end{eqnarray} 
in the full range of pion masses. Formulae (\ref{eq:extrlightc}) and (\ref{eq:extrlightd}) for the kaon are equivalent to those in (\ref{eq:extrlight}) and (\ref{eq:extrlightb}) respectively for the pion. The presence of the constant term $c_1$ in Eqs.\,(\ref{eq:extrlightc}) and (\ref{eq:extrlightd}) is a reflection of the fact that the strange quark mass is fixed to its physical value.
To simplify the notation we have used the same symbols for the coefficients in Eqs.\,(\ref{eq:extrlight})-(\ref{eq:extrlightd}) but the reader should note that their values are different in each case. 
We do not have sufficient data to include terms proportional to $m_K^2\, m_\pi^2$ or  $m_\pi^4$ with logarithmic corrections in Eq.\,(\ref{eq:extrlightd}). 

%We found that the value of the form factors and of their derivative in $x_\gamma=0$ is only slightly modified with respect to a fit of the data to Eq.\,(\ref{eq:extrlightb}) and included this effect in the final uncertainty. 
\begin{figure}[!t]
\begin{center}
\includegraphics[width=0.5\textwidth]{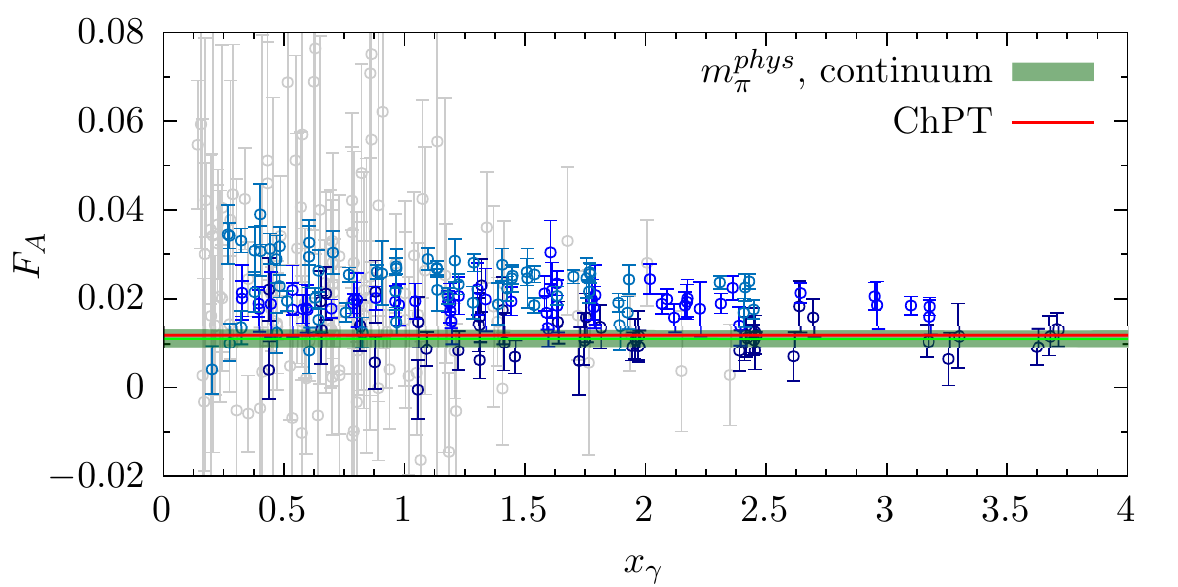}\hfill
\includegraphics[width=0.5\textwidth]{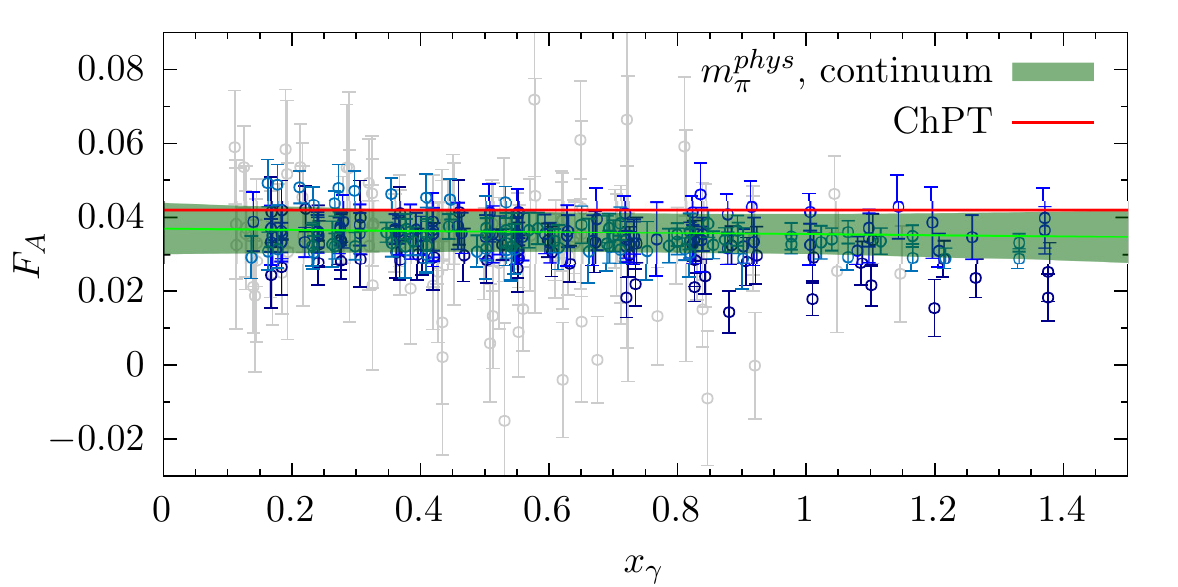}
\includegraphics[width=0.5\textwidth]{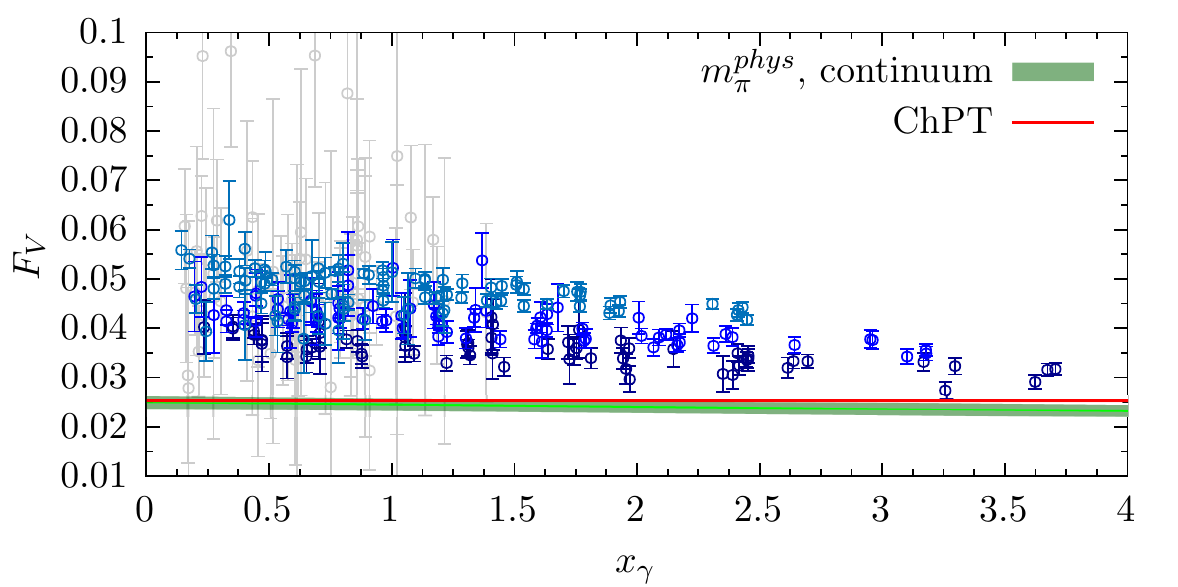}\hfill
\includegraphics[width=0.5\textwidth]{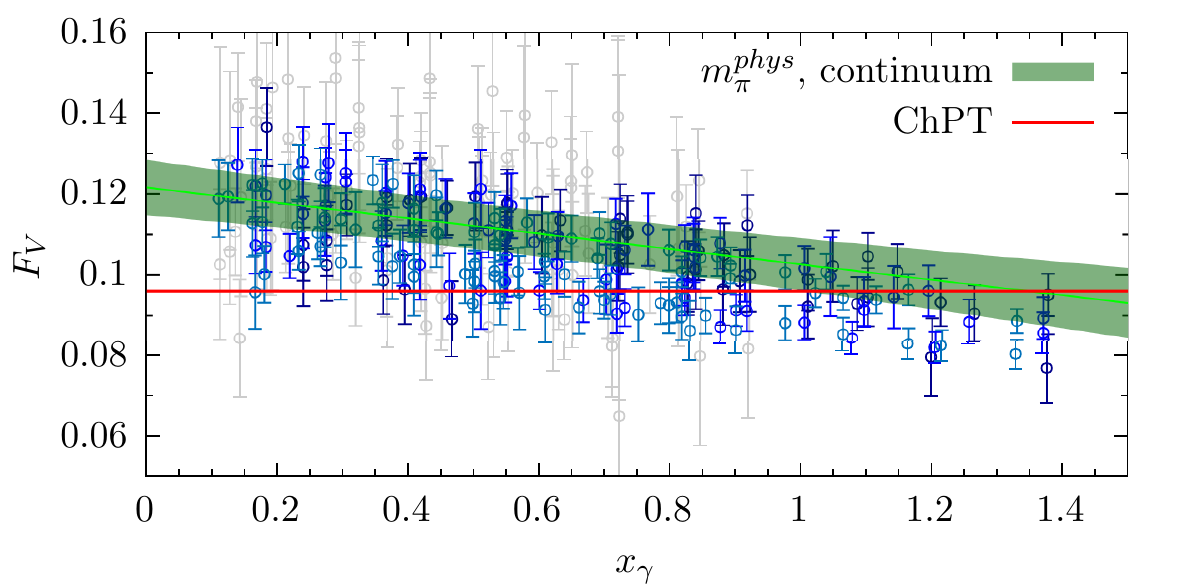}
\end{center}
\caption{
%\footnotesize  {
\it Extracted values of the pion (left) and kaon (right) form factors $F_A(x_\gamma)$ (upper) and $F_V(x_\gamma)$ (lower) as a function of $x_\gamma$  for the configurations at $a =0.0619$\,fm. The  horizontal red  lines correspond to the lowest order ChPT prediction in Eq.(\ref{eq:lochpt}).  The green lines and bands are the results of the fits, using  the formulae given in Eqs.\,(\ref{eq:extrlightb}) and (\ref{eq:extrlightd}),  after extrapolation to the continuum limit and physical quark masses,  together with the corresponding  uncertainties.  \hspace*{\fill}}
%}
\label{fig:FA}
\end{figure}
\begin{figure}[!t]
\begin{center}
\includegraphics[width=0.5\textwidth]{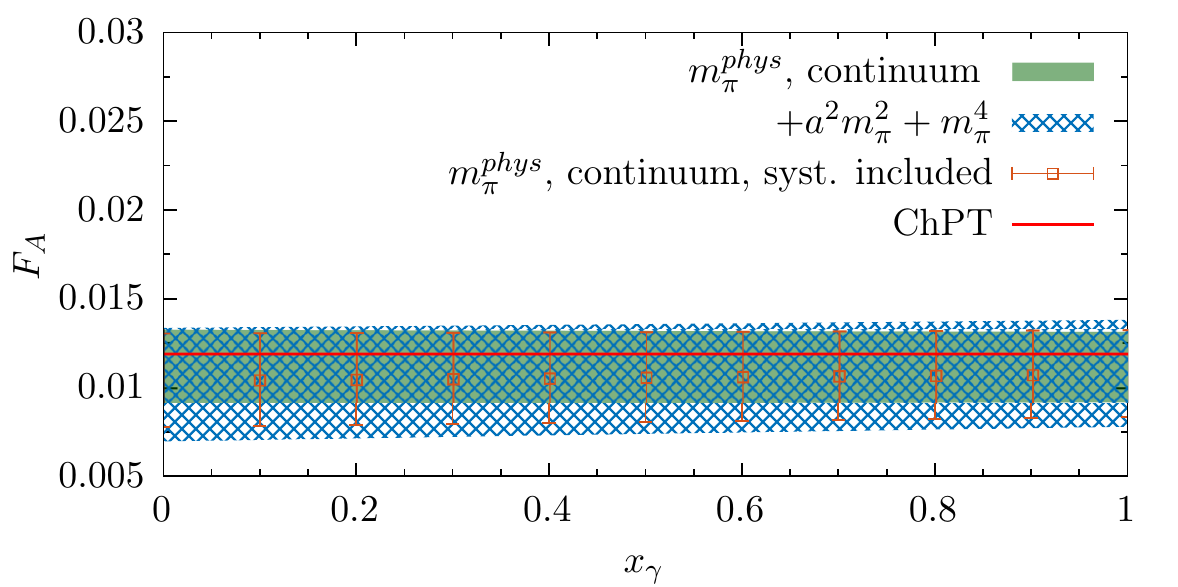}\hfill
\includegraphics[width=0.5\textwidth]{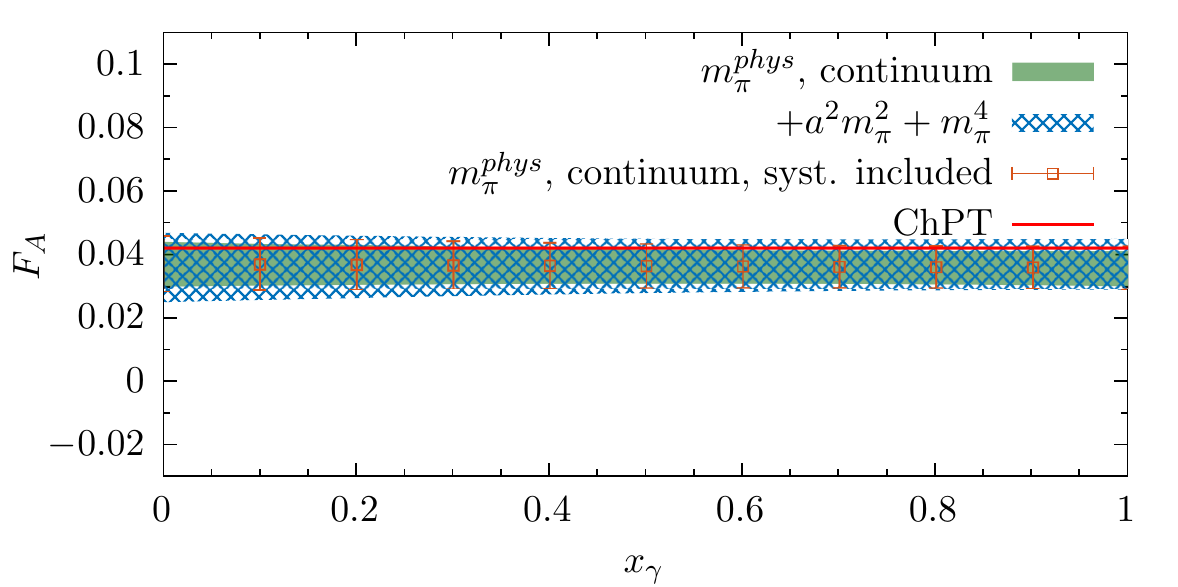}\hfill
\includegraphics[width=0.5\textwidth]{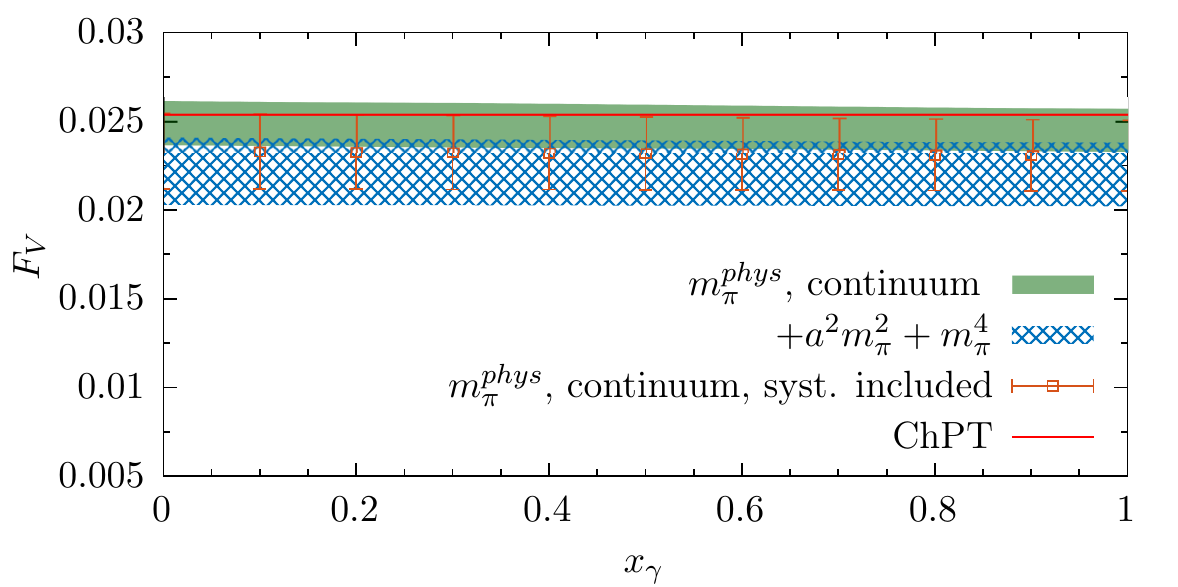}\hfill
\includegraphics[width=0.5\textwidth]{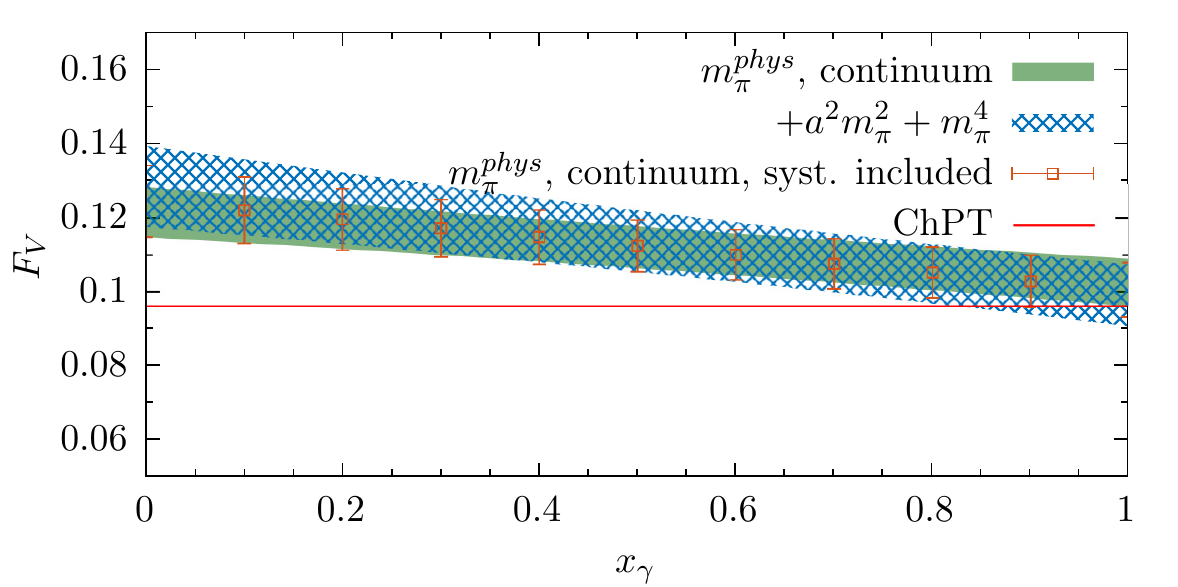}
\end{center}
\caption{
%\footnotesize  {
\it Extracted values of the pion (left) and  kaon (right) form factors $F_A(x_\gamma)$ (upper) and $F_V(x_\gamma)$ (lower) as a function of $x_\gamma$.  The horizontal red lines correspond to lowest order  ChPT  predictions in Eq.(\ref{eq:lochpt}). The full  green bands are the results of the fits after the continuum and chiral extrapolations obtained using Eqs.\,(\ref{eq:extrlight}) and (\ref{eq:extrlightc}) for the pion and kaon respectively
and the shaded blue bands are obtained using (\ref{eq:extrlightb}) or (\ref{eq:extrlightd}).  We also show the extrapolated form factors and the corresponding uncertainties (statistical and systematic) for selected values of $x_\gamma$.\hspace*{\fill}}
%}
\label{fig:FAbis}
\end{figure}
In Fig.\,\ref{fig:FA} we present the values of the pion (left panels) and  kaon  (right panels) form factors $F_A(x_\gamma)$ (upper panels) and $F_V(x_\gamma)$ (lower panels) as a function of $x_\gamma$  for the configurations at $a =0.0619$\,fm.   The plotted points with error bars  correspond to different values of the light-quark mass at several values of $x_\gamma$. The points with  large uncertainties ($\sigma_{F_{A,V}}\ge 0.01$ for the kaon or $\sigma_{F_{A,V}}\ge 0.008$ for the pion) are   shown with faint grey symbols.  These points are obtained for mesons with 
substantial non-zero momenta $\vec p\neq \vec 0$.  The results of our simulation are  compared  to the lowest  order in ChPT, given by \bea  F_A(x_\gamma)&=& {\rm const.} = \frac{8 m_P}{f_P}\, \left(L_9^r+L_{10}^r\right)
 \nonumber \\ F_V(x_\gamma)&=& {\rm const.} =  \frac{m_P }{4\pi^2 f_P}
\, ,  \label{eq:lochpt} \eea 
where $P$ represents $\pi$ or $K$ and we take $(L_9^r+L_{10}^r)\simeq 0.0017$\,\cite{Bijnens:2014lea}; this is indicated by the horizontal red  lines.  The blue lines and
green  bands are the results and uncertainties of the fits, obtained using Eqs.\,(\ref{eq:extrlightb}) and (\ref{eq:extrlightd}) after the extrapolation  to  physical quark masses and to zero lattice spacing has been performed.   
In Fig.\,\ref{fig:FAbis}  we show the  value of the pion (left) and the  kaon (right)  form factors $F_A(x_\gamma)$ (upper) and $F_V(x_\gamma)$ (lower) as a function of $x_\gamma$, extrapolated to the continuum at the physical point,   either  using  Eqs.\,(\ref{eq:extrlight}) and (\ref{eq:extrlightc}) for the pion and kaon respectively to fit  the data,  full green bands, or by using Eqs.\,(\ref{eq:extrlightb}) and (\ref{eq:extrlightd}),  shaded blue bands. In the figure we also show the values of the form factors for selected values of $x_\gamma$ extrapolated to the continuum and to the physical point, together with the corresponding statistical and systematic uncertainties.   The systematic uncertainties were estimated from the differences in the results coming from different fits of higher order terms in the meson masses,  the inclusion of different possible discretisation corrections and the functional forms of the fits, i.e. whether we use Eqs.\,(\ref{eq:extrlight}) and (\ref{eq:extrlightc}) for the pion and kaon respectively or Eqs.(\ref{eq:extrlightb}) and (\ref{eq:extrlightd}).  The results for the form factors  $F_{A,V}$ at selected values of $x_\gamma$, the corresponding uncertainties $\Delta_{F_{A,V}}$, and their correlation matrices are given for all the mesons in Appendix\,\ref{app:numres}. 

For heavy mesons $H$  we expect that the form factors scale as $m_h^{-3/2} \sim f_H/m_H$, where $m_h$ is the mass of the heavy quark contained in $H$
\begin{eqnarray} F_{A,V}(x_\gamma) = F_{A,V}^0 \, \frac{f_H}{m_H}\left( 1 +   O\left(\frac{\Lambda_{QCD}}{m_H}\right) +\dots\right)  +  O\left( a^2 \, m^2_H \right) \, , \label{eq:extrHeavy0}\end{eqnarray} 
where the constants $F_{A,V}^0 $ are a function of the light quark masses. 
 Since,  however, for this exploratory study,  we have  only two values of the heavy quark mass, both around the charm mass, we prefer to interpolate the values of the form factors  to the physical charm quark mass and then to fit  the  result with the simple formula 
%{\small
\begin{eqnarray} F_{A,V}(x_\gamma)&=& d_0+  d^\prime_0\, \frac{m_\pi^2}{\left(4\pi f_\pi\right)^2} +  \tilde d_0\, \frac{a^2}{r_0^2} 
%\nonumber \\ &&
 + \left( d_1+    d^\prime_1\, \frac{m_\pi^2}{\left(4\pi f_\pi\right)^2}+ \tilde d_1\, \frac{a^2}{r_0^2}\right) \, x_\gamma    \, . \label{eq:extrHeavy}\end{eqnarray} 
%}
%\beq  1+\left( \Delta_1+    \Delta^\prime_1\, \frac{m_\pi^2}{\left(4\pi f_\pi\right)^2}\right) x_\gamma \eeq
We have also performed fits with the {\it pole-like} formula 
\begin{eqnarray} F_{A,V}(x_\gamma)= \frac{d_0+  d^\prime_0\, \frac{m_\pi^2}{\left(4\pi f_\pi\right)^2} }{1+\left( \Delta_1+    \Delta^\prime_1\, \frac{m_\pi^2}{\left(4\pi f_\pi\right)^2}\right) x_\gamma  } + \tilde d_0\, \frac{a^2}{r_0^2}+ \tilde d_1\, \frac{a^2}{r_0^2} \,x_\gamma   \, . \label{eq:extrHeavyb}\end{eqnarray} 
%Sul punto 5) concordo. Prendendo FV il termine lorenziano e? $\epsilon_{\mu \alpha \gamma \beta} k_\gamma p_\beta$ e gli ipercubici rilevanti potrebbero essere $\epsilon_{\mu \alpha \gamma \beta}  k_\gamma^3 p_\beta, \epsilon_{\mu \alpha \gamma \beta}  k_\gamma^2 p_\beta^2$ e $\epsilon_{\mu \alpha \gamma \beta}  k_\gamma p_\beta^3$, ognuno con un loro f.f. (con $k = (ik, \vec{k})$ e $p = (iE, \vec{p})$ e metrica euclidea).
\vskip 1 cm 

%On the right hand plot  of Fig.\,\ref{fig:FA} we can compare the directly computed value of  $F_V$ to its ChPT prediction,   
%fixa_FA_Ds_compare
\begin{figure}[!t]
\begin{center}
\includegraphics[width=0.60\textwidth]{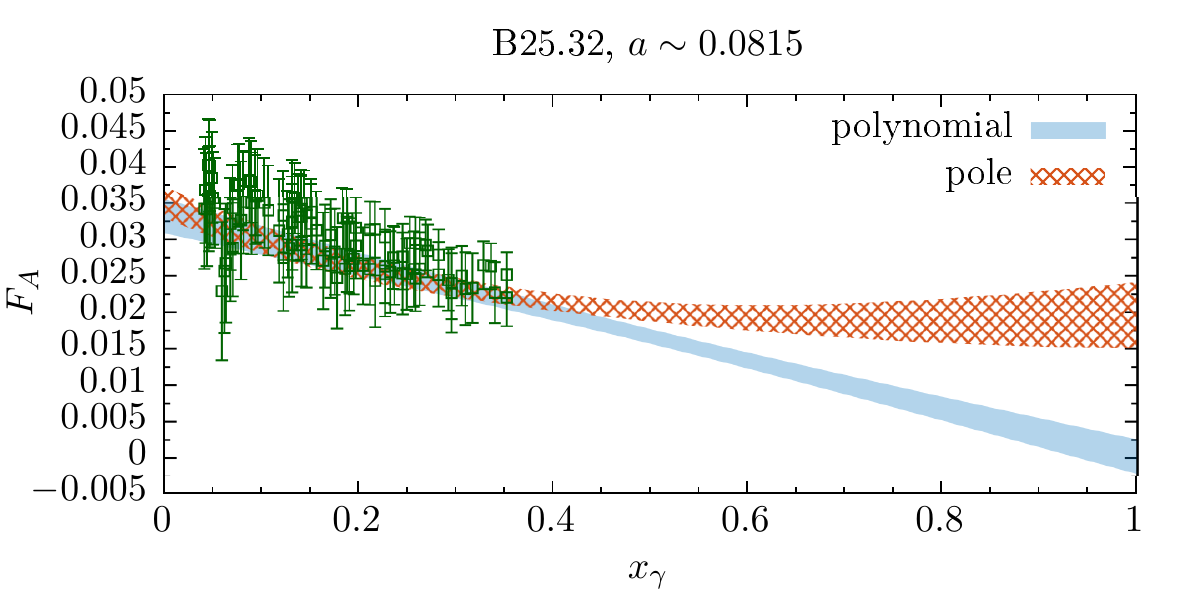}\hfill
\includegraphics[width=0.60\textwidth]{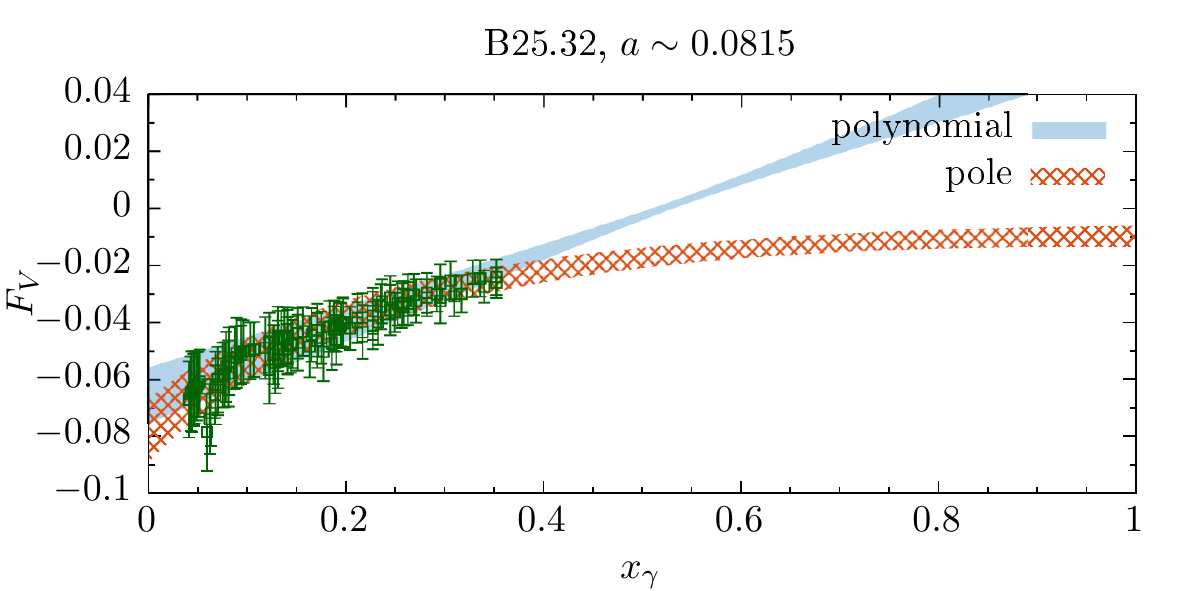}
\end{center}
\caption{
%\footnotesize  {
\it The form factors $F_A(x_\gamma)$ (upper) and $F_V(x_\gamma)$ (lower)  of  the $D_s$ meson as a function of $x_\gamma$ at fixed lattice spacing ($a=0.0815$\,fm) for the ensemble B25.32\,\cite{DiCarlo:2019thl}. The  full blue  and  shaded orange bands are the results of the fits  with the polynomial or pole  formulae given in Eqs.\,(\ref{eq:extrHeavy}) and~(\ref{eq:extrHeavyb})  respectively.   \hspace*{\fill}}
%}
\label{fig:Dscompare}
\end{figure}
In this first study, we only have results for the $D_{(s)}$ mesons in the range  $0\le x_\gamma\le 0.4$, corresponding to $E_\gamma \lesssim 400$\,MeV in the rest frame of the hadron.      
In Fig.\,\ref{fig:Dscompare} we give the results for the form factors of the $D_s$ meson, $F_A(x_\gamma)$ and $F_V(x_\gamma)$, at $a =0.0815$\,fm.   The  full blue  and shaded orange bands are the results of the fits  with the polynomial or pole  formula given in Eqs.\,(\ref{eq:extrHeavy}) and~(\ref{eq:extrHeavyb})  respectively. Since the lattice spacing is fixed, the coefficients $\tilde d_{0,1}$  are not included in the fit. We see that the both the fits give a good description of our results in the region where we have data, but differ significantly 
for $x_\gamma \ge 0.4$.  This means that, although both the linear and the pole fits describe accurately the form factors in the region in which we have data, it is not reliable to use these fits in the region $x_\gamma  \ge 0.4$. In our future investigations we plan to provide non-perturbative data for the form factors in the full kinematical range $0 \le x_\gamma \le 1-m_\ell^2/m_{D_{(s)}}^2$.

\begin{figure}[!t]
\begin{center}
\includegraphics[width=0.6\textwidth]{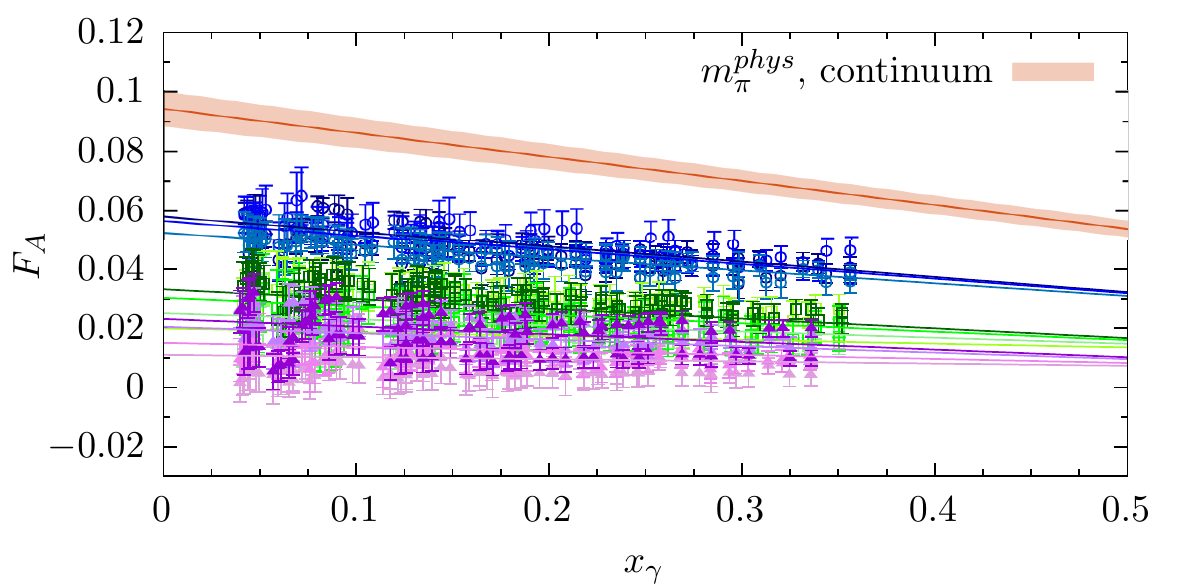}\hfill
\includegraphics[width=0.6\textwidth]{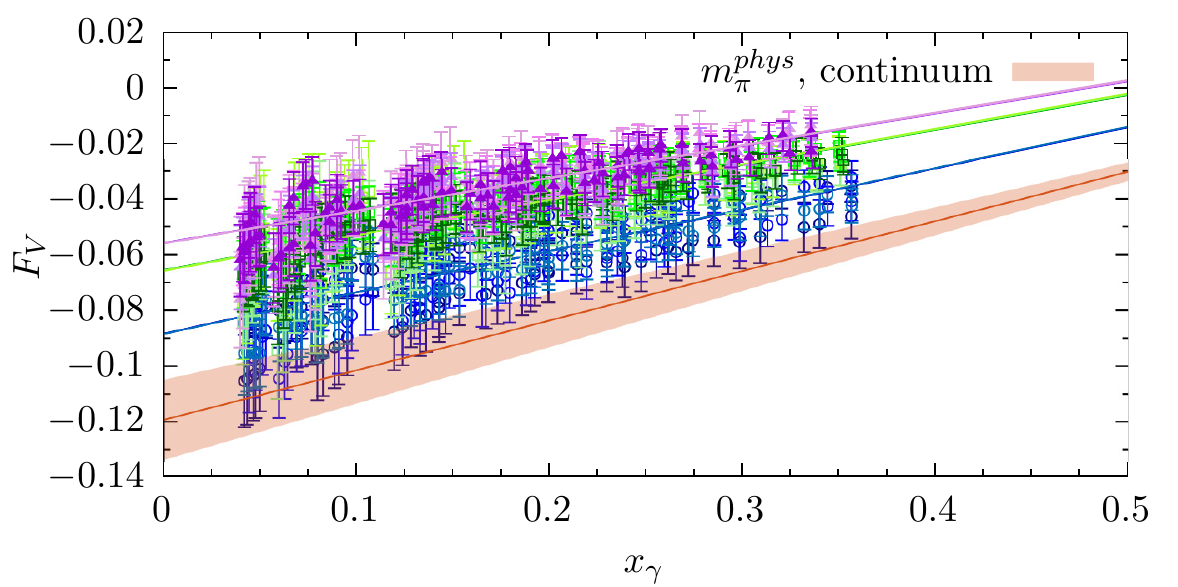}
\end{center}
\caption{
%\footnotesize  {
\it The form factors $F_A(x_\gamma)$ (upper) and $F_V(x_\gamma)$ (lower)  of  the $D_s$ meson as a function of $x_\gamma$ at three values of the lattice spacing with separate fits to the data using Eq.\,(\ref{eq:extrHeavy}) at each 
value of the lattice spacing.
The orange bands with their central red lines represent the result of a single fit to all the data extrapolated to the continuum limit and to physical quark masses. \hspace*{\fill}}
%}
\label{fig:DDd}
\end{figure}

\begin{figure}[!t]
\begin{center}
\includegraphics[width=0.46\textwidth]{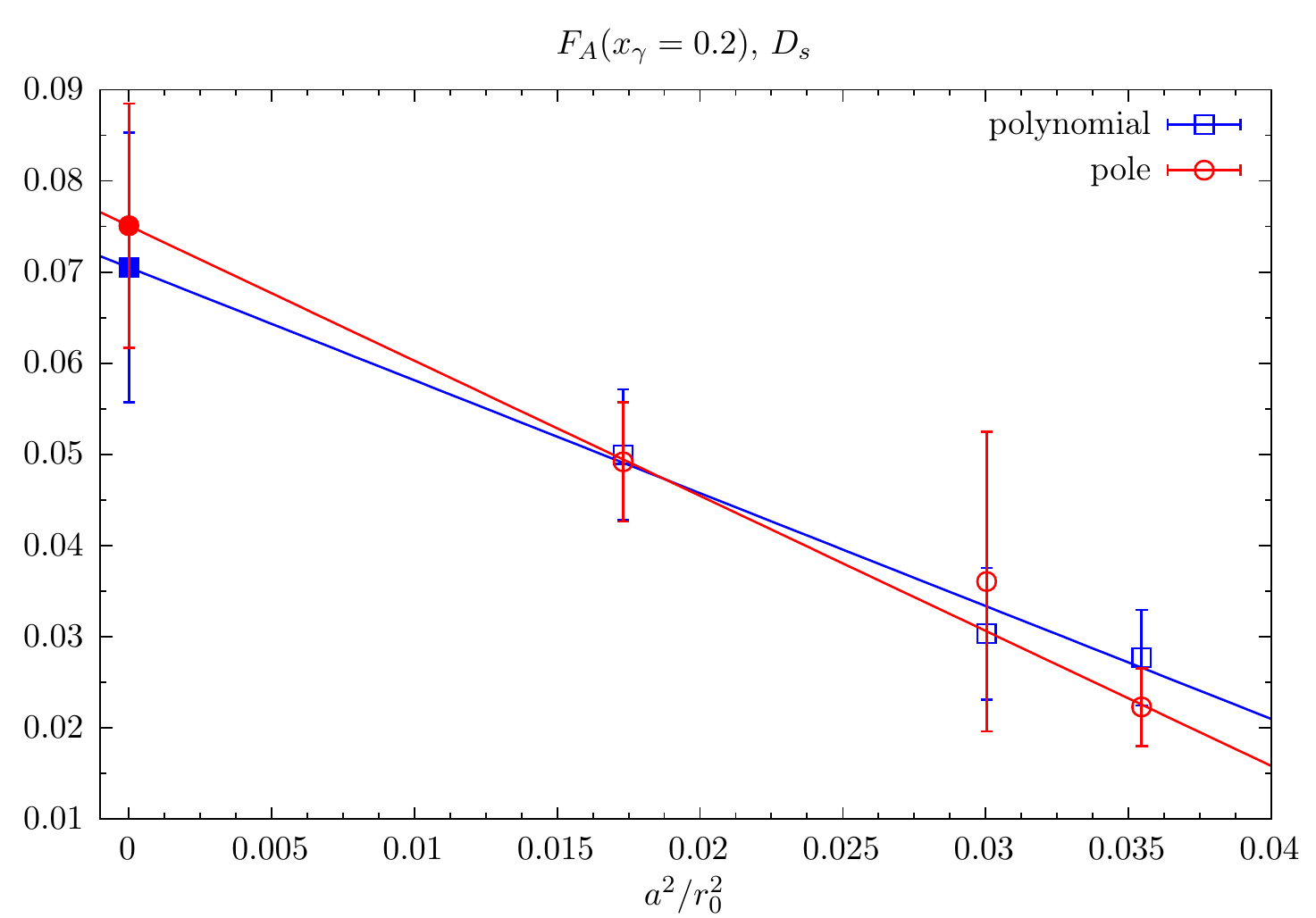}\hspace{0.3in}
\includegraphics[width=0.46\textwidth]{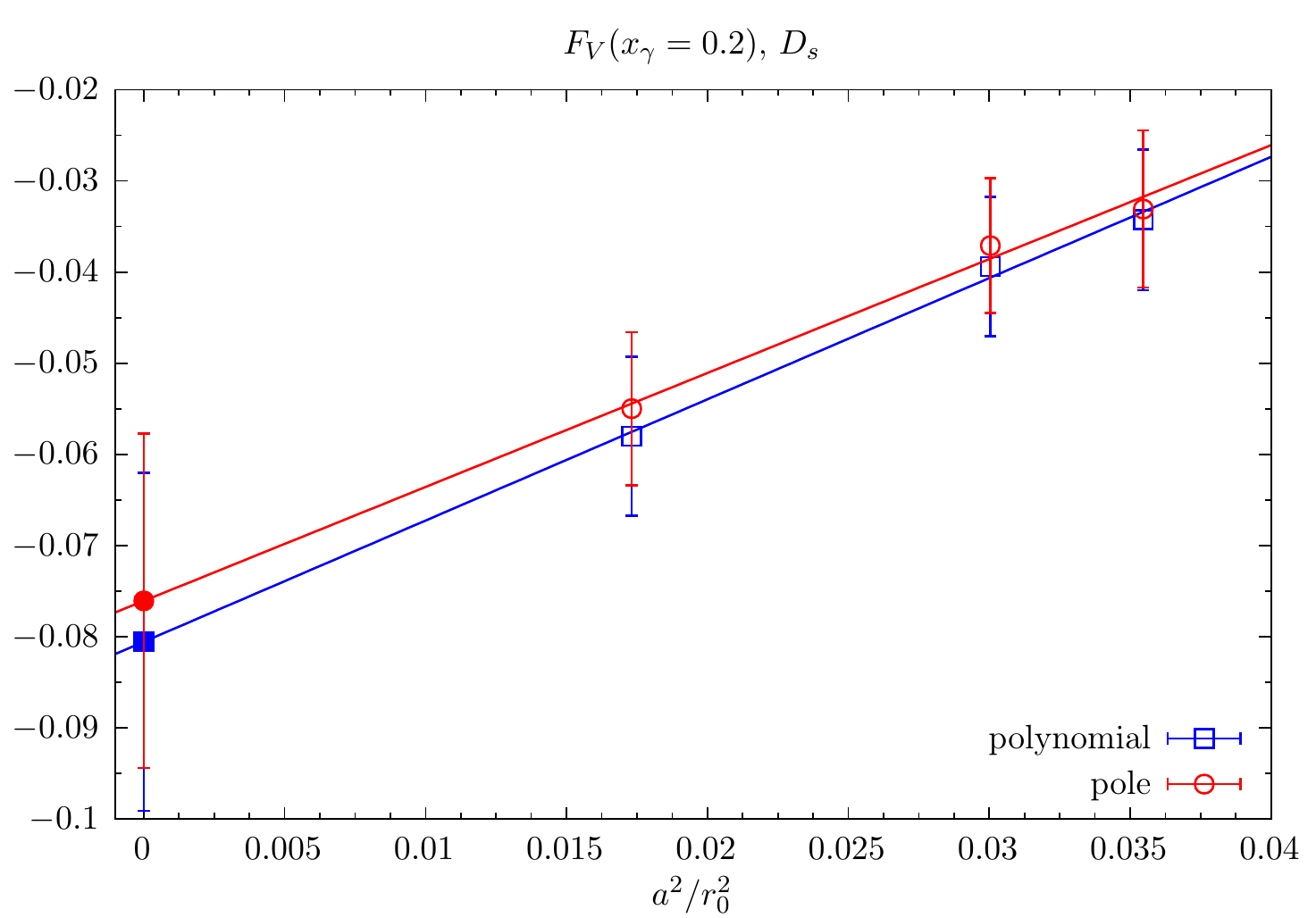}
\end{center}
\caption{
%\footnotesize  {
\it The\! form factors\! $F_A$ (left) and\! $F_V$ (right) of  the $D_s$ meson at $x_\gamma=0.2$ as functions of $a^2$. The polynomial and pole fits correspond to Eqs.\,(\ref{eq:extrHeavy}) and 
(\ref{eq:extrHeavyb}) respectively.
\hspace*{\fill}}
%}
\label{fig:FAFVxg.2}
\end{figure}

In Fig.\,\ref{fig:DDd}  we present the values of the form factors $F_A(x_\gamma)$ (upper) and $F_V(x_\gamma)$ (lower) for the $D_s$ meson as a function of $x_\gamma$.   We show the data obtained at the three different values of the lattice spacing, together with fits using Eq.\,(\ref{eq:extrHeavy}) at each value of the lattice spacing.  The orange bands with their central red lines are the results of a single fit to all the data after extrapolation to the continuum limit and to physical quark masses. The discretisation artefacts, which include ones of $O(m_c^2\,a^2)$, while approximately of the expected size, appear to be relatively large because the form factors are small. In fact the form factors at the three lattice spacings we have at our disposal  are fully consistent, within our uncertainties, with a linear behaviour in $a^2$, as illustrated in Fig.\,\ref{fig:FAFVxg.2} where the form factors at $x_\gamma=0.2$ are presented as a function of the lattice spacing. The points in the figure are obtained after extrapolation to physical quark masses either using a polynomial of pole ansatz corresponding to Eqs.\,(\ref{eq:extrHeavy}) or (\ref{eq:extrHeavyb}) at fixed lattice spacing.
In this first study, with only three lattice spacings at our disposal, we are unable to include corrections of higher order in $a^2$ beyond those present in Eqs.\,(\ref{eq:extrHeavy}) and~(\ref{eq:extrHeavyb}). In Appendix\,\ref{app:numres} we have estimated their effects in the uncertainties of our final results for the form factors.

We also study our physical results (i.e those obtained after the continuum and chiral extrapolations) as a function of $x_\gamma$ by fitting them to the following linear expressions: 
\begin{equation} F^{P}_{A,V}(x_\gamma) = C^{P}_{A,V} + D^{P}_{A,V}\, x_\gamma\,, 
\label{eq:linearfit}\end{equation}
where $P$ represents each of the pseudoscalar mesons, $\pi,\,K,\,D$ and $D_s$.

For the axial form factors we find:
\begin{align}
&\hspace{-0.5cm} C^{\pi}_A =0.010 \pm 0.003 \, ;
 &&\, D^{\pi}_A = 0.0004 \pm 0.0006 \, ;
 && \rho_{C^{\pi}_A , D^{\pi}_A}=-0.419\,; \nonumber \\
 &\hspace{-0.5cm} C^{K}_A =0.037 \pm 0.009 \, ;
 &&\, D^{K}_A =-0.001 \pm 0.007 \,;
 && \rho_{C^{K}_A , D^{K}_A}=-0.673 \,; \nonumber  \\
 &\hspace{-0.5cm} C^{D}_A = 0.109 \pm 0.009 \, ;
 &&\, D^{D}_A = -0.10\pm 0.03 \,;
 && \rho_{C^{D}_A , D^{D}_A}=-0.557 \,; \nonumber  \\
 &\hspace{-0.5cm} C^{D_s}_A =0.092 \pm 0.006 \, ;
 &&\, D^{D_s}_A = -0.07 \pm 0.01\, ;
 && \rho_{C^{D_s}_A , D^{D_s}_A}=-0.745\,.
   \label{eq:fittatiA}
  \end{align}
and for the vector form factors we obtain
\begin{align}
 &\hspace{-0.5cm} 
    C^{\pi}_V =0.023 \pm 0.002 \, ;
 &&\, D^{\pi}_V = -0.0003 \pm 0.0003 \,;
  && \rho_{C^{\pi}_V , D^{\pi}_V}=-0.570 \,;\nonumber \\
 &\hspace{-0.5cm} 
   C^{K}_V = 0.12 \pm 0.01 \, ;
 &&\, D^{K}_V = -0.02 \pm 0.01 \,;
 && \rho_{C^{K}_V , D^{K}_V}=-0.714\,;\nonumber  \\
 &\hspace{-0.5cm} 
   C^{D}_V =-0.15 \pm 0.02 \, ;
 &&\, D^{D}_V = 0.12 \pm 0.04 \,;
 && \rho_{C^{D}_V , D^{D}_V}=-0.580\,;\nonumber  \\
 &\hspace{-0.5cm}
   C^{D_s}_V =-0.12 \pm 0.02 \, ;
 &&\, D^{D_s}_V = 0.16 \pm 0.03 \,;
&& \rho_{C^{D_s}_V , D^{D_s}_V}=-0.900\,.
  \label{eq:fittatiV}
  \end{align}
In Eqs.\,(\ref{eq:fittatiA}) and (\ref{eq:fittatiV}), for each of the $C$'s and $D$'s, $\rho_{C,D}$ is the correlation between them, defined by
\begin{eqnarray}
\rho_{C,D}=\frac{\sum_i (C_i - \mu_C)(D_i-\mu_D)}{\sqrt{\sum_i (C_i - \mu_C)^2} \sqrt{\sum_i (D_i - \mu_D)^2}}\,,
\qquad \mu_{C}=\frac{1}{N}\sum_i C_i\,,\quad \mu_D=\frac{1}{N}\sum_i D_i\,,
\end{eqnarray}
where $C_i$ and $D_i$ are the jackknife samples and the sum runs over all the jackknifes following the procedure in Appendix A of Ref.\,\cite{DelDebbio:2007pz}.

For the pion and kaon we can compare the constants $C^{\pi,K}_{A,V}$ in Eqs.\,(\ref{eq:fittatiA}) and (\ref{eq:fittatiV}) with the constant (i.e. $x_\gamma$-independent) values obtained in ChPT using Eq.\,(\ref{eq:lochpt}):  $F^\pi_A=0.0119$,  $F^\pi_V=0.0254$, $F^K_A=0.042$, $F^K_V= 0.096$.

In the remainder of this section we present a brief comparison of our results with experimental data.   A more detailed phenomenological analysis will be presented in a separate paper.

For the pion the Particle Data Group (PDG)\,\cite{Tanabashi:2018oca} quotes the following results:  $F_A^\pi = 0.0119\, (1)$ (this value  comes from fixing the vector form factor at the CVC prediction from   $\pi^0 \to \gamma\gamma$ decays, $F_V^\pi(x_\gamma=0)= 1/\alpha \sqrt{2 \Gamma(\pi^0 \to \gamma\gamma)/(\pi  m_{\pi^0})}=0.0259\,(5)$) and $F_V^\pi = 0.025\, (2)$ in nice  agreement with our results, respectively $F_A^\pi =C^{\pi}_A =0.010\, (3)$ and $F_V^\pi =C^{\pi}_V =0.023\,(2)$.   Also the slope of $F_V^\pi(x_\gamma)$  has been measured from the expression $F_V^\pi(x_\gamma) =  F_V^\pi(1) \,  (1 + \lambda  (1 - x_\gamma))$ with the result  $\lambda = 0.10 \,(6)$, to be compared with our result   $\lambda=-D^{\pi}_V/(C^{\pi}_V+D^{\pi}_V)= 0.011(12)$.

For the kaon the PDG  quotes  the  two combinations  $F_V^K \pm F_A^K$. They present separate values obtained from $K\to e$ decays, $F^K_V + F^K_A=0.133\, (8)$, and from $K\to\mu$ decays, $F_V^K + F_A^K=0.165\, (13)$. Of course the results should be independent of whether the final-state charged lepton is an electron or muon. 
For this combination of form factors our value is $F_V^K + F_A^K=0.161 \pm 0.013$ at $x_\gamma=0$ and $F^K_A+F^K_V= 0.1363 \pm 0.0096$ at $x_\gamma=1$.   For the other combination of form factors the PDG quotes $F^K_A - F^K_V = - 0.21\, (6)$ obtained from $K\to\mu$ decays, which is quite different from our result $F_A^K - F^K_V =  -0.087 \pm 0.013$ at $x_\gamma=0$ or  $F^K_A-F^K_V= -0.06 \pm 0.01$  at $x_\gamma=1$.  From $K \to e$ decays there is only the upper bound $F^K_A(0) - F^K_V(0)< 0.49$. 
%There is also a measurement of the slope $\lambda=-0.38\, (20)\,(2)$ for be compared with our results $-D^{K}_V/(C^{K}_V+D^{K}_V)=0.24 (11)$.  
%A more detailed phenomenological analysis will be presented in a future publication.

The   results in Eqs.\,(\ref{eq:fittatiA})  and (\ref{eq:fittatiV}) can be combined with the values of the decays constants computed in Ref.\,\cite{Carrasco:2014cwa} and  \cite{Carrasco:2014poa}
\begin{align}  
f_\pi&=(130.41\pm 0.20 )\, {\rm MeV}     &&f_K=(155.0 \pm  1.9    )\, {\rm MeV} \nonumber \\f_D&=(207.4\pm 3.8 )\, {\rm MeV}   &&f_{D_s}=(247.2 \pm  4.1    )\, {\rm MeV}  \,,
%f_\pi&=&(129 \pm  4 )\, {\rm MeV} \qquad     f_K=(154 \pm  6    )\, {\rm MeV} \quad       (FLAG 155.7(3)\, {\rm MeV})
%\nonumber \\  f_D&=& (234\pm 10  )\, {\rm MeV} \quad       (FLAG 219(6)\, {\rm MeV})
%\qquad   f_{D_s}=  (258\pm  34   )\, {\rm MeV} \quad      (FLAG 246(3)\, {\rm MeV})\, , 
\end{align}
to compute the differential or total decay rate using the expressions given in Appendix\,\ref{sec:decayrateformulae}.
%f_D = 207.4(3.8) MeV, f_{D_s} = 247.2(4.1) MeV

For completeness, we also present the constants $\tilde C^{D_{(s)}}_{A,V}$ and $\tilde D^{D_{(s)}}_{A,V}$ which appear 
in the pole representation of the form factors for $D$ and $D_s$ mesons, 
\begin{eqnarray} F^{D_{(s)}}_{A,V}(x_\gamma)= \frac{\tilde C^{D_{(s)}}_{A,V} }{1+\tilde D^{D_{(s)}}_{A,V}\,  x_\gamma }  \, : \label{eq:extrHeavyc}\end{eqnarray}
 \begin{align}
  &\hspace{-0.5cm}\tilde  C^{D}_A =0.112 \pm 0.009 \, ;
 &&\, \tilde  D^{D}_A = 1.3 \pm 0.4 \,;\quad 
  && \rho_{\tilde C^{D}_V ,\tilde  D^{D}_V}=0.346\,; \nonumber  \\
 &\hspace{-0.5cm}
\,  \tilde  C^{D}_V =-0.15 \pm 0.02 \, ;
 &&\, \tilde  D^{D}_V= 1.2 \pm 0.4 \, ;
  && \rho_{\tilde  C^{D}_V ,\tilde  D^{D}_V}=-0.383\,;\label{eq:fittatiD}\\ 
&\hspace{-0.5cm}\tilde C^{D_{s}}_{A} =0.094 \pm 0.006 \, ;
&&\, \tilde D^{D_{s}}_{A} =1.1 \pm 0.2\, ;\quad 
&& \rho_{\tilde C^{D_s}_V ,\tilde  D^{D_s}_V}=0.546\,; \nonumber  \\
&\hspace{-0.5cm}
 \,  \tilde C^{D_{s}}_{V} =-0.12 \pm 0.02\, ;
&&\, \tilde D^{D_{s}}_{V}= 2.6 \pm 0.2 \, ;
&& \rho_{\tilde C^{D_s}_V ,\tilde  D^{D_s}_V}=-0.373\,.
\label{eq:fittatiDs}
\end{align}

%We are currently improving our lattice data and, after a detailed analysis of all the systematics,  we shall provide first-principles phenomenologically relevant results for the form factors in the full kinematical range  for both  light and  heavy  mesons. 
%The form factors for heavy mesons will represent in this respect a totally  unexplored field of investigation while,  in the case of light mesons, our first-principle results will make it possible to avoid  ChPT in phenomenological analyses.
\section*{Conclusions}
In conclusion we have shown that by using lattice QCD, even with moderate statistics, it is possible to predict with good precision  the structure dependent  form factors $F_A$ and $F_V$ relevant for $P \to \ell \bar \nu_\ell \gamma$ decays for both  light and  heavy mesons   and  that it is also  possible to extract their momentum dependence.    Previous determinations of these quantities relied either  on ChPT for light mesons  or on the heavy quark expansion and model-dependent assumptions for heavy mesons.  Our work shows that it is possible to compute the relevant  form factors from first principles. 
 
We found that the extraction of the axial form factor $F_A$ at small values of $x_\gamma$  is problematic because of the presence of very large discretisation  effects   of   $O\left(a^2 /\left(r_0^2 x_\gamma\right)\right)$ and we provided a procedure for the non-perturbative cancellation of these systematic errors. 
We also found that for charmed mesons the discretisation effects of $O(a^2 m_H^2)$, while of the expected order of magnitude, are large relative to the small size of the form factors. Nevertheless the results for the form factors at the three lattice spacings are consistent, within our uncertainties, with a linear behaviour in $a^2$. Simulations on one or more finer lattices would enable us to improve our estimates of the higher order artefacts and hence reduce the corresponding systematic uncertainty.
Such preliminary studies of charmed mesons are also essential in order to study radiative decays of $B$ mesons in the future. In this respect  the use of the ratio method may also be very useful\,\cite{Blossier:2009hg}. 

Although the present study clearly can and will be improved by, for example, increasing the statistics, covering the full range of $x_\gamma$ for $D$ and $D_s$ mesons or simulating on a finer lattice, the results presented in this work already allow for an accurate  comparison of  the  theoretical predictions with experimental measurements and we will discuss the phenomenological implications of our results in a forthcoming paper.
 
In future we also plan to study the emission of off-shell photons ($k^2\neq 0$), computing all four form factors appearing in Eq.\,(\ref{eq:ffdef}), which would allow us to predict the rates for processes in which the pseudoscalar meson decays into four leptons. These processes are very interesting in the search of physics beyond the Standard Model\,\cite{Barger:2011mt,Batell:2011qq,Albrecht:2019zul}.  %ooooooooooooooooooooooooooooooooooooooooooooooooooooooooooooooooooooooooooooooooooooooooooooo
\begin{acknowledgments}
We gratefully acknowledge helpful discussions with M.\,Testa. We acknowledge PRACE for awarding us access to Marconi at CINECA, Italy under the grant Pra17-4394.
We also acknowledge use of CPU time provided by CINECA under the
specific initiative INFN-LQCD123.
V.L., G.M. and S.S. thank MIUR  (Italy)  for  partial  support  under  the  contract  PRIN 2015. 
C.T.S. was partially supported by STFC (UK) grant\,ST/P000711/1 and by an Emeritus Fellowship from the Leverhulme Trust.
N.T.   and R.F. acknowledge the University of Rome Tor Vergata for the support granted to the project PLNUGAMMA.
F.S and S.S are supported by the Italian Ministry of Research (MIUR) under grant PRIN 20172LNEEZ. F.S is supported by INFN under GRANT73/CALAT.
\end{acknowledgments}
%%
%%ooooooooooooooooooooooooooooooooooooooooooooooooooooooooooooooooooooooooooooooooooooooooooooo

\newpage
\appendix

%%%
%%%
\section{Expressions for the decay rates in terms of $\bm{F_V}$ and $\bm{F_A}$}
\label{sec:decayrateformulae}
%%%
%%%
%In Ref.\,\cite{Carrasco:2015xwa} we proposed a strategy to compute non-perturbatively  the e.m.\,corrections to the processes $P\to \ell \bar \nu_\ell (\gamma)$  at order $\alpha_{em}$.    
In this appendix we present the explicit formulae  needed to evaluate the total and differential decay rates  at order $\alpha_{em}$, combining the non-perturbative determination of the virtual corrections computed with the approach of Ref.\,\cite{Carrasco:2015xwa} with the calculation  of the structure-dependent (SD) form factors $F_A$ and $F_V$  determined with the method proposed in this paper.    These formulae can be used to compute the  double differential decay rates $d^{\hspace{1pt}2}\Gamma/(dx_\gamma dx_\ell)$, the single differential decay rates, $d\Gamma/ dx_\ell$ or  $d\Gamma/dx_\gamma $,   as well as the integrated decay rate $\Gamma(\Delta E_\gamma) = \int_0^{2 \Delta E_\gamma/m_P} \,  dx_\gamma \, \left(d\Gamma/dx_\gamma\right)$ ($\Delta E_\gamma$ is the upper limit on the energy of the emitted photon in the meson rest-frame).

The exchange of a virtual photon depends on the hadron  structure,  since all momentum modes are included, and the amplitude must therefore be computed non-perturbatively.   On the other hand, the non-perturbative evaluation of the amplitude for the emission of a real photon is not strictly necessary\,\cite{Carrasco:2015xwa}.   Indeed, it is possible to compute the amplitudes for real-photon emission in perturbation theory when $x_\gamma$ is sufficiently small that the internal structure of the decaying meson is not resolved.  The infrared divergences in the non-perturbatively computed amplitude with the exchange of a virtual photon are cancelled in the decay rates by those present in the emission of a real photon, even when the latter is computed perturbatively. The reason for this cancellation is the universality of the infrared behaviour of the theory (i.e. the infrared divergences do not depend on the structure of the decaying hadron).  For large photon energies, for example those present in the decays of heavy mesons, a full non-perturbative determination of the relevant amplitudes is necessary.

To calculate the partial rates for the emission of a hard real photon it is sufficient to know the SD form factors, $F_A$ and $F_V$, and the meson's decay constant $f_P$.  For the integrated rate  $\Gamma(\Delta E_\gamma)$ instead, in the intermediate steps of the calculation it is necessary to introduce an infrared regulator.   To this end, in order to work with quantities that are finite when the infrared regulator is removed, it  is very useful to organise the  inclusive rate $\Gamma(\Delta E_\gamma)=\Gamma(P^- \to \ell^- \bar\nu_\ell (\gamma))\vert_{E_\gamma\le \Delta E_\gamma} $ as follows
\bea 
%     \Gamma(P^\pm \to \ell^\pm \nu_\ell [\gamma])  
     \Gamma(\Delta E_\gamma)&=& 
%      \Gamma_0 + \Gamma_1^{\textrm{pt}}(\Delta E_\gamma) \nonumber \\  & & = 
       \displaystyle \lim_{L \to \infty} \left[ \Gamma_0(L) - \Gamma_0^{\textrm{pt}}(L) \right] +     \displaystyle \lim_{\mu_\gamma \to 0} \left[ \Gamma_0^{\textrm{pt}}(\mu_\gamma) +  \Gamma_1^{\textrm{pt}}(\Delta E_\gamma, \mu_\gamma) \right] \nonumber \\  & & \hspace{0.4in}+   
         %\displaystyle \lim_{\mu_\gamma \to 0} 
       \left[ \Gamma_1(\Delta E_\gamma) -  \Gamma_1^{\textrm{pt}}(\Delta E_\gamma) \right] \, , 
     \label{eq:Gamma}
 \eea
where the subscripts $0, 1$ indicate the number of photons in the final state, while the superscript $\mathrm{pt}$ denotes the point-like approximation of the decaying meson and $\mu_\gamma$ is an infrared regulator.
On the right-hand side of Eq.\,(\ref{eq:Gamma}) the quantities $\Gamma_0(L)$  and  $\Gamma_1(\Delta E_\gamma)$  are evaluated on the lattice.   

The terms in the first parentheses on the right-hand side of Eq.\,(\ref{eq:Gamma}), $\Gamma_0(L)$ and $\Gamma_0^{\textrm{pt}}(L)$, have the same infrared divergences which therefore cancel in the difference.  Here we   use the lattice size $L$ as the intermediate infrared regulator by working in the QED$_\mathrm{L}$ formulation of QED in a finite volume\,\cite{Hayakawa:2008an}   but any other  consistent formulation of QED on the lattice can also be used. The difference $\left[ \Gamma_0 - \Gamma_0^{\textrm{pt}} \right]$ is independent of the regulator as this is removed\,\cite{Lubicz:2016xro}. 
 $\Gamma_0(L)$ depends on the structure of the decaying meson  and is  computed non-perturbatively\,
 Refs.\,\cite{Lubicz:2016xro,Lubicz:2016mpj,Tantalo:2016vxk,Giusti:2017dwk,DiCarlo:2019thl}. 

In the terms in the second parentheses on the right-hand side of Eq.\,(\ref{eq:Gamma}) the decaying meson  is taken to be a point-like charged particle and both $\Gamma_0^{\textrm{pt}}(\mu_\gamma)$ and $\Gamma_1^{\textrm{pt}}(\Delta E_\gamma, \mu_\gamma)$ can be computed  directly in infinite volume, in perturbation theory, using  some  infrared regulator, for example a photon mass $\mu_\gamma=m_\gamma$. Each term is infrared divergent, but the sum is convergent\,\cite{Bloch:1937pw} and independent of the infrared regulator. In Refs.\,\cite{Carrasco:2015xwa} and \cite{Lubicz:2016xro} the explicit perturbative calculations of $\left[ \Gamma^{\textrm{pt}}_0(\mu_\gamma)+\Gamma^{\textrm{pt}}_1(\Delta E_\gamma,\mu_\gamma) \right]$ and $\Gamma_0^{\textrm{pt}}(L)$ have been performed  with a small photon mass $\mu_\gamma$ or using the finite volume respectively, as the infrared regulators. 

Finally, the term on second line of the right-hand side of Eq.\,(\ref{eq:Gamma}) is infrared finite. It can be computed in the infinite-volume limit requiring only knowledge of the
structure dependent form factors,  $F_A(x_\gamma)$ and $F_V(x_\gamma)$ and of the meson's decay constant $f_P$
\beq  \left[ \Gamma_1(\Delta E_\gamma) -  \Gamma_1^{\textrm{pt}}(\Delta E_\gamma) \right]=  
\Gamma_{\mathrm{SD}}(\Delta E_\gamma) +\Gamma_{\mathrm{INT}}(\Delta E_\gamma) \, , \eeq
where $\Gamma_{\mathrm{SD}}$ is the structure-dependent contribution and $\Gamma_{\mathrm{INT}}$ is that from the interference between the SD and point-like components of the amplitudes. Both $\Gamma_\mathrm{SD}$ and $\Gamma_\mathrm{INT}$ are separately infrared finite and there is no need to introduce an infrared regulator in this term.

We express the differential  decay rate in terms of the following quantities:
\begin{itemize} 
\vspace{-0.1in}\item the two dimensionless kinematical variables
\begin{flalign} 
x_\gamma= \frac{2p\cdot k}{m_P^2}\;,
\qquad
x_\ell= \frac{2p\cdot p_\ell-m_\ell^2}{m_P^2}\;,
\end{flalign} 
where  $m_\ell$ the mass of the lepton $\ell$, $1-x_\gamma+x_\gamma r_\ell^2/(1-x_\gamma)\le x_\ell \le 1$ and    $0\le x_\gamma \le 1-r_\ell^2$, with   $
r_\ell=m_\ell/m_P$;
\item   the decay constant of the meson $f_P$; 
 \item the two SD axial and vector form factors $F_A$ and $F_V$.
 \end{itemize}

 The differential decay rate is given by the sum of three contributions,
\begin{flalign}
\frac{d^2\Gamma}{dx_\gamma dx_\ell}=
\frac{\alpha_{em}\, \Gamma^{(0)}}{4\pi}\, 
\left\{\frac{d^2\Gamma_{\mathrm{pt}}}{dx_\gamma dx_\ell}+
\frac{d^2\Gamma_{\mathrm{SD}}}{dx_\gamma dx_\ell}+
\frac{d^2\Gamma_{\mathrm{INT}}}{dx_\gamma dx_\ell}
\right\}\, , 
\label{eq:dgamma}
\end{flalign}
where  $\Gamma^{(0)}$ is the leptonic decay rate in the  absence of electromagnetic corrections. This is given by
\begin{flalign}
\Gamma^{(0)}= \frac{G_F^2\vert V_{CKM}\vert^2 f_P^2}{8\pi} m_P^3 r_\ell^2\left(1-r_\ell^2 \right)^2\;,
\end{flalign}
where $G_F$ is the Fermi's constant and $V_{CKM}$ the relevant CKM matrix element. 

The quantities in the braces on the right-hand side of Eq.\,(\ref{eq:dgamma}) are given by
\begin{eqnarray}
&&\frac{d^{\hspace{1pt}2}\Gamma_{\mathrm{pt}}}{dx_\gamma dx_\ell}
= 
\frac{
2\, f_{\mathrm{pt}}(x_\gamma,x_\ell)
}{(1-r_\ell^2)^2}\;,
\nonumber \\
%\nonumber \\
\rule[30pt]{0pt}{0pt}
&&\frac{d^{\hspace{1pt}2}\Gamma_{\mathrm{SD}}}{dx_\gamma dx_\ell}= 
\frac{
m_P^2\,
\left\{
\left[F_V(x_\gamma)+F_A(x_\gamma)\right]^2\,f_{\mathrm{SD}}^+(x_\gamma,x_\ell)
+
\left[F_V(x_\gamma)-F_A(x_\gamma)\right]^2\,f_{\mathrm{SD}}^-(x_\gamma,x_\ell)
\right\}
}{2f_P^2\, r_\ell^2(1-r_\ell^2)^2}\,,
\nonumber \\
%\nonumber \\
\rule[30pt]{0pt}{0pt}&&\frac{d^{\hspace{1pt}2}\Gamma_{\mathrm{INT}}}{dx_\gamma dx_\ell}
= -
\frac{
2m_P\, 
\left\{
\left[F_V(x_\gamma)+F_A(x_\gamma)\right]\,f_{\mathrm{INT}}^+(x_\gamma,x_\ell)
+
\left[F_V(x_\gamma)-F_A(x_\gamma)\right]\,f_{\mathrm{INT}}^-(x_\gamma,x_\ell)
\right\}
}{f_P\, (1-r_\ell^2)^2}\label{eq:d2Gammas}
\end{eqnarray}
and correspond  to the contribution of the point-like approximation, to the SD contribution and to the interference between point-like and SD terms respectively. The kinematical functions appearing in Eq.\,(\ref{eq:d2Gammas}) are given by
\begin{flalign}
&
f_{\mathrm{pt}}(x_\gamma,x_\ell)
=
\frac{1-x_\ell}{x_\gamma^2(x_\gamma+x_\ell-1)}
\left[
x_\gamma^2+2(1-x_\gamma)(1-r_\ell^2)-
\frac{2x_\gamma r_\ell^2(1-r_\ell^2)}{x_\gamma+x_\ell-1}
\right] \;,
\nonumber \\
\nonumber \\
&
f_{\mathrm{SD}}^+(x_\gamma,x_\ell)
=
(x_\gamma+x_\ell-1)\, \left[
(x_\gamma+x_\ell-1+r_\ell^2)(1-x_\gamma)-r_\ell^2
\right] \;,
\nonumber \\
\nonumber \\
&
f_{\mathrm{SD}}^-(x_\gamma,x_\ell)
=
-(1-x_\ell)\, \left[
(x_\ell-1+r_\ell^2)(1-x_\gamma)-r_\ell^2
\right] \;,
\nonumber \\
\nonumber \\
&
f_{\mathrm{INT}}^+(x_\gamma,x_\ell)
=
-\frac{1-x_\ell}{x_\gamma  \, (x_\gamma+x_\ell-1)}
\left[
(x_\gamma+x_\ell-1+r_\ell^2)(1-x_\gamma)-r_\ell^2
\right] \;,
\nonumber \\
\nonumber \\
&
f_{\mathrm{INT}}^-(x_\gamma,x_\ell)
=
\frac{1-x_\ell}{x_\gamma \, (x_\gamma+x_\ell-1)}
\left[
x^2_\gamma+(x_\gamma+x_\ell-1+r_\ell^2)(1-x_\gamma)-r_\ell^2
\right] \;.
\end{flalign}
%
%Note that the functions $x_\gamma^2 g_{SD}^\pm(x_\gamma,x_\ell)$ and $x_\gamma g_{INT}^\pm(x_\gamma,x_\ell)$ correspond respectively to 
The distribution with respect to the photon's momentum is obtained after integrating over the lepton's momentum
\bea   \frac{d\Gamma}{dx_\gamma}=\int^1_{x^{\rm min}_\ell(x_\gamma)} \, dx_\ell\,  \,  \frac{d^2\Gamma}{dx_\gamma dx_\ell}\,  . \eea   
As $x_\gamma \to 0$ the allowed kinematical range for $x_\ell$ is squeezed around its maximum,  $x^{\rm min}_\ell(x_\gamma)=1-x_\gamma+x_\gamma r_\ell^2/(1-x_\gamma) \le x_\ell \le 1$.  Thus, with  the  exception of  the contribution proportional to $f_{\mathrm{pt}}(x_\gamma,x_\ell)\sim 1/x_\gamma^2$,  all the other  contributions  vanish in the soft-photon region, which is consequently  dominated by the pointlike (eikonal) result
\bea   \frac{d\Gamma}{dx_\gamma} \sim \int^1_{x^{\rm min}_\ell(x_\gamma)} \, dx_\ell\,  \,  \frac{d^2\Gamma_{pt}}{dx_\gamma dx_\ell}\sim 1/x_\gamma \,  . \label{eq:dGammair}\eea 
%The functions $f_{SD}^\pm(x_\gamma,x_\ell)$ and $f_{INT}^\pm(x_\gamma,x_\ell)$ where defined in appendix B of  Ref.\,\cite{Carrasco:2015xwa}.
%We prefer to use $ g_{SD}^\pm(x_\gamma,x_\ell)$ and $ g_{INT}^\pm(x_\gamma,x_\ell)$ for the following reason. 
The $1/x_\gamma$ behaviour of the differential rate at small $x_\gamma$ leads to a logarithmic infrared divergence in the total rate. It is cancelled by the infrared divergence in the $O(\alpha_{em})$ virtual corrections to the inclusive decay rate. The SD and INT contributions vanish at small $x_\gamma$.

Eqs.\,(\ref{eq:dgamma})-(\ref{eq:dGammair}) allow us to compute the spectrum $d\Gamma/dx_\gamma$. 
We advocate organising the determination of the integrated rate in terms of the three sets of parentheses on the right-hand side of Eq.\,(\ref{eq:Gamma}). The procedure to evaluate the term in the first parentheses, $\Gamma_0(L) - \Gamma_0^{\textrm{pt}}(L)$, is explained in detail in Ref.\,\cite{Carrasco:2015xwa}, where the explicit expression for the term in the second parentheses, $\Gamma_0^{\textrm{pt}}(\mu_\gamma) +  \Gamma_1^{\textrm{pt}}(\Delta E_\gamma, \mu_\gamma)$, can also be found. The third term on the right-hand side of Eq.\,(\ref{eq:Gamma}), $\Gamma_1(\Delta E_\gamma) -  \Gamma_1^{\textrm{pt}}(\Delta E_\gamma)=\Gamma_{\mathrm{SD}}+\Gamma_{\mathrm{INT}}$, is the subject of this paper. As explained above, both $\Gamma_{\mathrm{SD}}$ and $\Gamma_{\mathrm{INT}}$  
are infrared finite and are obtained by integrating the differential rates over the physical range of $x_\gamma$,
\begin{equation}   
\Gamma_{\mathrm{SD}}(\Delta E_\gamma)=\int_0^{2 \Delta E_\gamma/m_P} \hspace{-10pt}dx_\gamma \, \frac{d\Gamma_{\mathrm{SD}}}{dx_\gamma}\,,
\qquad\quad\Gamma_{\mathrm{INT}}(\Delta E_\gamma)=\int_0^{2 \Delta E_\gamma/m_P} \hspace{-10pt}dx_\gamma \, \frac{d\Gamma_{\mathrm{INT}}}{dx_\gamma}
\,. \end{equation}

%%%
%%%
\section{Calculating matrix elements from finite Euclidean lattices}
\label{sec:eucorr}
%%%
%%%
In this appendix we derive some useful  formulae for the extraction  of the two relevant form factors, $F_{A,V}$, from the Euclidean correlation   functions expressed in terms of lattice operators  on a  lattice with finite  time extent
$T$. 
%We found it  convenient to derive these formulae by introducing a photon source at $t=T/2$
%and then reducing the external photon leg in complete analogy with the procedure  usually adopted for the external hadron states. Although not strictly necessary, this is very helpful to take correctly into account of the periodicity conditions in time. 
%
%Our starting point  is the following correlation function 
%%
%\begin{flalign}
%\mathcal{C}^{\alpha r}_W(t,T/2;\vec k,\vec p)=\sum_{i,j=1}^3
%\sum_{y,\vec z}\, e^{-i\vec k\cdot (\vec z+\vec{\hat i}/2)} \epsilon_i^r(k)\, D_{ij}(T/2,\vec z;y)\, \mathcal{M}_W^{\alpha j}(y,t;\vec p)\;.
%\label{eq:correlator}
%\end{flalign}
%%
%In the previous formula $\mathcal{M}_W^{\alpha\nu}$, which will be described in all details in the following,  is the quantity that contains all the hadronic information of the radiative amplitude,  while $D_{\mu\nu}(T/2,\vec z;y)$ is the
%photon propagator. In order to avoid the propagation of unphysical degrees of freedom, the photon propagator can be conveniently evaluated in Coulomb gauge.  In this gauge $D_{0\mu}(T/2,\vec z;y)=D_{\mu0}(T/2,\vec z;y)=0$ and   the sums over the Lorentz indexes appearing in the previous formula are limited to the spatial components $i$ and $j$.   With this  choice  it will straightforward to project the amplitude on the physical photon elicities to extract $F_{A,V}$.
% where the photon propagator has the other end-point and where the hadronic electromagnetic current is located.

In  order to construct the finite $T$ equivalent of $ C^{\alpha r}_W(t;\vec k,\vec p)$ in Eq.(\ref{eq:correlatorinf}), that we will denote as $\mathcal{C}^{\alpha r}_W(t,T/2;\vec k,\vec p)$,  it is convenient to define the following  hadronic  correlation function   at fixed $t$ and $t_y$ 
%, after having specified all the details concerning the photon propagator.
\beq   M_W^{\alpha r}(t_y,t;\vec k,\vec p)=\sum_{i=1,2,3}
\epsilon^r_i(\vec k)\, \sum_{\vec y} \sum_{\vec x}\, e^{-i\vec k \cdot (\vec y +\vec{\hat i}/2)+i\vec p\cdot \vec x}\, \mathtt{T}
 \langle j_W^\alpha(t) j^i_{em}(t_y,\vec y) P(0,\vec x) \rangle_{LT}\,  ,  \label{eq:MW}
\eeq
where $ \langle \dots \rangle_{LT}$ denotes the  average over the gauge field configurations at finite $L$ and $T$ and  we  introduced suitable independent  vectors    $\vec \epsilon^r(\vec k)$, $r=1,2$, corresponding to the physical polarisations  of the emitted  photon. A  possible simple choice, and one in which the unphysical polarisations vanish explicitly, is  given by
\begin{flalign}
\epsilon^1_\mu(\vec k)\equiv \left(0, \frac{-k_1 k_3}{\vert \vec k\vert\sqrt{k_1^2+k_2^2}},
\frac{-k_2 k_3}{\vert \vec k\vert\sqrt{k_1^2+k_2^2}}, \frac{\sqrt{k_1^2+k_2^2}}{\vert \vec k\vert}\right)\;,
\quad 
\epsilon^2_\mu(\vec k)\equiv \left(0,\frac{k_2}{\sqrt{k_1^2+k_2^2}},
-\frac{k_1}{\sqrt{k_1^2+k_2^2}}, 0\right)\; . 
\label{eq:pol1}
\end{flalign}
%
%In this gauge the unphysical polarisation vectors vanish identically, i.e. $\epsilon_\mu^0(\vec k)=\epsilon_\mu^3(\vec k)=0$. 
The    polarisation vectors satisfy
%polarisation vectors are such that
%
\begin{flalign}
\sum_{i=1}^3\epsilon^r_i(\vec k) k_i=0\;,
\qquad
\sum_{i=1}^3\epsilon^r_i(\vec k) \epsilon^s_i(\vec k)=\delta_{rs}\, . 
\label{eq:pol2}
\end{flalign}
Since in our simulations we always use $ \vec k =(0,0, \vert \vec k\vert)$,   the polarisation vectors  reduce to 
\beq  \epsilon_\mu^1=\left(0, -\frac{1}{\sqrt{2}},
-\frac{1}{\sqrt{2}}, 0\right)  \qquad \epsilon_\mu^2=\left(0, \frac{1}{\sqrt{2}}, -\frac{1}{\sqrt{2}}, 0\right)\, . \eeq

The ``topology'' of the correlation function in Eq.\,(\ref{eq:MW})  is explained in Figure\,\ref{fig:correlator}: 
\begin{itemize} \vspace{-0.2cm}
\item \vspace{-0.1cm}the incoming meson is interpolated at fixed spatial momentum $\vec p$  by the pseudoscalar operator $P$ placed at time $t=0$
\begin{flalign}
P(0)=\sum_{\vec x}\, e^{i\vec p\cdot \vec x}\, P(0,\vec x)\, ; 
\end{flalign}
\item   the hadronic weak current $j_W^\alpha(t)$ is placed at the generic time $t$. 
We  used a local discretisation of the weak current that, in the Twisted-Mass discretisation of the fermionic action used in this work\,\cite{Frezzotti:2000nk}, is explicitly given by  
\begin{flalign}
j_W^\alpha(t) = j_V^\alpha(t)-j_A^\alpha(t)\;,
\quad
j_V^\alpha(t)=Z_A\, \bar \psi_{U}(t)\gamma^\alpha\psi_{D}(t)\;,
\quad
j_A^\alpha(t)=Z_V\, \bar \psi_{U}(t)\gamma^\alpha\gamma_5\psi_{D}(t)\, , \label{eq:jWVA}
\end{flalign}
%
%{\bf ho scambiato $U$ e $D$ perché la corrente debole deve annichilire un $P^-$}
 where $j_V^\alpha(t)$ and $j_A^\alpha(t)$ are the vector and axial components that include the corresponding renormalisation factors. Note that the renormalisation factors to be used in Twisted-Mass at maximal twist are chirally-rotated with respect to the ones of standard Wilson fermions\,\cite{Frezzotti:2003ni}. In Eq.\,(\ref{eq:jWVA})
$\psi_{U}$ indicates the field of an up-type quark that, for the mesons considered in this study, can be either an up or a charm. Similarly, $\psi_{D}$ can be either a down or a strange quark field. The actions of the up-type and down-type quark fields have been discretised with opposite values of the chirally-rotated Wilson term in order to numerically
suppress O($a^2$) lattice artefacts in the meson masses\,\cite{Frezzotti:2003ni,Frezzotti:2004wz};
\item   the electromagnetic current   $j_{em}^\mu(t_y,\vec y)$,  carrying a three-momentum $\vec k$  is inserted at   $y=(t_y,\vec y)$.  This current is  defined by 
\bea j_{em}^\mu(t_y,\vec y) =  \sum_f \, q_f\, j_f^\mu(t_y,\vec y) \, , \eea
where $f$ is the flavour index, $q_f$ is equal to $2/3$ for up-type quarks and to $-1/3$ for down-type quarks.  and
\begin{flalign}
j_f^\mu(x)=-\left\{
\bar \psi_f(x)\frac{\pm i\gamma_5-\gamma^\mu}{2}\, U^\mu(x)\psi_f(x+\hat \mu)
-
\bar \psi_f(x+\hat \mu)\frac{\pm i\gamma_5+\gamma^\mu}{2}U^\mu(x)^\dagger \psi_f(x)
\right\}\,  .\label{eq:jfdef}
\end{flalign}
In Eq.\,(\ref{eq:jfdef}) $U_\mu(x)$ are the QCD link variables and the signs $\pm$ are induced by the choice made in the case of the flavour $f$ for the sign of the chirally-rotated Wilson term\,\cite{deDivitiis:2013xla}. 

We have used $e^{-i\vec k \cdot (\vec y +\vec{\hat i}/2)}$, rather than  the simpler, standard exponent $e^{-i\vec k \cdot \vec y }$, for the Fourier transform of the  current appearing in Eq.\,(\ref{eq:MW})
\begin{flalign}
j^r(t_y, \vec k) = \sum_{i=1}^3\,\epsilon^r_i(k)\, \sum_{\vec y} e^{-i\vec k\cdot (\vec y+\vec{\hat i}/2)}\, j_{em}^i(t_y,\vec y) \,  .
\end{flalign}
Our  choice of the exponent, which is  equivalent  to standard one in the continuum limit, is more convenient for the discussion of the lattice WIs 
since  we have used the point-split exactly conserved electromagnetic current in our simulations.
\item A technical subtlety needs to be stressed here. As discussed in the main text, in order to choose arbitrary (non-discretised) values of the spatial momenta for the meson and for the photon, we have introduced a ``flavoured'' extension of the electromagnetic current (see the  explanation in the caption of  Figure\,\ref{fig:disctheta}). In practice, in order to have two quarks ($\psi_0$ and $\psi_t$, where $0$ and $t$ are labels for the quark fields) having the same mass, the same electric charge, the same sign of the chirally-rotated Wilson term but different boundary conditions\,\cite{Boyle:2007wg}, the expression to be used in the numerical calculation is
\begin{flalign}
&
e^{-i\vec k\cdot (\vec x+\vec{\hat i}/2)}\, j_f^i(x)=
\nonumber \\
&
-\left\{
\bar \chi_t(x)\frac{\pm i\gamma_5-\gamma^i}{2}\, e^{\frac{i \pi(\theta^i_t+\theta^i_0)}{L}} U^i(x)\chi_0(x+\hat i)
-
\bar \chi_t(x+\hat i)\frac{\pm i\gamma_5+\gamma^i}{2}\, e^{-\frac{i \pi(\theta^i_t+\theta^i_0)}{L}}
U^i(x)^\dagger \chi_0(x)
\right\}\;,
\end{flalign}
where we have used the fact that (see Eq.\,(\ref{eq:momenta}))
\begin{flalign}
\vec k = \frac{2\pi(\vec \theta_0-\vec \theta_t)}{L}\;,
\qquad
\psi_{\{0,t\}}(x+\vec{\hat i} L) = e^{2\pi i \vec {\hat i}\cdot \vec \theta_{\{0,t\}}} \psi_{\{0,t\}}(x)\;,
\end{flalign}
and we have defined, as  usually done in implementing twisted boundary conditions\,\cite{deDivitiis:2004kq}, the periodic fields
\begin{flalign}
\chi_{\{0,t\}}(x) = e^{-\frac{2\pi i \vec x\cdot \vec \theta_{\{0,t\}}}{L}} \psi_{\{0,t\}}(x)\;.
\end{flalign}
%
%%%

\end{itemize}
 In  all the formulae that will follow the range of the time parameters is extended over the full lattice extension, $0\le t<T$ and $0\le t_y<T$.

We are now ready to define the finite $T$ correlation function
\begin{eqnarray}
\mathcal{C}^{\alpha r}_W(t,T/2;\vec k,\vec p)&=&\nonumber\\
&&\hspace{-1.2in}-i\,\theta\left(T/2-t\right) \sum_{t_y=0}^T \, \left(\theta\left(T/2-t_y\right)\, e^{E_\gamma t_y} +\theta\left(t_y-T/2\right)\, e^{-E_\gamma\, (T- t_y)}\, \right) M_W^{\alpha r}(t_y,t;\vec k,\vec p)\nonumber\\
&&\hspace{-1.2in}
-i\,\theta\left(t-T/2\right) \sum_{t_y=0}^T \, \left(\theta(T/2-t_y)  e^{-E_\gamma t_y}+\theta\left(t_y-T/2\right) \,e^{-E_\gamma\,(t_y-T) } \, \right) M_W^{\alpha r}(t_y,t;\vec k,\vec p) \, . 
\label{eq:correlator22}
\end{eqnarray}

 %

%this can be commented up to %%%

In the continuum and large-$T$ limits one can readily show that for $0\ll t \ll T/2$
\bea \mathcal{C}^{\alpha r}_W(t,T/2;\vec k,\vec p) 
\to C^{\alpha r}_W(t;\vec k,\vec p)=  H^{\alpha r}_W(k,\vec p)\, \frac{e^{-t(E-E_\gamma)}\, \bra{P}P\ket{0} }{2E} + \cdots\;,
\eea
where $C^{\alpha r}_W(t;\vec k,\vec p)$ is the correlation function introduced in the main text defined in Eq.\,(\ref{eq:correlatorinf}),
$H^{\alpha r}_W(k,\vec p)$ is the \emph{physical} matrix element defined in Eq.\,(\ref{eq:starting}) and the ellipsis represent sub-leading exponentials. 

For {\it negative} time $t$ on the other hand, i.e. for time separations such that  $T/2 \ll t \ll T$, in the continuum and large-$T$ limits we have 
\begin{flalign}
\mathcal{C}^{\alpha r}_W(t,T/2;\vec k,\vec p) \to 
\left[H^{\alpha r}_W(k,\vec p)\right]^\dagger\, \frac{
e^{-(T-t)(E-E_\gamma)}\, \bra{0}P\ket{P} }{2E}
+ \cdots\, 
\label{eq:whereplateaux}
\end{flalign}
with the ellipsis again representing the sub-leading exponentials. 

It is useful to note that, in order to separate the axial and vector form factors, it is enough to compute separately the correlation functions corresponding to the vector, $\mathcal{C}^{\alpha r}_V(t,T/2;\vec k,\vec p)$, and the axial, $\mathcal{C}^{\alpha r}_A(t,T/2;\vec k,\vec p)$, components of the weak current. Moreover, from the properties
\begin{flalign}\label{eq:Hermitian}
\left[H^{\alpha r}_A(k,\vec p)\right]^\dagger = H^{\alpha r}_A(k,\vec p)\;,
\qquad
\left[H^{\alpha r}_V(k,\vec p)\right]^\dagger = -H^{\alpha r}_V(k,\vec p)\;,
\end{flalign}
we deduce the following properties of the corresponding correlation functions under time reversal:
\begin{flalign}
\mathcal{C}^{\alpha r}_A(T-t,T/2;\vec k,\vec p) =\mathcal{C}^{\alpha r}_A(t,T/2;\vec k,\vec p)\;,
\qquad
\mathcal{C}^{\alpha r}_V(T-t,T/2;\vec k,\vec p) =-\mathcal{C}^{\alpha r}_V(t,T/2;\vec k,\vec p)\;.
\label{eq:symmetry}
\end{flalign}
Using these relations, the quantities  
\begin{flalign}
H^{ir}_A(k,\vec p)=
\epsilon_i^r\, p\cdot k\, \left[\frac{F_A(p\cdot k)}{m_P}+\frac{f_P}{p\cdot k}\right]\;,
\qquad
H^{ir}_V(k,\vec p)=i \left(E_\gamma\, \vec \epsilon^r \wedge \vec p-E\, \vec \epsilon^r \wedge \vec k\right)^i\, 
\frac{F_V(p\cdot k)}{m_P}
\end{flalign}
were extracted from the ratios of the correlation functions averaged over the two temporal halves of the lattice
\begin{flalign}
R^{i r}_{A,V}(t,T/2;\vec k,\vec p) 
&= \frac{2 \, E \, \mathcal{C}^{\alpha r}_{A,V}(t,T/2;\vec k,\vec p)}{e^{-t(E-E_\gamma)}\, \bra{P}P\ket{0}}\,
=
H^{ir}_{A,V}(k,\vec p) + \dots  \label{eq:plateaux}
\end{flalign}

In all the formulae of this Appendix we have used continuum notation for the four vectors but the momentum  and energy carried by the current (including the associated projectors)  have to be read by performing the following substitutions
\begin{flalign}
k^i \to \hat k^i=\frac{2}{a}\sin\left(\frac{ak^i}{2}\right)\;,
\qquad
\vert \vec k\vert \to \vert \hat{ \vec k} \vert = \sqrt{\hat{\vec k}_1^2+\hat{\vec k}_2^2+\hat{\vec k}_3^2}\;,
\qquad E_\gamma = \frac{2}{a}\sinh^{-1}\left(\frac{a\vert \hat{ \vec k} \vert}{2}\right)\; .
\label{eq:egamma}
\end{flalign}

In the lattice regularisation that we are using (i.e.\,Wilson quarks at maximal twist), Eqs.\,(\ref{eq:Hermitian}) and (\ref{eq:symmetry}) 
hold for given values of the indices $\alpha$ and $r$ $\in\{1,2\}$ only up to O($a^{2n+1}$) 
lattice artefacts (for integer $n$). One can show however, that, as a consequence of exact lattice 
symmetries (see e.g. Refs.\,\cite{Frezzotti:2003ni,Frezzotti:2004wz}) and the choice of momenta and polarisation vectors 
given  in Eqs.\,(\ref{eq:momenta}) and (\ref{eq:epsilons}), these O($a^{2n+1}$) cutoff effects cancel if one evaluates 
appropriate combinations of the relevant correlation functions, namely
 \begin{equation}\label{eq:CAjr}
 \frac14 \sum_{r=1,2} \sum_{j=1,2}  C_A^{jr} (t, T /2; {\bf k}, {\bf p}) / \epsilon^r_j
\end{equation}
and
\begin{equation}\label{eq:CVjr}
\frac14 \sum_{r=1,2} \sum_{j=1,2}  C_V^{jr} (t, T /2; {\bf k}, {\bf p})  /  F_{r,j}(E_\gamma, E) ,
\qquad  F_{r,j}(E_\gamma, E) = i ( E_\gamma \epsilon^r \wedge {\bf p} - E \epsilon^r
\wedge {\bf k} )_j \,,
\end{equation}
which are precisely those that occur in Eqs.\,(\ref{eq:estimators3b}) and~(\ref{eq:estimators4}) of the main text.
For the terms in Eqs.\,(\ref{eq:CAjr}) and (\ref{eq:CVjr}), the time-reflection properties of Eq.\,(\ref{eq:symmetry}) hold and the 
derived matrix elements, in addition to satisfying the Hermiticity properties of Eq.\,(\ref{eq:Hermitian}), allow for the extraction of 
the form factors $F_A$ and $F_V$ with no O($a^{2n+1}$) lattice artefacts. 
Our analysis of lattice correlators leading to the results in this paper has been
based on data obtained from automatically O($a$) improved combinations of the form (\ref{eq:CAjr}) and (\ref{eq:CVjr}).
%
%%%
\section{Exploiting the electromagnetic Ward identity to relate the matrix element $\bm{H_A^{\alpha r}(k,\vec p)}$ to the decay constant $\bm{f_P}$}
\label{sec:acorr}
%%%
%%%
In this appendix we study the Ward Identity (WI) that relates the axial correlation function  $C^{\alpha r}_A(t;\vec k,\vec p)$ to the axial-pseudoscalar correlation function and, consequently, the matrix element $H_A^{\alpha r}(k,\vec p)$ to the decay constant of the meson $f_P$. As discussed in the main text, a careful analysis of the cut-off effects reveals that the WI does not exclude the possibility of  different $O(a^2)$ artefacts appearing in the decay constant extracted from the three-point function and that from the two-point function. 

 We start with a remark about the matrix element of the axial current, determined at a finite lattice spacing $a$ and using a particular lattice discretisation, which we write in the form 
\begin{equation}   \langle 0\vert j_A^\alpha(0)  \vert P(\vec p) \rangle=f_P^L p_L^\alpha  \, , \label{eq:axialcurrent} \end{equation}
where the combination $f_P^L p_L^\alpha$ is a vector under the orthogonal group $H(4)$\,
\footnote{In this appendix, as in Sec.\,\ref{sec:fsf}, the label $L$ stands for ``Lattice", as the discussion concerns the Ward Identity in a discrete space-time. It should not be confused here with the spatial extent of the Lattice.}.  At finite $a$ the definition of the lattice decay constant   $f_P^L$ depends  on the definition that we assume for the lattice momentum $p_L^\alpha$, for example we may choose $p_L^\alpha=p^\alpha$ or $p_L^\alpha=2/a \sin[a p^\alpha/2]$, where $p^\alpha$ is the continuum value of the momentum.  In  particular we  define $\hat f_P$ by  $\langle 0\vert j_A^\alpha(0)  \vert P(\vec p) \rangle = \hat f_P(\vec p) p^\alpha$. Note that $\hat{f}_P= f_P + O(a^2)$, where $f_P$ is the continuum value of the decay constant. 

%In  general, the  lattice vector WI at finite lattice spacing reads 
%\begin{equation} \langle \hat O \frac{\delta S_F^L}{\delta\alpha^f_V(x)}\vert_{\alpha^f_V(x)=0} \rangle + \langle \frac{ \delta \hat O}{\delta\alpha^f_V(x)}\vert_{\alpha^f_V(x)=0} \rangle =0\, ,\label{eq:ward} \end{equation}
%where $S_F^L$ is the lattice fermion action,  $\hat O$ a generic operator, $\langle \dots\rangle$ represents the matrix element of the operators on the vacuum, which is invariant under vector-like rotations controlled by the continuous  parameter $\alpha^f_V(x)$.  The variation of the action is given by (a definition of all the quantities can be found in ~\cite{Bochicchio:1985xa})
%\begin{equation}  \frac{\delta S_F^L}{i \, \delta\alpha^f_V(x)}=-\nabla_\mu \hat V^C_\mu - \bar \psi\left[ \lambda^f, M_0\right]\psi \, ,\end{equation}
%where repeated indexes are summed up,  the (extended) conserved current, for Wilson-like Fermions, is given by
%\begin{equation}\hat V^C_\mu(x)=\frac{1}{2} \left[  \psi(x) \lambda^f U_\mu(x)(\gamma_\mu-r) \psi (x+\hat \mu) +\psi (x+\hat \mu )  \lambda^f U^\dagger_\mu(x)(\gamma_\mu+r) \psi (x) \right]\, , \end{equation}
%and the right-derivative is defined  by
%\begin{equation}\nabla_\mu f(x) =\frac{f(x) -f(x-\hat \mu)}{a} \, . \end{equation}
%Note that neither $\nabla_\mu$ nor $V^C_\mu$ are  vectors under the hyper-cubic group $H_4$ (exchange of axes and change in sign of a component  if you invert one of the axis), although $\nabla_\mu \hat V^C_\mu$ is a scalar under this group.  

Consider the following correlation function which is relevant to our study, 
 \begin{flalign}
\int \dfour y\, \dthree \vec{x}\, e^{-ik \cdot y-i\vec p\cdot \vec x}\, 
 \bra{0}\mathtt{T}\left[j_A^\alpha(0) j_{em}^\mu(y) P(-t,-\vec x) \right]\ket{0}\; .
\end{flalign}
With Wilson-like Fermions, such as those used in our study, at fixed lattice spacing (for simplicity  in the $T\to \infty$ limit), the electromagnetic WI implies that\,\cite{Bochicchio:1985xa} 
\begin{eqnarray}
&&\frac{1}{a}\, 
\sum_{\mu=0}^3\int \dfour y\, \dthree x\, e^{-ik \cdot y-i\vec p\cdot \vec x}\, 
 \bra{0}\mathtt{T}\left[j_A^\alpha(0) \left\{j_{em}^\mu(y)-j_{em}^\mu(y-\hat \mu) \right\} P(-x) \right]\ket{0}
\nonumber \\
&&= -
\int \dfour y\, \dthree x\, e^{-ik \cdot y-i\vec p\cdot \vec x}
\left\{\delta^4(y)-\delta^4(y+x)\right\}
 \bra{0}\mathtt{T}\left[j_A^\alpha(0) P(-x) \right]\ket{0}\;,\label{eq:ward1}
\end{eqnarray}
where  integrals over the spatial coordinates have to be read as lattice sums and,  in the case of a real photon,  $k^0=iE_\gamma(\vec k)=i\vert\vec k \vert$. 
%The WI has been obtained  by rotating the two operators that carry a (an opposite)  charge  namely weak axial current, $j_A^\alpha$,  and the pseudo scalar density, $P$,  according to the formula 
%\begin{equation} \frac{ \delta \hat O}{\delta\alpha^f_V(x)}\vert_{\alpha^f_V(x)=0}= i \,  Q_{\hat O} \hat O  \,  ,  \end{equation}
%where $Q_{\hat O}$ is the charge of the operator $ \hat O$.  

The WI  can be rewritten in the form
\begin{flalign}
\sum_{i=1}^3 \frac{2\sin( a k_i  /2)}{a} \, C^{\alpha i}_A(t;\vec k,\vec p)=
C_A^{\alpha}(t;\vec p)-C_A^{\alpha}(t;\vec k, \vec p) \;,
\label{eq:wi}
\end{flalign}
%
%{ \bf per Nazario ho un segno di differenza quando metto a posto le i  }
%
where we have defined (note the shift in the exponent with respect to Eq.\,(\ref{eq:ward1}))
\bea
C^{\alpha \mu}_A(t;\vec k,\vec p)=-i \, 
\int \dfour y\,\dthree x\, e^{-ik \cdot (y+\hat \mu/2)-i\vec p\cdot \vec x}\, 
 \bra{0}\mathtt{T}\left[j_A^\alpha(0) j_{em}^\mu(y) P(-t,-\vec x) \right]\ket{0}\; ,
\label{eq:ward11}\eea 
%To this end we can conveniently set $k^0=0$ and, by defining
%
and 
\bea C_A^{\alpha}(t; \vec p)&=&\int \dfour y\, \dthree x\, e^{-ik \cdot y-i\vec p\cdot \vec x}\, 
\delta^4(y)
 \bra{0}\mathtt{T}\left[ j_A^\alpha(0) P(-t,-\vec x) \right]\ket{0}\nonumber \\&=&\int \dthree x\, e^{-i\vec p\cdot \vec x}\, 
\bra{0}\mathtt{T} \left[j_A^\alpha(0) P(-t, -\vec x)\right] \ket{0}\nonumber \\
 C_A^{\alpha}(t; E_\gamma,  \vec p-\vec k)&=&\int \dfour y\, \dthree x\, e^{-ik \cdot y-i\vec p\cdot \vec x}\, 
\delta^4(y+x)\,
 \bra{0}\mathtt{T}\left[j_A^\alpha(0) P(-t,-\vec x)  \right]\ket{0}\nonumber \\&=&\int  \, \dthree x\, e^{E_\gamma\,t  -i(\vec p-\vec k) \cdot \vec x}\, 
 \bra{0}\mathtt{T}\left[j_A^\alpha(0) P(-t,-\vec x)  \right]\ket{0}
 \;. \label{eq:ward2}
\eea
%
%\begin{flalign}
%C^{\alpha}(t;\vec p)=
%\int d^3x\, e^{-i\vec p\cdot \vec x}\, 
%\bra{0}\mathtt{T} \left[j_A^\alpha(0) P(-t, -\vec x)\right] \ket{0}\;,
%\end{flalign}
%%
%
%%
%\begin{eqnarray}&&
%\frac{2\sin(k_\mu a /2)}{a}\, C^{\alpha \mu}_A(t;\vec k, \vec p)=-i \, 
%\int d^4y\, d^3x\, e^{-ik \cdot y-i\vec p\cdot \vec x}\, 
%\delta^4(y)
% \bra{0}\mathtt{T}\left[ j_A^\alpha(0) P(-x) \right]\ket{0}\nonumber \\&& +i \, 
%\int \dfour y\, d^3x\, e^{-ik \cdot y-i\vec p\cdot \vec x}\, 
%\delta^4(y+x)\,
% \bra{0}\mathtt{T}\left[j_A^\alpha(0) P(-x) \right]\ket{0}\;,  \label{eq:ward2old}
%\end{eqnarray}
%we have
We can derive the  Ward identity for the matrix element itself by going onto the mass shell of the pseudoscalar meson,  which in the Euclidean corresponds to selecting the energy of the external  hadronic state to be $E_P$  as $\vert - t\vert$   becomes  very large. 

Consider first the case with $\vec  k \neq 0$. In this case  the second term on the right-hand side of Eq.\,(\ref{eq:wi}) does not contribute because it corresponds to a different energy.   Thus, in this case, we have the following  identity which is true  at all orders in $a$:
\begin{eqnarray}\frac{2\sin(k_\mu a /2)}{a}\,  H^{\alpha \mu}_{L}(k,\vec p) &=&-i\,  \frac{2\sin(k_\mu a /2)}{a}\, \int \, \dfour y\, e^{-ik \cdot (y+\hat \mu/2) }\left[ \langle 0\vert\mathtt{T}\left[j_A^\alpha(0) \, j^\mu_{em}\left( y \right)\right)\right]\vert P(\vec p) \rangle
\nonumber \\ &=&  \langle 0\vert j_A^\alpha(0)  \vert P(\vec p) \rangle \, .\end{eqnarray}
Note that to arrive at this identity we do not need to specify the choice of $f^L_P$.  As $a\to 0$ the discretised derivative in Fourier space $2/a \, \sin(a  k_\mu/2) \to  k_\mu$  and we recover the continuum WI in Eq.\,(\ref{eq:contWI}). 
%Thus, at finite lattice spacing  $a, $  the lattice tensor, defined  as 
%\begin{equation} H^{\alpha \mu}_{L}(k,\vec p)= -i \, \int \, d^4y\, e^{-ik \cdot (y+\hat \mu/2) }\left[ \langle 0\vert\mathtt{T}\left[j_A^\alpha(0) \, j^\mu_{em}\left( y \right)\right)\right]\vert P(\vec p) \rangle\, ,  \end{equation}
%obeys to the WI 
%\begin{equation} \frac{2\sin(k_\mu a /2)}{a}\,  H^{\alpha \mu}_{L}(k,\vec p) = -  \langle 0\vert j_A^\alpha(0)  \vert P(\vec p) \rangle \, , \end{equation}
%   
We can now proceed in analogy to the continuum and separate $H^{\alpha \mu}_{L}(k,\vec p)$ into  a point-like and a structure-dependent   tensor,
$H^{\alpha \mu}_{L}(k,\vec p)=  H^{\alpha \mu}_{L\textrm{-}pt}(k, \vec p)+H^{\alpha \mu}_{L\textrm{-}SD}(k,\vec p) $  such that 
\begin{equation}\frac{2\sin(a k_\mu  /2)}{a}\,  H^{\alpha \mu}_{L\textrm{-}SD}(k,\vec p) =0  \,   \end{equation}
at fixed $a$. Even in the continuum, the separation of $H^{\alpha\mu}$ into a point-like and a structure-dependent  component has an ambiguity in the terms, starting at $O(k^2)$, which are not constrained by the Ward Identity or the equations of motion.  Moreover, there are an infinite number of possible point-like lattice-regularised versions of $H^{\alpha \mu}_{L\textrm{-}pt}(k,\vec p)$ which tend to the chosen  continuum one as $a \to 0$.  
We choose to define $H^{\alpha \mu}_{L\textrm{-}pt}$ by 
\begin{equation} H^{\alpha \mu}_{L\textrm{-}pt}(k,\vec p)= 
 f^L_P \left( A(k,\vec p) \delta^{\alpha\mu} + \frac{ T^{ \alpha\mu}(k,\vec p)}{\Delta}\right) \, , \label{eq:latte}\end{equation}
where $\Delta^{-1}$ is some version of a lattice boson propagator, for example
\begin{equation}\Delta^{-1}=  \frac{1}{4/a^2 \sum_\rho  \sin^2[(p-k)_\rho  a/2] + m_P^2} \to \frac{1}{-2 p \cdot k  +k^2} + O(a^2) \,   \end{equation}
as $a\to 0$,  $A(k,\vec p) = 1+ O(a^2)$  and $T^{\alpha\mu}(k, \vec p)=(2 p-k)^\mu \, (p-k)^\alpha+ O(a^2)$ are  functions of the momenta  which depend on the lattice regularisation  and $f^L_P$ is  the meson decay constant extracted from the matrix element in Eq.\,(\ref{eq:axialcurrent}). 
%$\langle 0\vert j_A^\alpha\left(0\right) \vert P(\vec p) \rangle $ at $a \neq  0$ 
%\bea \langle 0\vert j_A^\alpha\left(0\right) \vert P(\vec p) \rangle = f^L_P\,  p_L^\alpha\, . \eea 
We therefore have
\begin{equation} \lim_{a\to 0} H^{\alpha \mu}_{L\textrm{-}pt}(k,\vec p)=  H^{\alpha \mu}_{pt}(k,\vec p)  \, . \eeq
At fixed lattice spacing the only condition that must be   satisfied is that in applying the WI,
\bea &&\frac{2\sin(a k_\mu  /2)}{a}\,  H^{\alpha \mu}_{L\textrm{-}pt}(k,\vec p) = f_P^L \left(\frac{2\sin(k^\alpha a /2)}{a} \, A(k,\vec p) +\frac{2\sin(k_\mu a /2)}{a}\, \frac{T^{\alpha\mu}(k,\vec p)}{\Delta} \right)\nonumber \\ &&  = \, f^L_P\,  p_L^\alpha  = \, f_P\, p^\alpha + O(a^2)\,,  \label{eq:WIH}\eea
 the denominator $\Delta$ of  Eq.\,(\ref{eq:latte}) disappears. 
 % and  only $p^\alpha$ appears  in the limit $a\to 0$ (obviously we also have  $f_P^L \to f_P$  and $ p_L^\alpha \to  p^\alpha$ as $a\to 0$).  
The WI guarantees that the right-hand side of Eq.\,(\ref{eq:WIH}) is the matrix element 
$\langle 0\vert j_A^\alpha\left(0\right) \vert P(\vec p) \rangle$ including all orders in $a$.  

%We can show  that for a {\it physical } photon with $k^2=0$, $\vec p \equiv (0,0,p)$ and $\vec k\equiv (0,0,k)$ we have $\epsilon_\mu^r(\vec k) V(k,\vec p)^\mu=0$, where $\epsilon_\mu^r(\vec k) $ is one of the polarisation vectors with component only in the $x-y$ directions.
%%
By iterating order by order in $a$,  we may find solutions of the form
\begin{eqnarray}  A(k,\vec p) &=& 1 + a^2  \tilde A(p^2)  + O(a^4)\nonumber \\ 
T^{\alpha\mu}(k,\vec p) &=& (2 p-k)^\mu(p-k)^\alpha  + a^2   \tilde T^{\alpha\mu}(k,\vec p) + O(a^4) \end{eqnarray}
that satisfy the WI, where the coefficients of the expansion  are  not unique. The only relevant
term for the extraction of the form factor $F_A$ however, is the coefficient $A(k,\vec p)$ (since $\epsilon_\mu^r(\vec k) T^{\alpha\mu}(k,\vec p) =0$), which may differ from one by terms of $O(a^2)$ thus giving an {\it effective} decay constant which is different from the one naively expected from the WI. The absence of lattice artefacts of $O(a^{2n+1})$ is a consequence of our use of the
combinations of lattice correlations functions in Eqs.\,(\ref{eq:CAjr}) and (\ref{eq:CVjr}) 
and the resulting $H_L^{\alpha\mu}$ matrix elements (see Eq.\,(\ref{eq:latte})). Alternatively one 
might work with lattice formulations which preserve chiral symmetry, such as those based on Overlap or 
Domain Wall fermions. For O($a$) improved Wilson fermion lattice actions, instead, 
corrections of O($a^3$) will in general occur.

 From the above discussion we conclude that  the lattice $H^{jr}_A(k,\vec p)$    has the form
% \begin{equation}
%H^{ir}_A(k,\vec p)=
%\epsilon^i_r\, \frac{m_P}{2}\, \left[\left(x_\gamma \, F_A(x_\gamma)\right)^L+ \left(\frac{2\, \tilde  f^L_P}{m_P}\right)^L\right]
%\end{equation}
%may have corrections of $O(a^2)$ both in the effective  decay constant $\tilde  f_P$ which differs, in spite of the WI,  from the one extracted from the matrix element of the axial current at finite $a$,  and in $F_A$.  Thus we should fit the data and find  that $\lim_{a\to 0} 
%\tilde f^L_P\to f_P$ ($f^L_P \to f_P$)  where $f_P$ is the continuum decay constant. From the same fit we may also extract the $a\to 0$ limit of $ F_A^L$, $F_A$
\begin{flalign}
H^{jr}_A(k,\vec  p)=
\epsilon_j^r\, \frac{m_P}{2}\, \left[ x_\gamma \left(F_A(x_\gamma) + a^2 \Delta F_A(x_\gamma)\right) +\frac{2}{m_P}\left(f_P+a^2 \Delta f_P \right) \right]+ \cdots\:,
\label{eq:estimators2}
\end{flalign}
where the dots represent higher order discretisation corrections.

In order to implement the strategy described in Eq.\,(\ref{eq:fsf}),  we need to perform a 
direct calculation of $H^{jr}_A(0, \vec p)$ and hence to study the $k\to 0$ limit  of the WI.
The problem is non-trivial because from the spectral analysis of $C^{\alpha \mu}_A(t;\vec k,\vec p)$ it follows that 
\begin{flalign}
C^{\alpha \mu}_A(t;\vec k,\vec p)= c_1^{\alpha \mu} e^{-tE_P(\vec p)} + c_2^{\alpha \mu} e^{-t\left\{E_P(\vec p-\vec k)+E_\gamma(\vec k)\right\}} + \cdots\;,
\end{flalign}
where the ellipsis represent exponentially suppressed contributions,  with a gap that is $O(m_\pi)$. The first exponential corresponds to the on-shell external meson $P$ and represents the state we are interested in. The second exponential corresponds to the state $P+\gamma$ where both the meson and the photon are on-shell and have a total momentum $\vec p$ and a relative momentum $\vec k$.  A similar time dependence also appears in the WI from the rotation of the $P$ source; this is the second term on the right-hand side of Eq.\,(\ref{eq:wi}).  As discussed above, when $\vec  k\neq  0$ it is possible to isolate the matrix element corresponding to the state $P$.  

The problem we now address is to study the limit $\vec k\to \vec 0$, paying special attention to the leading cut-off effects.  This can be done by using  the  exact WI satisfied by $C^{\alpha \mu}_A(t;\vec k, \vec p)$ at finite lattice spacing; in particular  we aim to understand the structure of the correlation function $C^{\alpha \mu}_A(t;\vec k, \vec p)$ at $\vec k=0$.   To this end we consider  the two-point correlation functions  on the right-hand side of Eq.\,(\ref{eq:wi}) when $\alpha$ is a spatial index (the case $\alpha=0$ is similar, but in the following we shall concentrate on the case $\alpha=1,2,3$). By setting  $E_\gamma=0$ in the last term of Eq.\,(\ref{eq:ward2}), we have
\begin{flalign}
&C_A^{j}(t;\vec p) = \frac{p^j \hat f_P(\vec p)\hat G_P(\vec p)}{2\hat E_P(\vec p)} e^{-t\hat E_P(\vec p)} + \dots\;,
\nonumber \\
\nonumber \\
&C_A^{j}(t;E_\gamma=0, \vec p-\vec k)= C_A^{j}(t , \vec p-\vec k) = \frac{(p-k)^j \hat f_P(\vec p-\vec k) \hat G_P(\vec p-\vec k)}{2\hat E_P(\vec p-\vec k)} e^{-t\hat E_P(\vec p-\vec k)} + \dots\;,
\end{flalign}
where the ellipsis represent sub-leading exponentials (the gap is at least $2 m_\pi$). In the previous expressions $\hat f_P(\vec p)= f_P +O(a^2)$, $\hat G_P(\vec p)=G_P+O(a^2)$   and $\hat E_P(\vec p)=  E_P(\vec p) +O(a^2)$ where $f_P$, $G_P$ and $E_P(\vec p)$ are respectively the continuum decay constant, the continuum matrix element of the pseudoscalar density used as  interpolating operator, $G_P=\bra{0}P\ket{P(\vec p)}$, and the continuum energy of the meson.  
%{\bf I think that in all the formulae by Nazario the sign of the exponent in p must change sign and ha a doubt also with our formulae in the main text.}

By using the previous two expressions and by differentiating Eq.\,(\ref{eq:wi}) with respect to the component $k^i$ of  $\vec k$ and then setting $\vec k=\vec 0$ we obtain
\begin{eqnarray}
C^{j i}_A(t;\vec 0,\vec p)&=&\frac{\hat f_P(\vec p)\hat G_P(\vec p)}{2\hat E_P(\vec p)} e^{-t\hat E_P(\vec p)} \times
\nonumber \\
&&\hspace{-0.8in}
\left\{
\delta^{ij} + p^j\left[ 
\frac{1}{\hat f_P(\vec p)}\frac{\partial \hat f_P(\vec p)}{\partial p^i}
+\frac{1}{\hat G_P(\vec p)}\frac{\partial \hat G_P(\vec p)}{\partial p^i}
-\left(t+\frac{1}{\hat E_P(\vec p)} \right)\frac{\partial \hat E_P(\vec p)}{\partial p^i}
\right]
\right\}
+ \dots\;,\label{eq:CjiA}
\end{eqnarray}
where the ellipsis again represents the sub-leading exponentials and we have used the fact that $-\partial_{k_i} g(\vec p-\vec k)=\partial_{p_i} g(\vec p-\vec k)$.

As   can be seen, the structure of $C^{j i}_A(t;\vec 0,\vec p)$ is highly non trivial. Note in particular the term linear in $t$ that is a manifestation of the singular behaviour at large distances of the correlation function (this generates a double pole in momentum space that is at the original the infrared divergence).
An important consequence of the strategy proposed in section\,\ref{sec:fsf} which we have used in our  calculations,  is that the  terms in squared brackets in Eq.\,(\ref{eq:CjiA}) disappear at any value of $\vec p$ when we contract the correlation function with the physical polarisation vectors of the photon. With our choice of kinematics, these satisfy the relation
\begin{flalign}
\sum_{\mu=0}^3 \epsilon_\mu^r(\vec k) p_\mu = \sum_{i=1}^3 \epsilon_i^r(\vec 0) p_i =0\;. 
\end{flalign}
Indeed, the $H(3)$ symmetry implies that 
\begin{flalign}
\frac{\partial \hat f_P(\vec p)}{\partial p^i} = p^i\times O(a^2)\;,
\qquad
\frac{\partial \hat G_P(\vec p)}{\partial p^i} = p^i\times O(a^2)\;,
\qquad
\frac{\partial \hat E_P(\vec p)}{\partial p^i}  =   \frac{ p^i}{E_P(\vec p)} + O(a^2)\; , 
\end{flalign}
and thus 
\begin{flalign}
C^{j r}_A(t;\vec 0,\vec p)=
\sum_{i=1}^3 \epsilon_i^r(\vec 0) C^{j i}_A(t;\vec 0,\vec p)
=
\epsilon_j^r(\vec 0)
\frac{\hat f_P(\vec p)\hat G_P(\vec p)}{2\hat E_P(\vec p)} e^{-t\hat E_P(\vec p)} 
+ \dots\;. 
\end{flalign}
We conclude that $C^{j r}_A(t;\vec 0,\vec p)$ can be analyzed as expected to extract the coefficient of the leading exponential.  We stress that  the above demonstration shows that from $C^{j r}_A(t;\vec 0,\vec p)$ we can extract precisely the decay constant  which appears in the lattice matrix element of the axial current in Eq.\,(\ref{eq:axialcurrent}), without to have to make a choice for the lattice momentum $p^\alpha_L$. 
%%%
%%%
\section{}
\label{app:numres}
In this Appendix we present some numerical information that may be useful to the reader. We start by listing in 
Tables\,\ref{tab:masses&simudetails} and \ref{tab:MpiL} the parameters used in our  numerical simulations: the values of $\beta$ and the corresponding lattice spacings, the volumes, the quark mass parameters and the corresponding \emph{pion} masses, $m_\pi$, and $m_\pi\,L$, the numbers of configurations, and the twisting angles introduced  to inject momenta in the correlation functions. 

Given the smooth behaviour that we find for the form factors as functions of $x_\gamma$ in the region where we have data, for most phenomenological purposes it is sufficient to use form factors obtained using the ansatze and coefficients given in Section\,\ref{sec:numerical}.  
However, in the tables in Secs.\,\ref{subsec:FAFVpi}-\ref{subsec:FAFVDs}
below, we also present the values of the form factors, $F_A$ and $F_V$   at selected values of the photon energy $x_\gamma$, for the  pion, kaon, $D$ and $D_s$ mesons, together with the corresponding uncertainties, $\Delta_{F_A}$ and $\Delta_{F_V}$. The results have been extrapolated to the continuum and to physical quark masses. We also give the correlation matrices of these results. For the $D$ and $D_s$  mesons, for which we only have data in a limited range of $x_\gamma$, the results in Secs.\, \ref{subsec:FAFVD} and \ref{subsec:FAFVDs} were obtained by averaging the results obtained using
Eqs.\,(\ref{eq:linearfit}) and (\ref{eq:extrHeavyc}) and including the difference between the two anzatze in the estimate of the uncertainties. As might be expected from Fig.\,\ref{fig:Dscompare} and the accompanying discussion, the extrapolations using the two anzatze diverge significantly at larger $x_\gamma$ which is reflected in the growing uncertainties in the results in the tables in Secs.\,\ref{subsec:FAFVD} and \ref{subsec:FAFVDs}.
\begin{table}[t!]
\begin{center}
\footnotesize
\renewcommand{\arraystretch}{1.20}
\begin{tabular}{||c|c|c|c|c|c|c|c||}
\hline
ensemble & $\beta$ & $a$(fm) & $V / a^4$ &$a\mu_{sea}=a\mu_\ell$&$a\mu_\sigma$&$a\mu_\delta$&$N_{cfg}$ \\
\hline
$A30.32$ & $1.90$ &$0.885(18)$ & $32^{3}\times 64$ &$0.0030$ &$0.15$ &$0.19$ &$150$ \\
$A40.32$ & & & & $0.0040$ & & & $100$  \\
%$A50.32$ & & & $0.0050$ & & &  $150$ \\
\hline
%$A40.24$ & $1.90$ & $24^{3}\times 48 $ & $0.0040$ &$0.15$ & $0.19$& $150$  \\
$A60.24$ &  $1.90$ &$0.885(18)$ & $24^4\times 48$ & $0.0060$ & 0.15 & 0.19 &  $150$   \\
$A80.24$ & & & &$0.0080$ & & &  $150$   \\
%$A100.24$ &   & & $0.0100$ & & &  $150$ \\
\hline
$B25.32$ & $1.95$ &$0.815(30)$& $32^{3}\times 64$ &$0.0025$&$0.135$ &$0.170$& $150$ \\
$B35.32$ & & & & $0.0035$  & & & $150$  \\
$B55.32$ & & & & $0.0055$ & & & $150$  \\
$B75.32$ &  & & & $0.0075$ & & & $80$    \\
\hline
%$B85.24$ & $1.95$ & $24^{3}\times 48 $ & $0.0085$ &$0.135$ &$0.170$ & $150$  \\
%\hline
$D15.48$ & $2.10$ &$0.619(18)$& $48^{3}\times 96$ &$0.0015$&$0.12$ &$0.1385 $& $100$ \\
$D20.48$ & & & &$0.0020$  &  &  & $100$  \\
$D30.48$ & & & &$0.0030$ & & & $100$  \\
 \hline
\end{tabular}
\renewcommand{\arraystretch}{1.0}
\end{center}
\caption{\it Values of the simulated sea and valence quark bare masses for each ensemble used in this work.  The table is the same as in Ref.\,\cite{DiCarlo:2019thl} except for $\mu_s$ and  $\mu_c$ which are given in Table\,\ref{tab:MpiL}. 
\hspace*{\fill}}
\label{tab:masses&simudetails}
\end{table} 

\FloatBarrier
\begin{table}[htbp!]
\begin{center}
\footnotesize
\renewcommand{\arraystretch}{1.20}
\begin{tabular}{||c||c|c|c|c|c|c|c|c||}
\hline
ensemble& $\beta$ & $L$(fm) &$m_\pi$(MeV) & $m_\pi L$ & $a\mu_{\rm sea}=a\mu_\ell$ & $a\mu_s$ &$a\mu_c$ &$\theta_{i=0,s,t}$ 
%\\  &&&&&&& &   
%$p_z = \frac{2\pi}{L}\left(  \theta_0- \theta_s\right)$ and  $ k_z = \frac{2\pi}{L}\left( \theta_0- \theta_t\right)$, 
 \\  \hline 
$A30.32$ & $1.90$&$ 2.84$&$273$&$3.9$          & 0.0030 & 0.02363,	 & 0.27903,		& 0,   0.2288, 0.3432,  \\
$A40.32$ &            &            &$315$&$4.5$ & 0.0040 &0.02760&0.29900& 0.6864,  0.8580 \\
 \hline
%$A40.24$ & $1.90$&$ 2.13$&$282$&$3.05$\\
$A60.24$ &    $1.90$&$  2.13 $         &$383$&$4.1$ & 0.0060 &&&\\
$A80.24$ &            &            &$441$&$4.7$ & 0.0080 &&&\\
%$A100.24$ &          &            &$443$&$4.78$\\
 \hline
$B25.32$ & $1.95$&$ 2.61$&$256$&$3.4$         & 0.0025  &0.02094,	 &	0.24725,	 & 0,   0.2107,   0.3160,   \\
$B35.32$ &            &            &$300$&$4.0$ & 0.0035&0.0239&0.267300&0.6321,  0.7901\\
$B55.32$ &            &            &$373$&$4.9$ & 0.0055&&&\\
$B75.32$ &            &            &$436$&$6.1$ & 0.0075&&&\\
 \hline
%$B85.24$ & $1.95$&$ 1.96$&$435$&$4.32$\\
% \hline
$D15.48$ & $2.10$&$ 2.97$&$228$&$3.4$   & 0.0015 & 0.01612,	 &	0.19037,	& 0,   0.2400,   0.3601,  	\\
$D20.48$ &            &            &$252$&$3.8$ & 0.0020 &0.01910&0.20540 & 0.7201,   0.9002\\
$D30.48$ &            &            &$309$&$4.7$ & 0.0030 &&&\\
 \hline
\end{tabular}
\renewcommand{\arraystretch}{1.20}
\end{center}
\caption{\it Central values of the pion mass $m_\pi$, of the lattice size $L$ and of the product $m_\pi L$ for the various ensembles used in this work.  We also give the values of the {\it angles} use to define the $z$-component of the meson and photon momenta, $\vec p = \left(0,0,\frac{2\pi}{L}\left(  \theta_0- \theta_s\right)\right)$  and $\vec k =\left(0,0, \frac{2\pi}{L}\left( \theta_0- \theta_t\right)\right)$ respectively.   \hspace*{\fill} }
%The values of $M_\pi$ are extrapolated to the continuum and infinite volume limits, according to the ChPT fit (\ref{eq:cptmpi2Ch}), described in Section \ref{sec:r0}.}
\label{tab:MpiL}
\end{table}
\FloatBarrier
\mbox{}
\newpage
\subsection{Results for $\bm{F_A(x_\gamma)}$ and $\bm{F_V(x_\gamma)}$ of the pion}
\label{subsec:FAFVpi}
\begin{table}[h!]
\begin{eqnarray} && \qquad \qquad \qquad \qquad   \qquad \qquad \qquad \qquad \qquad   F_A \quad {\rm Correlation} \quad  {\rm  Matrix} \nonumber \\ &&\footnotesize{
\begin{tabular}{|c|c|c|}
\hline
 $x_\gamma$  & $F_A $  & $\Delta_{F_A} $\\ \hline
0 &  0.0104088 &  0.00262483 \\ \hline 
0.1 &  0.0104435 &  0.00260149 \\ \hline 
0.2 &  0.0104782 &  0.0025792 \\ \hline 
0.3 &  0.0105129 &  0.00255799 \\ \hline 
0.4 &  0.0105477 &  0.00253788 \\ \hline 
0.5 &  0.0105824 &  0.00251889 \\ \hline 
0.6 &  0.0106171 &  0.00250106 \\ \hline 
0.7 &  0.0106519 &  0.00248441 \\ \hline 
0.8 &  0.0106866 &  0.00246897 \\ \hline 
0.9 &  0.0107213 &  0.00245474 \\ \hline 
1 &  0.010756 &  0.00244177 \\ \hline 
\end{tabular}
\quad \left(\begin{tabular}{c|c|c|c|c|c|c|c|c|c|c}1.000& 1.000& 0.999& 0.998& 0.997& 0.995& 0.992& 0.989& 0.986& 0.982& 0.977\\ 
1.000& 1.000& 1.000& 0.999& 0.998& 0.997& 0.995& 0.992& 0.989& 0.985& 0.981\\ 
0.999& 1.000& 1.000& 1.000& 0.999& 0.998& 0.996& 0.994& 0.992& 0.989& 0.985\\ 
0.998& 0.999& 1.000& 1.000& 1.000& 0.999& 0.998& 0.996& 0.994& 0.992& 0.988\\ 
0.997& 0.998& 0.999& 1.000& 1.000& 1.000& 0.999& 0.998& 0.996& 0.994& 0.991\\ 
0.995& 0.997& 0.998& 0.999& 1.000& 1.000& 1.000& 0.999& 0.998& 0.996& 0.994\\ 
0.992& 0.995& 0.996& 0.998& 0.999& 1.000& 1.000& 1.000& 0.999& 0.998& 0.996\\ 
0.989& 0.992& 0.994& 0.996& 0.998& 0.999& 1.000& 1.000& 1.000& 0.999& 0.998\\ 
0.986& 0.989& 0.992& 0.994& 0.996& 0.998& 0.999& 1.000& 1.000& 1.000& 0.999\\ 
0.982& 0.985& 0.989& 0.992& 0.994& 0.996& 0.998& 0.999& 1.000& 1.000& 1.000\\ 
0.977& 0.981& 0.985& 0.988& 0.991& 0.994& 0.996& 0.998& 0.999& 1.000& 1.000\\ 
\end{tabular}\right) \nonumber
  }\end{eqnarray}
 
\vspace{0.3in}
\begin{eqnarray} && \qquad \qquad \qquad \qquad   \qquad \qquad \qquad \qquad \qquad   F_V \quad {\rm Correlation} \quad  {\rm  Matrix} \nonumber \\ &&\footnotesize{
\begin{tabular}{|c|c|c|}
\hline
 $x_\gamma$  & $F_V $  & $\Delta_{F_V} $\\ \hline
0 &  0.0233352 &  0.00214581 \\ \hline 
0.1 &  0.023309 &  0.0021304 \\ \hline 
0.2 &  0.0232828 &  0.00211523 \\ \hline 
0.3 &  0.0232566 &  0.00210031 \\ \hline 
0.4 &  0.0232303 &  0.00208563 \\ \hline 
0.5 &  0.0232041 &  0.00207121 \\ \hline 
0.6 &  0.0231779 &  0.00205705 \\ \hline 
0.7 &  0.0231517 &  0.00204315 \\ \hline 
0.8 &  0.0231254 &  0.00202952 \\ \hline 
0.9 &  0.0230992 &  0.00201617 \\ \hline 
1 &  0.023073 &  0.0020031 \\ \hline 
\end{tabular}
\vspace{0.4in}
\quad \left(\begin{tabular}{c|c|c|c|c|c|c|c|c|c|c}1.000& 1.000& 1.000& 0.999& 0.999& 0.999& 0.998& 0.997& 0.996& 0.995& 0.994\\ 
1.000& 1.000& 1.000& 1.000& 0.999& 0.999& 0.998& 0.998& 0.997& 0.996& 0.995\\ 
1.000& 1.000& 1.000& 1.000& 1.000& 0.999& 0.999& 0.998& 0.998& 0.997& 0.996\\ 
0.999& 1.000& 1.000& 1.000& 1.000& 1.000& 0.999& 0.999& 0.998& 0.998& 0.997\\ 
0.999& 0.999& 1.000& 1.000& 1.000& 1.000& 1.000& 0.999& 0.999& 0.998& 0.998\\ 
0.999& 0.999& 0.999& 1.000& 1.000& 1.000& 1.000& 1.000& 0.999& 0.999& 0.998\\ 
0.998& 0.998& 0.999& 0.999& 1.000& 1.000& 1.000& 1.000& 1.000& 0.999& 0.999\\ 
0.997& 0.998& 0.998& 0.999& 0.999& 1.000& 1.000& 1.000& 1.000& 1.000& 0.999\\ 
0.996& 0.997& 0.998& 0.998& 0.999& 0.999& 1.000& 1.000& 1.000& 1.000& 1.000\\ 
0.995& 0.996& 0.997& 0.998& 0.998& 0.999& 0.999& 1.000& 1.000& 1.000& 1.000\\ 
0.994& 0.995& 0.996& 0.997& 0.998& 0.998& 0.999& 0.999& 1.000& 1.000& 1.000\\ 
\end{tabular}\right) \nonumber
  }\end{eqnarray}
 
\end{table}\FloatBarrier
\mbox{}
\newpage
\subsection{Results for $\bm{F_A(x_\gamma)}$ and $\bm{F_V(x_\gamma)}$ of the kaon}
\label{subsec:FAFVK}
\begin{table}[h]
\begin{eqnarray} && \qquad \qquad \qquad \qquad   \qquad \qquad \qquad \qquad \qquad   F_A \quad {\rm Correlation} \quad  {\rm  Matrix} \nonumber \\ &&\footnotesize{
\begin{tabular}{|c|c|c|}
\hline
 $x_\gamma$  & $F_A $  & $\Delta_{F_A} $\\ \hline
0 &  0.0370382 &  0.00876335 \\ \hline 
0.1 &  0.0369171 &  0.00828189 \\ \hline 
0.2 &  0.0367961 &  0.00784116 \\ \hline 
0.3 &  0.0366751 &  0.00744839 \\ \hline 
0.4 &  0.0365541 &  0.00711155 \\ \hline 
0.5 &  0.0364331 &  0.00683889 \\ \hline 
0.6 &  0.0363121 &  0.00663833 \\ \hline 
0.7 &  0.0361911 &  0.00651654 \\ \hline 
0.8 &  0.0360701 &  0.00647795 \\ \hline 
0.9 &  0.0359491 &  0.00652404 \\ \hline 
1 &  0.035828 &  0.00665305 \\ \hline 
\end{tabular}
\quad \left(\begin{tabular}{c|c|c|c|c|c|c|c|c|c|c}1.000& 0.998& 0.990& 0.975& 0.951& 0.916& 0.869& 0.808& 0.736& 0.654& 0.566\\ 
0.998& 1.000& 0.997& 0.988& 0.970& 0.941& 0.899& 0.845& 0.779& 0.703& 0.620\\ 
0.990& 0.997& 1.000& 0.997& 0.985& 0.963& 0.929& 0.883& 0.823& 0.753& 0.676\\ 
0.975& 0.988& 0.997& 1.000& 0.996& 0.982& 0.957& 0.918& 0.867& 0.805& 0.734\\ 
0.951& 0.970& 0.985& 0.996& 1.000& 0.995& 0.979& 0.950& 0.909& 0.855& 0.793\\ 
0.916& 0.941& 0.963& 0.982& 0.995& 1.000& 0.994& 0.976& 0.946& 0.902& 0.849\\ 
0.869& 0.899& 0.929& 0.957& 0.979& 0.994& 1.000& 0.994& 0.975& 0.943& 0.900\\ 
0.808& 0.845& 0.883& 0.918& 0.950& 0.976& 0.994& 1.000& 0.993& 0.974& 0.943\\ 
0.736& 0.779& 0.823& 0.867& 0.909& 0.946& 0.975& 0.993& 1.000& 0.994& 0.975\\ 
0.654& 0.703& 0.753& 0.805& 0.855& 0.902& 0.943& 0.974& 0.994& 1.000& 0.994\\ 
0.566& 0.620& 0.676& 0.734& 0.793& 0.849& 0.900& 0.943& 0.975& 0.994& 1.000\\ 
\end{tabular}\right) \nonumber
  }\end{eqnarray}
 
\vspace{0.3in}
\begin{eqnarray} && \qquad \qquad \qquad \qquad   \qquad \qquad \qquad \qquad \qquad   F_V \quad {\rm Correlation} \quad  {\rm  Matrix} \nonumber \\ &&\footnotesize{
\begin{tabular}{|c|c|c|}
\hline
 $x_\gamma$  & $F_V $  & $\Delta_{F_V} $\\ \hline
0 &  0.12439 &  0.00960998 \\ \hline 
0.1 &  0.121998 &  0.00891388 \\ \hline 
0.2 &  0.119606 &  0.00828371 \\ \hline 
0.3 &  0.117214 &  0.0077356 \\ \hline 
0.4 &  0.114821 &  0.0072881 \\ \hline 
0.5 &  0.112429 &  0.00696062 \\ \hline 
0.6 &  0.110037 &  0.00677062 \\ \hline 
0.7 &  0.107645 &  0.00672974 \\ \hline 
0.8 &  0.105253 &  0.00684066 \\ \hline 
0.9 &  0.102861 &  0.00709626 \\ \hline 
1 &  0.100469 &  0.00748173 \\ \hline 
\end{tabular}
\quad \left(\begin{tabular}{c|c|c|c|c|c|c|c|c|c|c}1.000& 0.997& 0.985& 0.961& 0.921& 0.860& 0.777& 0.674& 0.558& 0.435& 0.316\\ 
0.997& 1.000& 0.996& 0.980& 0.949& 0.898& 0.825& 0.731& 0.622& 0.506& 0.391\\ 
0.985& 0.996& 1.000& 0.994& 0.974& 0.935& 0.874& 0.791& 0.692& 0.583& 0.474\\ 
0.961& 0.980& 0.994& 1.000& 0.993& 0.967& 0.921& 0.852& 0.765& 0.666& 0.565\\ 
0.921& 0.949& 0.974& 0.993& 1.000& 0.991& 0.961& 0.909& 0.837& 0.752& 0.661\\ 
0.860& 0.898& 0.935& 0.967& 0.991& 1.000& 0.989& 0.957& 0.903& 0.834& 0.756\\ 
0.777& 0.825& 0.874& 0.921& 0.961& 0.989& 1.000& 0.989& 0.956& 0.905& 0.843\\ 
0.674& 0.731& 0.791& 0.852& 0.909& 0.957& 0.989& 1.000& 0.989& 0.958& 0.914\\ 
0.558& 0.622& 0.692& 0.765& 0.837& 0.903& 0.956& 0.989& 1.000& 0.990& 0.964\\ 
0.435& 0.506& 0.583& 0.666& 0.752& 0.834& 0.905& 0.958& 0.990& 1.000& 0.992\\ 
0.316& 0.391& 0.474& 0.565& 0.661& 0.756& 0.843& 0.914& 0.964& 0.992& 1.000\\ 
\end{tabular}\right) \nonumber
  }\end{eqnarray}
 
\end{table}
\FloatBarrier
\mbox{}
\newpage
\subsection{Results for $\bm{F_A(x_\gamma)}$ and $\bm{F_V(x_\gamma)}$ of the $\bm{D}$ meson}
\label{subsec:FAFVD}
\begin{table}[h]
\begin{eqnarray} && \qquad \qquad \qquad \qquad   \qquad \qquad \qquad \qquad \qquad   F_A \quad {\rm Correlation} \quad  {\rm  Matrix} \nonumber \\ &&\footnotesize{
\begin{tabular}{|c|c|c|}
\hline
 $x_\gamma$  & $F_A $  & $\Delta_{F_A} $\\ \hline
0 &  0.11029 &  0.00885428 \\ \hline 
0.1 &  0.0988347 &  0.00740568 \\ \hline 
0.2 &  0.0887281 &  0.00727782 \\ \hline 
0.3 &  0.079585 &  0.00791868 \\ \hline 
0.4 &  0.0711547 &  0.0092365 \\ \hline 
0.5 &  0.063267 &  0.0111973 \\ \hline 
0.6 &  0.0558021 &  0.0137148 \\ \hline 
0.7 &  0.0486731 &  0.0166835 \\ \hline 
0.8 &  0.0418154 &  0.0200096 \\ \hline 
0.9 &  0.0351801 &  0.02362 \\ \hline 
1 &  0.0287293 &  0.027459 \\ \hline 
\end{tabular}
\quad \left(\begin{tabular}{c|c|c|c|c|c|c|c|c|c|c}1.000& 0.907& 0.704& 0.491& 0.314& 0.187& 0.102& 0.049& 0.015& -0.006& -0.020\\ 
0.907& 1.000& 0.931& 0.776& 0.598& 0.439& 0.314& 0.223& 0.156& 0.108& 0.072\\ 
0.704& 0.931& 1.000& 0.948& 0.827& 0.688& 0.564& 0.465& 0.387& 0.327& 0.281\\ 
0.491& 0.776& 0.948& 1.000& 0.961& 0.875& 0.781& 0.697& 0.627& 0.570& 0.524\\ 
0.314& 0.598& 0.827& 0.961& 1.000& 0.974& 0.921& 0.863& 0.810& 0.764& 0.726\\ 
0.187& 0.439& 0.688& 0.875& 0.974& 1.000& 0.985& 0.954& 0.920& 0.888& 0.859\\ 
0.102& 0.314& 0.564& 0.781& 0.921& 0.985& 1.000& 0.991& 0.974& 0.953& 0.934\\ 
0.049& 0.223& 0.465& 0.697& 0.863& 0.954& 0.991& 1.000& 0.995& 0.985& 0.972\\ 
0.015& 0.156& 0.387& 0.627& 0.810& 0.920& 0.974& 0.995& 1.000& 0.997& 0.991\\ 
-0.006& 0.108& 0.327& 0.570& 0.764& 0.888& 0.953& 0.985& 0.997& 1.000& 0.998\\ 
-0.020& 0.072& 0.281& 0.524& 0.726& 0.859& 0.934& 0.972& 0.991& 0.998& 1.000\\ 
\end{tabular}\right) \nonumber
  }\end{eqnarray}
 
\vspace{0.3in}
\begin{eqnarray} && \qquad \qquad \qquad \qquad   \qquad \qquad \qquad \qquad \qquad   F_V \quad {\rm Correlation} \quad  {\rm  Matrix} \nonumber \\ &&\footnotesize{
\begin{tabular}{|c|c|c|}
\hline
 $x_\gamma$  & $F_V $  & $\Delta_{F_V} $\\ \hline
0 &  -0.150466 &  0.0144033 \\ \hline 
0.1 &  -0.135916 &  0.0119914 \\ \hline 
0.2 &  -0.123034 &  0.0114497 \\ \hline 
0.3 &  -0.111365 &  0.0119456 \\ \hline 
0.4 &  -0.100606 &  0.0132415 \\ \hline 
0.5 &  -0.0905481 &  0.0152892 \\ \hline 
0.6 &  -0.0810413 &  0.0180286 \\ \hline 
0.7 &  -0.071976 &  0.0213636 \\ \hline 
0.8 &  -0.0632697 &  0.0251904 \\ \hline 
0.9 &  -0.0548596 &  0.0294171 \\ \hline 
1 &  -0.0466964 &  0.0339692 \\ \hline 
\end{tabular}
\quad \left(\begin{tabular}{c|c|c|c|c|c|c|c|c|c|c}1.000& 0.925& 0.749& 0.552& 0.377& 0.239& 0.140& 0.072& 0.027& -0.003& -0.023\\ 
0.925& 1.000& 0.941& 0.807& 0.647& 0.494& 0.364& 0.263& 0.187& 0.130& 0.087\\ 
0.749& 0.941& 1.000& 0.956& 0.851& 0.721& 0.596& 0.488& 0.401& 0.331& 0.276\\ 
0.552& 0.807& 0.956& 1.000& 0.966& 0.885& 0.789& 0.696& 0.615& 0.547& 0.492\\ 
0.377& 0.647& 0.851& 0.966& 1.000& 0.975& 0.918& 0.852& 0.788& 0.732& 0.684\\ 
0.239& 0.494& 0.721& 0.885& 0.975& 1.000& 0.983& 0.946& 0.904& 0.862& 0.824\\ 
0.140& 0.364& 0.596& 0.789& 0.918& 0.983& 1.000& 0.989& 0.966& 0.939& 0.913\\ 
0.072& 0.263& 0.488& 0.696& 0.852& 0.946& 0.989& 1.000& 0.993& 0.979& 0.962\\ 
0.027& 0.187& 0.401& 0.615& 0.788& 0.904& 0.966& 0.993& 1.000& 0.996& 0.987\\ 
-0.003& 0.130& 0.331& 0.547& 0.732& 0.862& 0.939& 0.979& 0.996& 1.000& 0.997\\ 
-0.023& 0.087& 0.276& 0.492& 0.684& 0.824& 0.913& 0.962& 0.987& 0.997& 1.000\\ 
\end{tabular}\right) \nonumber
  }\end{eqnarray}
 
\end{table}
\FloatBarrier
\mbox{}
\newpage
\subsection{Results for $\bm{F_A(x_\gamma)}$ and $\bm{F_V(x_\gamma)}$ of the $\bm{D_s}$ meson}
\label{subsec:FAFVDs}
\begin{table}[h]
\begin{eqnarray} && \qquad \qquad \qquad \qquad   \qquad \qquad \qquad \qquad \qquad   F_A \quad {\rm Correlation} \quad  {\rm  Matrix} \nonumber \\ &&\footnotesize{
\begin{tabular}{|c|c|c|}
\hline
 $x_\gamma$  & $F_A $  & $\Delta_{F_A} $\\ \hline
0 &  0.09307 &  0.00598514 \\ \hline 
0.1 &  0.0849608 &  0.00486963 \\ \hline 
0.2 &  0.0776688 &  0.00422644 \\ \hline 
0.3 &  0.0709899 &  0.00389515 \\ \hline 
0.4 &  0.0647833 &  0.00400167 \\ \hline 
0.5 &  0.0589483 &  0.00469896 \\ \hline 
0.6 &  0.0534115 &  0.00595597 \\ \hline 
0.7 &  0.0481175 &  0.00763605 \\ \hline 
0.8 &  0.0430239 &  0.00962076 \\ \hline 
0.9 &  0.0380979 &  0.0118319 \\ \hline 
1 &  0.0333133 &  0.0142191 \\ \hline 
\end{tabular}
\quad \left(\begin{tabular}{c|c|c|c|c|c|c|c|c|c|c}1.000& 0.959& 0.862& 0.725& 0.537& 0.338& 0.186& 0.091& 0.034& 0.001& -0.020\\ 
0.959& 1.000& 0.966& 0.861& 0.664& 0.423& 0.224& 0.091& 0.006& -0.048& -0.084\\ 
0.862& 0.966& 1.000& 0.957& 0.802& 0.569& 0.357& 0.206& 0.105& 0.037& -0.010\\ 
0.725& 0.861& 0.957& 1.000& 0.937& 0.768& 0.583& 0.437& 0.334& 0.261& 0.208\\ 
0.537& 0.664& 0.802& 0.937& 1.000& 0.942& 0.824& 0.713& 0.626& 0.561& 0.512\\ 
0.338& 0.423& 0.569& 0.768& 0.942& 1.000& 0.966& 0.906& 0.848& 0.802& 0.764\\ 
0.186& 0.224& 0.357& 0.583& 0.824& 0.966& 1.000& 0.984& 0.956& 0.928& 0.903\\ 
0.091& 0.091& 0.206& 0.437& 0.713& 0.906& 0.984& 1.000& 0.993& 0.979& 0.965\\ 
0.034& 0.006& 0.105& 0.334& 0.626& 0.848& 0.956& 0.993& 1.000& 0.996& 0.989\\ 
0.001& -0.048& 0.037& 0.261& 0.561& 0.802& 0.928& 0.979& 0.996& 1.000& 0.998\\ 
-0.020& -0.084& -0.010& 0.208& 0.512& 0.764& 0.903& 0.965& 0.989& 0.998& 1.000\\ 
\end{tabular}\right) \nonumber
  }\end{eqnarray}
 
\begin{eqnarray} && \qquad \qquad \qquad \qquad   \qquad \qquad \qquad \qquad \qquad   F_V \quad {\rm Correlation} \quad  {\rm  Matrix} \nonumber \\ &&\footnotesize{
\begin{tabular}{|c|c|c|}
\hline
 $x_\gamma$  & $F_V $  & $\Delta_{F_V} $\\ \hline
0 &  -0.120018 &  0.0155225 \\ \hline 
0.1 &  -0.0989568 &  0.0117214 \\ \hline 
0.2 &  -0.0824261 &  0.00951697 \\ \hline 
0.3 &  -0.0684115 &  0.00783577 \\ \hline 
0.4 &  -0.05594 &  0.00806476 \\ \hline 
0.5 &  -0.0444834 &  0.0110619 \\ \hline 
0.6 &  -0.03373 &  0.0158477 \\ \hline 
0.7 &  -0.0234841 &  0.0215682 \\ \hline 
0.8 &  -0.0136163 &  0.0278383 \\ \hline 
0.9 &  -0.00403801 &  0.0344741 \\ \hline 
1 &  0.00531392 &  0.0413737 \\ \hline 
\end{tabular}
\quad \left(\begin{tabular}{c|c|c|c|c|c|c|c|c|c|c}1.000& 0.933& 0.886& 0.898& 0.782& 0.543& 0.379& 0.288& 0.236& 0.205& 0.184\\ 
0.933& 1.000& 0.989& 0.935& 0.667& 0.317& 0.107& -0.004& -0.066& -0.103& -0.127\\ 
0.886& 0.989& 1.000& 0.955& 0.679& 0.314& 0.093& -0.024& -0.090& -0.130& -0.157\\ 
0.898& 0.935& 0.955& 1.000& 0.862& 0.571& 0.369& 0.255& 0.188& 0.147& 0.119\\ 
0.782& 0.667& 0.679& 0.862& 1.000& 0.908& 0.787& 0.707& 0.656& 0.623& 0.600\\ 
0.543& 0.317& 0.314& 0.571& 0.908& 1.000& 0.973& 0.938& 0.912& 0.893& 0.879\\ 
0.379& 0.107& 0.093& 0.369& 0.787& 0.973& 1.000& 0.993& 0.982& 0.972& 0.965\\ 
0.288& -0.004& -0.024& 0.255& 0.707& 0.938& 0.993& 1.000& 0.998& 0.993& 0.990\\ 
0.236& -0.066& -0.090& 0.188& 0.656& 0.912& 0.982& 0.998& 1.000& 0.999& 0.997\\ 
0.205& -0.103& -0.130& 0.147& 0.623& 0.893& 0.972& 0.993& 0.999& 1.000& 1.000\\ 
0.184& -0.127& -0.157& 0.119& 0.600& 0.879& 0.965& 0.990& 0.997& 1.000& 1.000\\ 
\end{tabular}\right) \nonumber
  }\end{eqnarray}
 
\end{table}
\FloatBarrier
\mbox{}
\FloatBarrier
\newpage
%ooooooooooooooooooooooooooooooooooooooooooooooooooooooooooooooooooooooooooooooooooooooooooooo

%ooooooooooooooooooooooooooooooooooooooooooooooooooooooooooooooooooooooooooooooooooooooooooooo
\end{document}